\documentclass[aps,prd,preprint,nofootinbib,showpacs]{revtex4-1}
\usepackage[colorlinks]{hyperref}
\usepackage{amsmath,amssymb}
\usepackage{url,color}
\usepackage{graphicx}
\usepackage{slashed}
\usepackage{dsfont}
\usepackage{float}
\usepackage{booktabs} 
\usepackage{caption}

\usepackage{array}
\newcolumntype{M}[1]{>{\centering\arraybackslash}m{#1}}
\newcolumntype{N}{@{}m{0pt}@{}}
\makeatletter
\newif\if@preliminary
\@preliminaryfalse
\def\preliminary{\@preliminarytrue}
\def\preprintno#1{\def\@preprintno{#1}}
\def\address#1{\def\@address{#1}}
\def\email#1#2{\thanks{\tt #1@{}#2}}
\long\def\@makecaption#1#2{%
  \vskip\abovecaptionskip
  \sbox\@tempboxa{#1: \emph{#2}}%
  \ifdim \wd\@tempboxa >\hsize
    #1: \emph{#2}\par
  \else
    \hbox to\hsize{\hfil\box\@tempboxa\hfil}%
  \fi
  \vskip\belowcaptionskip}
\makeatother

\newcommand{\ii}{\mathrm{i}}

\newcommand{\LL}{\mathcal{L}}
\newcommand{\com}[2]{\left [ #1 , #2 \right ] }

\newcommand{\vW}{\mathbf{W}}

\newcommand{\vH}{\mathbf{H}}

\newcommand{\tr}[1]{\operatorname{tr}\left[#1\right]}

\newcommand{\vB}{\mathbf{B}}
\newcommand{\vD}{\mathbf{D}}

\newcommand{\GeV}{\text{GeV}}
\newcommand{\TeV}{\text{TeV}}
\newcommand{\ab}{\text{ab}}

\newcommand{\amp}{\mathcal{A}}
\newcommand{\bss}{\begin{tiny}}
\newcommand{\ess}{\end{tiny}}

\begin{document}

\preprint{DESY 15-183}
\preprint{SI-HEP-2015-22}
\preprint{KA-TP-19-2015}
\preprint{KEK Preprint 2015-1869}

 \title{%
   Resonances at the LHC beyond the Higgs boson:\\The scalar/tensor Case}

 \author{Wolfgang Kilian}
 \email{kilian}{physik.uni-siegen.de}
 \affiliation{Department of Physics, %
   University of Siegen, %
   D--57068 Siegen, Germany} 
 \author{Thorsten Ohl}
 \email{ohl}{physik.uni-wuerzburg.de}
 \affiliation{Faculty of Physics and Astronomy, %
   W\"urzburg University, %
   D--97074 W\"urzburg, Germany}
 \author{J\"urgen Reuter}
 \email{juergen.reuter}{desy.de}
 \affiliation{DESY Theory Group, %
   D--22603 Hamburg, Germany}
 \author{Marco Sekulla}
 \email{marco.sekulla}{kit.edu}
 \affiliation{Department of Physics, %
   University of Siegen, %
   D--57068 Siegen, Germany}
 \affiliation{High Energy Accelerator Research Organization (KEK), %
   Tsukuba, Ibaraki 305-0801, Japan}
 \affiliation{Institute for Theoretical Physics, %
   Karlsruhe Institute of Technology, %
   D--76128 Karlsruhe, Germany}


\begin{abstract}
  We study in a bottom-up approach the
  theoretically consistent description of additional
  resonances in the electroweak sector beyond the discovered Higgs
  boson as simplified models.  We focus on scalar 
  and tensor resonances. Our formalism is suited for strongly
  coupled models, but can also be applied to weakly interacting
  theories. The
  spurious degrees of freedom of tensor resonances that would lead to
  bad high-energy behavior are treated using a generalization of the
  St\"uckelberg formalism.  We calculate scattering
  amplitudes for vector-boson and Higgs boson pairs. The high-energy
  region is regulated by the T-matrix unitarization procedure, leading
  to amplitudes that are well behaved on the whole phase
  space.  We
  present numerical results for complete partonic processes that involve
  resonant vector-boson scattering, for the current and
  upcoming runs of LHC.
\end{abstract}


\maketitle



\tableofcontents


\section{Introduction}
\label{sec:intro}
	
After the discovery of a $125\;\GeV$ Higgs boson, phenomenological
high-energy physics has entered a new era.  The new particle fits the
expectation of the minimal Standard Model (SM).  This model is thus
established as an effective field theory (EFT) that correctly
describes all current particle data (except for still missing possible
particle signals for dark matter and additional CP violation).  We
know about high-energy scales where 
the effective theory eventually breaks down --- the scale of neutrino
mass generation, the Planck scale --- but those are far outside the
reach of collider physics.  The hierarchy between those scales and the
electroweak symmetry breaking scale, combined with the fact that all
known elementary particles are weakly interacting, puzzles us due to
the apparent
fine-tuning in perturbative renormalization.  However, the hierarchy
puzzle as such has no phenomenological consequences.  In principle,
the SM may provide a complete description of all present and future
collider data, limited just by our ability to do calculations.

Nevertheless, the apparent success of the SM does not imply that we
have full control over the spectrum at presently accessible energies,
say between 100 GeV as the electroweak mass scale and a few TeV.
First of all, there is the possibility of extra light weakly
interacting particles which escape detection at the LHC.
We will not consider this in the present work but investigate new
physics above the mass scale of $W$, $Z$, and Higgs.

The SM is complete as a renormalizable theory and weakly interacting.
Hence, it provides a mechanism for suppressing the impact
of new physics on observables.  This fact is generally expressed by the
decoupling theorem~\cite{Appelquist:1974tg}:  All heavy particles (heavy
compared to the masses of $W,Z$, Higgs) can be integrated out, and
their physical effects are suppressed by powers of $m/M$ or $E/M$,
where $E$ is the effective energy of the measured elementary
interaction, and $M$ is the mass scale associated with new physics.
The EFT approach, which has been widely adopted for precision LHC
analyses, encodes this in a Lagrangian which contains operators of
dimension six and, in some cases, eight or even
higher~\cite{Hagiwara:1993ck}.  Decoupling of scalar particles in the
case of Two-Higgs doublet models (2HDM) has been considered
in~\cite{Gunion:2002zf}, as well as in~\cite{Haber:1989xc,Haber:2013mia}.

For a new particle with a mass of $1\;\TeV$, the leading corrections to SM
particle properties are at the percent level and below.  This is a
challenge for LHC analyses.  On the other hand, in scattering
processes at the LHC, the partonic energy $E$ can easily enter the
$\TeV$ range, so direct detection is favored.  Various classes of
new-physics models with extended fermion and gauge sectors can be
excluded up to several $\TeV$.  However, the current experimental
sensitivity on details of the Higgs/Nambu-Goldstone sector is still
marginal.  This is due to the fact that the effective energy available
for vector-boson scattering in LHC collisions, for instance, is
severely suppressed by steeply falling quark and $W/Z$ structure
functions.

In this paper we study new physics that is coupled to the
Higgs/Nambu-Goldstone sector and manifests itself in scattering
processes of $W$, $Z$, and Higgs particles.  The Higgs particle does
not occur in the initial state and has its own experimental issues, so
we restrict the discussion to Nambu-Goldstone
bosons~\cite{Nambu:1960xd,Goldstone:1961eq,Goldstone:1962es}, which the
Nambu-Goldstone boson equivalence
theorem~\cite{Vayonakis:1976vz,Chanowitz:1985hj,Gounaris:1986cr,Yao:1988aj,Bagger:1989fc,He:1992nga,He:1993qa,He:1993yd,He:1996cm}
relates to longitudinally 
polarized $W$ and $Z$ bosons.  That is, we investigate processes of
the class $V^*V^*\to VV$ ($V=W^\pm,Z,H$), where the initial vector
bosons are radiated almost on-shell and collinear off initial
energetic quarks in the colliding protons.

\subsection{New effects in vector-boson scattering}

Vector-boson scattering (VBS) as a physical process in hadronic
collisions has been observed recently by the ATLAS and CMS
collaborations~\cite{Aad:2014zda,ATLAS:2014rwa,CMS:2015jaa}.
The SM prediction has been confirmed, but the initial limits on extra
interactions are still rather weak, probing an energy scale close to
the pair-production threshold of $\sim 200\;\GeV$.  With higher energy
and better precision becoming available at the LHC, and at future
lepton and hadron colliders, data will become much more sensitive to
new effects in this sector.  There is no reason to
restrict the modelling to weak interactions.  In fact, the initially
limited experimental resolution and energy reach encourages us to consider
new strong interactions, as such deviations from the SM are
experimentally most accessible.

For decades, the theory of VBS processes has been the subject of a
vast literature, first in the disguise of the low-energy
theorem~\cite{Weinberg:1966kf,Chanowitz:1987vj}, for questions of
unitarity~\cite{Lee:1977yc,Chanowitz:1985hj,Espriu:2014jya,Delgado:2015kxa}
and as a means of phenomenological
studies~\cite{Appelquist:1980vg,Longhitano:1980iz,Dicus:1990fz,Barger:1990py,Berger:1991uj,Gupta:1993tq,Appelquist:1993ka,Chanowitz:1995tn,Bagger:1995mk,Butterworth:2002tt,Eboli:2003nq,Eboli:2006wa,Distler:2006if,Han:2009em,Freitas:2012uk,Espriu:2012ih,Doroba:2012pd,Chang:2013aya,Delgado:2014ixa,Fabbrichesi:2015hsa}. A
review of recent work can be found
in~\cite{Szleper:2014xxa}. Most of those studies were 
tailored to the Higgs-less case, which is by now excluded.  In the
presence of a light Higgs, in the SM, all VBS processes are
perturbative and
respect unitarity at all energies.  This situation changes drastically
once non-SM interactions are present.

Regarding the possible scenarios of new physics affecting VBS, there
are no significant restrictions from low-energy data or from the
absence of LHC discoveries.  Asymptotically, the process is
determined by the amplitudes of Nambu-Goldstone boson scattering,
where the initial state contains an even number of Nambu-Goldstone bosons
and thus no half-integer representations of $SU(2)_L$.  Any bosonic
excitation coupling to this state also has integer $SU(2)_L$ quantum numbers
and thus cannot couple left-handed with right-handed SM fermions.
In the limit of exact electroweak symmetry,
VBS processes and ordinary SM (fermionic) processes thus probe
distinct areas of new physics.
Electroweak symmetry breaking mixes those
sectors, but the mixing terms are again suppressed by the electroweak
scale (in operators, by additional factors of the Higgs doublet), and
are therefore subleading.

The only important constraint is quantum-mechanical unitarity, which
is severely violated in a perturbative calculation if we naively
insert the dimension-eight operators of the EFT.  We have
discussed this fact in detail in
Ref.~\cite{Alboteanu:2008my,Kilian:2014zja} and proposed a 
framework of unitarization which allows us to augment the SM in an
arbitrary way, while maintaining high-energy unitarity and
simultaneously matching the new effects to the low-energy EFT.  We
will adopt this framework, the \emph{T-matrix scheme}, for the concrete
models below.

\subsection{Outline of the present paper}

Extending the work of~\cite{Kilian:2014zja}, in the present paper we
consider a wider class of scenarios beyond the SM and beyond the
electroweak mass scale.  Instead of just extrapolating the EFT, which
generically leads to asymptotic saturation of amplitudes, we add new
states.  The quantum numbers of the new states are chosen such that
they retain unsuppressed interactions with the VBS system in the limit
of vanishing gauge couplings.  As mentioned above, this implies a certain set
of quantum-number assignments and, incidentally, suppresses their
couplings to the SM fermion sector.  We may consider strongly coupled
states, which we would classify as resonances in analogy with mesons
in QCD, or weakly coupled states which we would call new elementary
particles.  There is a continuous transition between these
extremes, such that we can cover all cases on equal footing.

We defer the discussion of vector resonances to a future publication,
since those states mix, after EWSB, with $W$ and $Z$ bosons and thus
exhibit a possibly different phenomenology.  This limits the model to
four distinct cases, namely scalar and tensor resonances with two
different assignments of electroweak quantum numbers, respectively.
We embed these states in an extended EFT and match this to the
low-energy EFT where the resonances are integrated out.  For the
high-energy limit, we apply the T-matrix scheme which keeps the model
within unitarity bounds when it eventually becomes strongly
interacting at energies above the resonance.

The case of a tensor resonance requires special considerations.  While
renormalizable weakly interacting theories cannot include elementary
tensor particles, it is nevertheless possible to set up an effective
theory which contains a tensor particle and remains weakly interacting
over a considerable range of energies.  This has been observed in the
context of gravity in extra
dimensions~\cite{Antoniadis:1990ew,ArkaniHamed:1998rs,Antoniadis:1998ig,Han:1998sg}, where massive tensor particles arise in the low-energy effective theory. 
Massive gravitons provide a very specific pattern of couplings
to the Higgs doublet, gauge bosons and fermions.  We will set up a
more generic model where such relations are absent, and construct a
Lagrangian description of St\"uckelberg type, where we can separate
the genuine tensor resonance with a controlled high-energy behavior
from unrelated higher-dimensional operators that become relevant
asymptotically.  The massive-graviton model emerges as a special case.
(Massive) higher-spin fields have been discussed e.g. in~\cite{Bouatta:2004kk,Singh:1974qz,Buchbinder:2005ua,Huang:2005js}.  

Given the observation that new resonances cannot necessarily be
distinguished from asymptotic saturation if the resonance energy is high and
event rates are low, we may ask the question whether the two cases
are distinguishable, i.e., whether a resonance model yields a
different prediction from a EFT extrapolation with specific
coefficients.  We will discuss this issue in an exemplary way for
specific parameter sets.  Furthermore, the new model allows for weakly
coupled resonances that do not leave a significant trace in the
low-energy EFT, but could nevertheless lead to a visible signal in
collider data.

To obtain numerical results, we take the unitarized model, which is
originally formulated in the gaugeless limit, re-insert gauge
couplings and continue the amplitudes off-shell along the lines
of~\cite{Kilian:2014zja}.  This allows us to set up a model definition for a
Monte-Carlo integrator and event generator, which we use to generate
partonic event samples for the LHC, cross sections and physical
distributions. A more detailed elaboration of the 
calculations can be found in \cite{Sekulla:2015}.

\section{Extended Effective Field Theory (EFT)}

\subsection{Low-Energy EFT}

We are going to develop models for the high-energy behavior of
scattering amplitudes of SM particles.  This cannot be done without
precisely stating the assumptions that go into those models, and to
cast them into convenient notation and parameterization.

First of all, we assume that the SM is a reasonable
low-energy effective theory.  That is, a weakly interacting
(Lagrangian) gauge field theory with spontaneous $SU(2)_L\times
U(1)_Y\to U(1)_{EM}$ symmetry breaking mediated by a complex Higgs
doublet, supplemented by the standard sets of quarks and leptons,
describes all particle-physics data at and below the electroweak scale
to a good approximation.

Regarding the interactions of fermions and vector bosons, this
conclusion can be drawn from the impressive success in fitting
electroweak and flavour data to the SM.  We cannot yet be so sure in
the Higgs sector proper.  While the Higgs boson was discovered in
accordance with the mass range that the precision analysis of
electroweak observables suggests, there is still room for sizable
deviations from the SM predictions for its couplings.  In particular,
the Higgs
self-couplings have not been measured at all.  Nevertheless, we will
assume that those couplings are close to their SM values, such that
deviations can be attributed to higher-dimensional terms in the EFT.
Future data from LHC and beyond will tell whether this is true.  If
not, we may generalize our findings to a nonlinearly realized Higgs
sector.  We have set up our parameterization such that this would
cause few modifications in the calculations.

A second assumption regards the low-energy spectrum: we assume that
there are no additional light particles, such as Higgs singlets or extra
doublets, below the EW scale.  If this was not true, it would not
invalidate the extended-EFT approach, but require the low-energy EFT
to be revised in order to include extra particles as building blocks.
Again, the model extensions discussed here would remain unchanged, but
we could expect a richer phenomenology of final states that emerge
from couplings to the extra light particles.  

\subsection{Including Resonances}

We want to describe massive tensor and scalar resonances as extensions
of the SM, coupled to the scattering channels accessible in VBS.  We
start from the low-energy EFT, the SM with higher-dimensional
operators included, and add a resonance with appropriate spin and
gauge quantum numbers to the Lagrangian.  Requiring the assumed
symmetries to be manifest, uniquely determines the form of the
couplings, again in an EFT sense, i.e.~as an power series expansion of
operators in some inverse mass scale~$\Lambda$.

It is tempting to identify $\Lambda$ with the resonance mass~$M$.
This would imply 
arbitrary strong interactions at the mass scale of the resonance.  The
form of couplings would be arbitrary since for $E\approx M=\Lambda$,
there is no
viable power expansion, and there are no reliable predictions.  While
this is a conceivable scenario, we rather consider a more economical
setup where the resonance at mass $M$ can be separated from other
effects which are attributed to an even higher scale $\Lambda$.  As we
will show below, it is possible and consistent to choose $\Lambda\gg M$,
both for scalar and tensor states.  $\Lambda$ is then the appropriate
scale for all higher-dimensional operators in the extended EFT.  In
the low-energy EFT, integrating out the resonance yields well-defined
higher-dimensional couplings suppressed by powers of $M$, which
combine with the undetermined $\Lambda$-suppressed coefficients
inherited from the extended EFT.  Depending on their relative
magnitude, we may --- or may not --- be able to relate the operator
coefficients in the low-energy EFT to the resonance couplings of the
extended EFT.

\section{Resonances: Spin classification}
\label{sec:resonances}

\subsection{Scalar Resonances}

A new massive spin-zero state might appear as another Higgs boson.
Indeed, a new Higgs singlet $\phi$ can couple to the SM
Higgs doublet
$\vH$ via the renormalizable
operators $\tr{\vH^\dagger \vH}\phi$ and $\tr{\vH^\dagger \vH}\phi^2$,
while a new Higgs doublet $\vH'$ can couple via $\tr{\vH^\dagger\vH'}^2$ and
$\tr{\vH^\dagger\vH}\tr{\vH'{}^\dagger\vH'}$.\footnote{%
  For notational conventions, cf.\ appendix~\ref{appendix:fields}.}
These terms contribute
to Higgs mixing and self-interactions, but not directly to VBS.  In
the EFT formalism, the observed Higgs boson is the only light scalar
by definition, and in the renormalizable part of the Lagrangian it saturates
the vector-boson couplings.  Coupling an extra scalar to VBS then
requires two Higgs-field derivatives $\vD_\mu\vH$ and thus introduces
an effective dimension-five operator.

In a renormalizable extension of the SM Higgs sector, after
diagonalization new Higgses may eventually appear in VBS processes.
However, we have just noted that in the EFT formalism, their couplings are
higher-dimensional and thus power-suppressed.  This is an incarnation
of the Higgs decoupling theorem~\cite{{Ball:1994ve}}.

Renormalizability corresponds to the existence of special trajectories
in parameter space, where all irrelevant (i.\,e.~higher-dimensional)
operators can be removed simultaneously from the Lagrangian by a
nonlinear field redefinition. Without a good reason {\em a priori} for
allowing only points on these trajectories, 
we consider the renormalizable (possibly weakly interacting) case as a
special case that is included in the general framework.  This applies,
in particular, to Higgs sector extensions by singlets and doublets, as
long as the extra scalars can be considered heavy in the sense of the
EFT formalism.

For our purposes, the phenomenology of generic scalar resonances is
then very similar to tensor resonances (see below), namely
breaking the renormalizability of the SM and inducing
higher-dimensional operators both in the low-energy EFT where they are
integrated out, and in the high-energy model where they appear
explicitly in the phenomenological Lagrangian.  We will have to apply
a unitarization framework in the energy range at and beyond the resonance.

\subsection{Tensor Resonances: Fierz-Pauli formalism}
\label{sec:fierz-pauli}

We now turn to massive spin-two particles, postponing spin-one for
later investigations, as stated above.

The physical particle corresponds to an irreducible
representation of the rotation group in its own rest frame and thus
consists of five component fields, mixing under rotation.
Strictly speaking, there is no reason to develop a relativistic field
theory for a generic interacting spin-two particle.  If there is no UV
completion of the interacting model, it is not possible to construct a
complete Hilbert space and unitary scattering matrix.  However, for
convenience of calculation, it is clearly advantageous to embed the
tensor particle in the usual relativistic field-theory context of the
EFT for the SM.  We therefore introduce extra
fields, coupled to currents built from SM fields in a Lorentz- and
gauge-invariant way, in a Lagrangian formalism.

For the scalar case, this is straightforward since a spin-zero
particle is represented by a Lorentz scalar field that also has a
single component.  In the tensor case, we have to deal
with the fact that the appropriate Lorentz representation has more
than five components.
In the rest frame, the Lorentz symmetry (or its universal cover
$SL(2,\mathbb{C})$) is kinematically broken down to its $SU(2)$
subgroup, the universal cover of
the rotation symmetry.  The Lorentz decuplet decomposes into the
irreducible spin states
\begin{equation}
\label{eq:decomposition}
  \text{symmetric tensor} \to \text{spin states}\ (2) + (1) + (0) +
  (0) \quad .
\end{equation}
Looking at the symmetric rest-frame polarization tensor
$\varepsilon^{\mu\nu}$, the irreducible parts correspond to the
components $\varepsilon^{ij}$ (traceless), $\varepsilon^{0i}$,
$\varepsilon^{00}$, and $\sum\varepsilon^{ii}$ (trace), respectively.  Under
the full Lorentz group, $\varepsilon^{\mu\nu}$ is also reducible and
decomposes into the traceless and trace parts.  However, in the
presence of interactions it is not straightforward to maintain this
decomposition for off-shell
amplitudes~\cite{Fronsdal:1958,Weinberg:1964cn,Huang:2005js}. 

Our model setup requires that, on-shell, only the pure spin-two state
propagates.  If we represent the resonance by a single field, the
tensor-field propagator must reduce to the form \cite{Weinberg:1964cn}
\begin{equation}
\label{eq:prop-FP}
  G_f^{\mu\nu,\rho\sigma}(k) =
  \frac{\ii\sum_\lambda\bar\varepsilon_{(\lambda)}^{\mu\nu} (k,m)
    \; \varepsilon_{(\lambda)}^{\rho\sigma} (k,m)}
  {k^2 - m_f^2 + i\epsilon} + \text{non-resonant}
\end{equation}
Here, $\lambda$ sums over a basis of five real-symmetric, mutually
orthogonal polarization tensors that satisfy the
constraints
\begin{equation}
\label{eq:epsilon-constraints}
  k_\mu \varepsilon_{(\lambda)}^{\mu\nu} (k,m) = 0,
  \qquad
  \varepsilon^{\;\;\;\mu}_{(\lambda)\;\mu} (k,m) = 0,
\end{equation}
as long as $k$ is an on-shell momentum vector, $k^2=m^2$.

The solution to this problem is unique up to the non-resonant part
\cite{Huang:2005js}, 
\begin{align}
\label{eq:propagatortensor}
  G_f^{\mu_1\mu_2,\nu_1\nu_2}
  &= \ii \frac{P^{\mu_1\mu_2,\nu_1\nu_2}(k,m)}{k^2-m^2+\ii \epsilon}
      + \text{non-resonant},
\end{align}
where the projection operator of spin-two can be written in terms of the
spin-one projection operator,
\begin{align}
  P^{\mu_1\mu_2,\nu_1\nu_2}(k,m)
  &=
  \sum_\lambda \bar{\varepsilon}_{(\lambda)}^{\mu_1\mu_2}(k,m) \,
  \varepsilon_{(\lambda)}^{\nu_1\nu_2}(k,m)
  \notag \\
  &=
  \frac{1}{2}\biggl[ P^{\mu_1\nu_1}(k,m)P^{\mu_2\nu_2}(k,m) +
  P^{\mu_1\nu_2}(k,m)P^{\mu_1\nu_2}(k,m)\biggr] 
  \notag \\
  &\quad
  \qquad\qquad - \frac{1}{3} P^{\mu_1\mu_2}(k,m)P^{\nu_1\nu_2}(k,m),
\end{align}
with
\begin{align}
  P^{\mu\nu}(k,m)
  &=
  \sum_\lambda \bar{\varepsilon}_{(\lambda)}^{\mu}(k,m) 
               \varepsilon_{(\lambda)}^{\nu}(k,m) 
  = g^{\mu\nu}-\frac{k^\mu k^\nu}{m^2}.
\end{align}
This propagator, with vanishing non-resonant part, can be obtained from the
free Fierz-Pauli Lagrangian~\cite{Fierz:1939ix,Singh:1974qz} coupled
to a tensor source $J_f^{\mu\nu}$
\begin{align}
\label{eq:lagrangiantensor}
  \LL =&\frac{1}{2} \partial_\alpha f_{\mu\nu}\partial^\alpha f^{\mu\nu}
      - \frac{1}{2} m^2 f_{\mu \nu}f^{\mu \nu} \notag\\
     & \;- \partial^\alpha f_{\alpha\mu} \partial_\beta f^{\beta\mu}
      - f^\alpha_{\;\alpha}\partial^\mu\partial^\nu f_{\mu\nu}
      - \frac{1}{2} \partial_\alpha f^\mu_{\;\mu}\partial^\alpha f^\nu_{\;\nu}
      + \frac{1}{2} m^2 f^\mu_{\;\mu} f^\nu_{\;\nu}
      + f_{\mu\nu}J_f^{\mu\nu}.
\end{align}
In the classical theory, the Lagrangian~\eqref{eq:lagrangiantensor}
enforces the conditions
\begin{equation}
  \partial_\mu f^{\mu\nu} = 0
  \quad\text{and}\quad
  f^\mu_{\;\mu} = 0\,.
\end{equation}
This is, in principle, a valid Lagrangian description of a tensor
resonance.  However, since we have to deal with off-shell amplitudes for an
effective theory, it will be useful to investigate the role of
various terms in more detail.  Returning to the
propagator~(\ref{eq:propagatortensor}), there are momentum factors
$k^\mu$ in different combinations that project out the proper spin-two
part on the pole.  Going to lower energies, these factors vanish
more rapidly than the $g^{\mu\nu}$ terms and therefore reduce to
operators of higher dimension.  Beyond the resonance, they will rise
more rapidly and therefore potentially provide the dominant part that
enters the unitarization prescription.

\subsection{Tensor Resonances: St\"uckelberg formulation}
\label{sec:stueckelberg}

As discussed above,
the extra momentum factors in the spin-two propagator represent the
mismatch between the $SO(3)$ little group representation of massive
on-shell particles and the full Lorentz-group off-shell
representations in a relativistic description.  This is in analogy 
with a massive spin-one boson, which in the relativistic case acquires an
extra zero component. In the following, we identify the extra degrees
of freedom for a propagating spin-two object and separate them for the
purpose of power-counting in an actual calculation.

To this end, inspired by the spin-one case, we will use the
so-called St\"uckelberg formulation for 
tensor resonances. This has been studied in the context of effective field
theories for massive
gravity~\cite{ArkaniHamed:2002sp,ArkaniHamed:2003vb,Schwartz:2003vj},
\cite{deRham:2011rn,Hassan:2012qv} and \cite{Noller:2013yja}. The work along
these lines has been nicely reviewed in~\cite{Hinterbichler:2011tt}.

Given an arbitrary symmetric polarization tensor
$\varepsilon^{\mu\nu}$ that is not restricted by auxiliary conditions,
we can subtract terms constructed from momenta, vector and scalar
polarizations
\begin{equation}
  \varepsilon^{\prime\;\mu\nu} = \varepsilon^{\mu\nu}
  -
  \frac1m(k^\mu\varepsilon_V^\nu + k^\nu\varepsilon_V^\mu)
  -
  \frac{k^\mu k^\nu}{m^2} \varepsilon_S
  -
  g^{\mu\nu}\varepsilon_T,
\end{equation}
and demand that (i) the Fierz-Pauli polarization tensor
$\varepsilon'{}^{\mu\nu}$ satisfies the on-shell
constraints~(\ref{eq:epsilon-constraints}), and (ii) the vector
polarization is transversal
$k_\mu\varepsilon_V^\mu=0$.  The resulting vector and scalar polarizations
$\varepsilon_V$, $\varepsilon_S$, $\varepsilon_T$ can be expressed as
contractions of the original $\varepsilon^{\mu\nu}$,
\begin{subequations}
\label{eq:vst-polarizations}
\begin{align}
  \varepsilon_V^\mu
  &=
  \frac{1}{m}\left(k_\nu\varepsilon^{\mu\nu}
    - \frac{1}{m^2}k^\mu k_\nu k_\rho\varepsilon^{\nu\rho}\right),
  \\
  \varepsilon_S
  &=
  \frac13\left(4\frac{k_\mu k_\nu}{m^2} - g_{\mu\nu}\right)
  \varepsilon^{\mu\nu},
  \\
  \varepsilon_T
  &=
  \frac13\left(g_{\mu\nu} - \frac{k_\mu k_\nu}{m^2}\right)
  \varepsilon^{\mu\nu}.
\end{align}
\end{subequations}
Formally, this subtraction removes the extra representations in the
decomposition~(\ref{eq:decomposition}).  We note that this
prescription naturally extends to off-shell wave functions.

For the purpose of calculation, we can reproduce the
effect of the propagator (\ref{eq:propagatortensor}) if we remove all
$k^\mu$ factors from the tensor-field propagator but add a vector and
two scalar fields with their respective propagators.  To enforce the on-shell
relations~(\ref{eq:vst-polarizations}) for their polarization (i.e.,
wave function) factors, their interactions must be prescribed
by the original tensor interactions.  In field theory, such relations
can be enforced by demanding a gauge invariance.   Since the momenta
have been banished from the numerators of the propagators this way, the
power-counting in the resulting Feynman rules will be explicit, in
analogy with the 't Hooft-Feynman gauge of a gauge
theory.

St\"uckelberg~\cite{Stueckelberg:1900zz,Stueckelberg:1941th,Ruegg:2003ps}
originally formulated the algorithm
that systematically introduces the compensating fields together with the
extra gauge invariance in the Lagrangian formalism.  Applying the
algorithm to the massive tensor case, we start with the Fierz-Pauli
Lagrangian which corresponds to the minimal single-field propagator of
the pure spin-two tensor.  After removing any explicit constraints from
the tensor field, we introduce first the St\"uckelberg vector $A^\mu$
that cancels the $f^{0\mu}$ components, by the replacement
\begin{equation}
  f^{\mu \nu} \rightarrow 
  f^{\mu \nu} + \frac{1}{m}\partial^\mu A^\nu + \frac{1}{m}\partial^\nu A^\mu, 
\end{equation}
and then cancel the extra unwanted $A^0$ components that this field
introduces, together with $f^{00}$, by a St\"uckelberg
scalar~$\sigma$,
\begin{equation}
  A^\mu \rightarrow A^\mu + \frac{1}{m}\partial^\mu \sigma
\end{equation}
Finally, we introduce another St\"uckelberg scalar $\phi$ for
cancelling the trace by
\begin{equation}
  f^{\mu \nu} \rightarrow f^{\mu \nu} + g^{\mu \nu} \phi
\end{equation}
This scheme guarantees that the interactions of
the new fields in the Lagrangian are correctly related to the original
interactions of the tensor field.  The resulting Lagrangian exhibits
the gauge invariances that reflect the redundancy of the
St\"uckelberg fields and there is a gauge (called unitary gauge) in which
all St\"uckelberg fields vanish and the original Fierz-Pauli Lagrangian is
recovered. The new Fierz-Pauli Lagrangian with the additional scalar and
vector modes reads 
\begin{equation}
  \label{eq:FullStueckelberg}
  \begin{aligned}
    \LL \; = 
    & \quad \frac{1}{2} \partial_\alpha f_{\mu\nu}\partial^\alpha f^{\mu\nu}
    - \frac{1}{2} m^2 f_{\mu \nu}f^{\mu \nu} 
    - \partial^\alpha f_{\alpha\mu} \partial_\beta f^{\beta\mu}
    - f^\alpha_{\;\alpha}\partial^\mu\partial^\nu f_{\mu\nu}\\
    & \; - \frac{1}{2} \partial_\alpha f^\mu_{\;\mu}\partial^\alpha f^\nu_{\;\nu}
    + \frac{1}{2} m^2 f^\mu_{\;\mu} f^\nu_{\;\nu}
    - \partial_\mu A_\nu \partial^\mu A^\nu
    + \partial_\mu A^\mu \partial_\nu A^\nu \\
    & \; - 2 m f_{\mu \nu} \partial^\mu A^\nu
    + 2  m f^\mu_{\;\mu} \partial_\nu A^\nu 
    + 6 m \phi \partial_\mu A^\mu \\
    & \; - 2 f_{\mu \nu} \partial^\mu \partial^\nu \sigma
    + 2  f^\mu_{\;\mu} \partial^2 \sigma 
    - 2 f_{\mu \nu} \partial^\mu \partial^\nu \phi
    + 2 f^\mu_{\;\mu} \partial^2 \phi \\
    & \; -3 \partial_\mu \phi \partial^\mu \phi
    + 6 m^2 \phi \phi  
    + 3 m^2 f^\mu_{\;\mu} \phi
    \\
    &  \; + \left ( f_{\mu \nu}+g_{\mu \nu} \phi+
      \frac{2}{m}\partial_\mu A_\nu +
      \frac{2}{m^2}\partial_\mu \partial_\nu \sigma 
    \right )J_f^{\mu \nu}  \qquad .
  \end{aligned}
\end{equation}
The scheme simplifies slightly since both scalars are related to the
original tensor, so their interactions are not independent.  We can
choose the gauge
\begin{equation}
  \phi = -\sigma
\end{equation}
and arrive at a minimal St\"uckelberg Lagrangian~\cite{Bonifacio:2015rea}
(adjusted by partial integration and simplified),
\begin{equation}
  \begin{aligned}
    \LL \; = 
    & \quad \frac{1}{2} \partial_\alpha f_{\mu\nu}\partial^\alpha f^{\mu\nu}
    - \frac{1}{2} m^2 f_{\mu \nu}f^{\mu \nu}  \\
    & \; - \left (\partial^\alpha f_{\alpha \mu} -
      \frac{1}{2} \partial_\mu f^\rho_{\;\rho} 
      - m A_\mu \right ) ^2 \\
    & \; - \frac{1}{4} \partial_\alpha f^\mu_{\;\mu}\partial^\alpha f^\nu_{\;\nu}
    + \frac{1}{4} m^2 f^\mu_{\;\mu} f^\nu_{\;\nu}
    - \partial_\mu A_\nu \partial^\mu A^\nu
    + m^2 A_\mu A^\mu\\
    & \; + \left (
      \partial_\mu A^\mu - 3 m \sigma + \frac{1}{2}m f^\mu_{\;\mu}
    \right )^2 \\
    & \; +3 \partial_\mu \sigma \partial^\mu \sigma
    - 3 m^2 \sigma \sigma 
    \\
    & \; + \left ( f_{\mu \nu} -  g_{\mu \nu} \sigma +
      \frac{2}{m}\partial_\mu A_\nu + \frac{2}{m^2}\partial_\mu
      \partial_\nu \sigma 
    \right )J_f^{\mu \nu}  \qquad .
  \end{aligned}
\end{equation}
For perturbative calculations we have to fix the gauge up to
residual gauge transformations $\lambda(x)$ that decouple on-shell,
i.e.~satisfy the harmonic condition $(\partial^2+m^2)\lambda=0$.  To
this end, we choose linear gauge conditions,
\begin{subequations}
\begin{align}
  \partial_\mu A^\mu - 3 m \sigma + \frac{1}{2}m f^\mu_{\;\mu} &= 0 \\
  \partial^\alpha f_{\alpha \mu} - \frac{1}{2} \partial_\mu f^\rho_{\;\rho}
  - m A_\mu &= 0 
\end{align}
\end{subequations}
and end with a diagonalized Lagrangian,
\begin{equation}
  \begin{aligned}
    \LL \; = 
    & \quad \frac{1}{2}  f_{\mu\nu}\left (-\partial^2 -  m^2 \right
    )f^{\mu \nu}  \\ 
    & \; + \frac{1}{2}  f^\mu_{\;\mu}\left (-\frac{1}{2} \left
        (-\partial^2 - m^2 \right ) \right )f^\nu_{\;\nu} \\ 
    & \; + \frac{1}{2} A_\mu \left ( -2 \left (-\partial^2 -  m^2 \right )
    \right) A^\mu\\ 
    & \; +\frac{1}{2} \sigma \left ( 6 \left 
        ( -\partial^2 - m^2 \right ) \right )\sigma 
    \\
    & \; + \left ( f_{\mu \nu} -  g_{\mu \nu} \sigma +
      \frac{1}{m}\left (\partial_\mu A_\nu + \partial_\nu A_\mu \right )
      + \frac{2}{m^2} 		\partial_\mu \partial_\nu \sigma
    \right )J_f^{\mu \nu} . 
  \end{aligned}
\end{equation}
Next, we normalize the fields canonically
\begin{equation}
  \begin{aligned}
    \LL \; = 
    & \quad \frac{1}{2}  f_{\mu\nu}\left (-\partial^2 -  m^2 \right
    )f^{\mu \nu}  \\ 
    & \; + \frac{1}{2}  f^\mu_{\;\mu}\left (-\frac{1}{2} \left
        (-\partial^2 - m^2 \right ) \right )f^\nu_{\;\nu} \\ 
    & \; + \frac{1}{2} A_\mu \left ( \partial^2 +  m^2 \right )  A^\mu\\
    & \; +\frac{1}{2} \sigma \left (  -\partial^2 - m^2 \right ) \sigma 
    \\
    & \; + \left ( f_{\mu \nu} -  \frac{1}{\sqrt{6}} g_{\mu \nu} \sigma +
      \frac{1}{\sqrt{2}m}\left (\partial_\mu A_\nu + \partial_\nu A_\mu \right )
      + \frac{\sqrt{2}}{\sqrt{3}m^2} 		\partial_\mu \partial_\nu \sigma
    \right )J_f^{\mu \nu} 
  \end{aligned}
\end{equation}
and find the canonical propagators
\begin{align}
  \label{eq:proptensor}
  \Delta_{\mu\nu,\rho\sigma} (f) 
  &= \frac{\ii}{k^2 - m^2}
     \left(\frac{1}{2} g_{\mu\rho}g_{\nu\sigma} 
       +\frac{1}{2}g_{\mu\sigma}g_{\nu\rho}
       -\frac{1}{2} g_{\mu\nu}g_{\rho\sigma} 
     \right ) 
  \\
  \Delta_{\mu\nu}  (A) 
  &= \frac{-\ii}{k^2 - m^2} g_{\mu\nu} 
  \\
  \Delta  (\sigma) 
  &= \frac{\ii}{k^2 - m^2}
\end{align}
for the resulting unconstrained tensor, vector, and scalar fields,
respectively.\footnote{%
  For a complete formulation at the quantum level, the gauge-fixed
  Lagrangian has to be embedded in a BRST formalism.  Introducing
  appropriate Faddeev-Popov ghosts and auxiliary Nakanishi-Lautrup
  fields, the classical action can be rendered BRST invariant.  The
  quantum effective action with resonance exchange is then defined as the
  solution to a Slavnov-Taylor equation, to all orders in the EW
  perturbative expansion.  The gauge-fixing terms become BRST
  variations which do not contribute to physical amplitudes, and the
  St\"uckelberg fields combine with the ghosts and auxiliary fields to
  BRST representations that can be consistently eliminated from the Hilbert
  space.  For free fields, this procedure is detailed
  in~\cite{Buchbinder:2005ua}.
} 
As desired, these propagators do not contain any
momentum factors.  This fact turns out to be essential for a
Monte-Carlo calculation for physical processes, where all bosons are
off-shell in a generic momentum configuration.

\subsection{Tensor Resonances: Summary}

Given this lengthy derivation, we may ask again whether the
St\"uckelberg formulation has any advantage over the original
Fierz-Pauli Lagrangian.  Algebraically, both are equivalent and result
in identical on-shell amplitudes.

This should be viewed in analogy with massive vector bosons, for which
the St\"uckelberg approach reproduces the usual reformulation as a
spontaneously broken gauge theory.  Again, this is mathematically
equivalent to the original model, as has been pointed out
repeatedly~\cite{Burgess:1992gz}.  However, once the accessible
energy in a process exceeds the resonance mass, there is a conceptual
difference.  In the gauge-theory version, there is no
higher-dimensional operator with a $1/M$ coefficient.  Any additional
effects would come with a new cutoff $1/\Lambda$.  Scattering
amplitudes are bounded beyond the resonance as long as $\Lambda$ is
considered large.  By contrast, in the formulation with massive vector
bosons, there are $k^\mu/M$ terms in the propagator which {\em a priori}
require the inclusion of a whole series of operators with $1/M$
factors.  The model is strongly interacting from the onset and has no
predictivity.  If actual data show that interactions are indeed weak,
this fact would be interpreted as a fine-tuned cancellation among
terms.

Turning this argument around, \emph{if} a vector boson is observed to
interact weakly over a significant range of scales above its mass, it
is natural to describe it as a gauge boson,
which in turn determines the allowed interaction pattern.
Analogously, \emph{if} we assume that a tensor resonance interacts
weakly over a significant range of scales above its mass, it is
natural to describe it by the St\"uckelberg approach.  We will
therefore adopt the St\"uckelberg Lagrangian as the basis of a
tensor-EFT with a minimum set of free parameters.

Clearly, we can always add extra interactions with further free
parameters.  Those interactions take the form of higher-dimensional
operators which do not contribute on the resonance.  They describe
unrelated new-physics effects.

\section{Lagrangian for the extended EFT}

We now combine the findings of the previous section in order to set up a
Lagrangian description of the resonances, as an extension of the low-energy
EFT which already (implicitly) includes the complete set of higher-dimensional
operators.  Apart from the Lorentz representations as scalar or tensor, we have
to consider the representation of the internal symmetry group.  As we will
argue in detail below, we take this as the Higgs-sector global symmetry
$SU(2)_L\times SU(2)_R$, where only the $SU(2)_L\times U(1)_Y$ subgroup is
gauged.  $SU(2)_R$ breaking terms can be systematically included, but we do
not consider those in the present work.

\subsection{Isospin}
\label{sec:multiplets}

In the literature on VBS, resonances have traditionally been
categorized in terms of weak isospin, i.e., custodial $SU(2)_C$
multiplets.  This is appropriate for a Higgsless scenario, where the
actual scale of EWSB is given by its natural value $4\pi v\approx
3\;\TeV$ (cf. e.g.~\cite{EWLag}).  Without a light Higgs boson, VBS
scattering at the LHC would probe the physics at energies below the true EWSB
scale, so the (approximate) low-energy symmetry applies.

However, since the discovery of the Higgs boson, we know that VBS
processes probe a scale above the masses of the physical Higgs and the
electroweak gauge bosons.  We have to impose the unbroken high-energy
symmetry on the theoretical description.  Neglecting hypercharge, this
is $SU(2)_L\times SU(2)_R$.  We therefore describe new resonances
coupled to the SM Higgs sector in terms of
$SU(2)_L\times SU(2)_R$ multiplets.

It is not obvious that new physics coupled to the Higgs sector
actually has this symmetry.  $SU(2)_L\times SU(2)_R$ is, first of all,
an accidental approximate symmetry of the SM EWSB sector.  There are no possible
terms in the dimension-four Higgs potential that break $SU(2)_R$, so
EWSB leaves the diagonal custodial $SU(2)$ symmetry untouched.
However, hypercharge and top-quark couplings are not consistent with
$SU(2)_R$.  Nevertheless,
in the gaugeless limit the hypercharge coupling vanishes,
and top quarks are irrelevant for VBS anyway, so $SU(2)_R$ remains a
good symmetry of VBS (at high $p_T$) in the SM.  Beyond the SM, new
effects in VBS are transmitted only via the Higgs doublet.  In the
low-energy EFT, they require higher-dimensional operators.
These would cause power corrections to the $\rho$ parameter and are therefore
constrained by the observed agreement of the measured $\rho$ parameter
with the pure SM prediction.  For our purposes, we thus adopt
$SU(2)_R$ as a symmetry of new physics in the Higgs sector, to keep
things simple.  We have to keep in mind that this need not be the
case, and leave the discussion of $SU(2)_R$ breaking in this context
to future work.

Resonances of even spin with unsuppressed couplings to a pair of
Higgs/Nambu-Goldstone bosons, must reside in the symmetric part of the
decomposition of the product representation of the $SU(2)_L\times
SU(2)_R$ symmetry, $(\frac{1}{2},\frac{1}{2})\otimes
(\frac{1}{2},\frac{1}{2})$.  In the effective interaction operator,
this representation appears as a $H\otimes H^\dagger$ factor.  There
are only two possibilities: 
\begin{enumerate}
\item $(0,0)$: a neutral singlet (isoscalar).
\item $(1,1)$: a $3\times 3$ matrix, which contains nine components.
  After EWSB, the multiplet decomposes into an isotensor (five
  components), an isovector (three components), and an isoscalar (one
  component).  In terms of the gauged $SU(2)_L \times U(1)_Y$ subgroup, the
  nonet decomposes into a complex $SU(2)_L$ triplet with a doubly
  charged component and a real $SU(2)_L$ triplet, as described
  in~\cite{Kaplan:1983fs}.  The relative mass splitting between these
  states is of order $(m_W/M)^2$, where $M$ is the average resonance
  mass.  For our purposes where we assume $M\gg m_W$, we ignore that
  splitting and thus deal with a nonet of degenerate resonance
  components.
\end{enumerate}
We note that due to the existence of the light Higgs, the close
analogy between spin and isospin is broken at this point: tensor
states have just five physical degrees of freedom, but an isotensor
resonance in VBS, given the symmetry assumptions of the present paper, does not
exist in isolation.  The distinction comes into play
once physical Higgs bosons are involved in a process.  In VBS
amplitudes, the symmetry relates, for any given resonance multiplet,
Nambu-Goldstone pairs with Higgs pairs, i.e., $VV (V=W,Z)$ to $HH$ production.

For simplicity of notation, we will continue to denote the $(0,0)$
case as isoscalar and the $(1,1)$ as isotensor, respectively, keeping
in mind that the latter case actually is always accompanied by
isovector and isoscalar components.

For a scalar isoscalar resonance $\sigma$, we may consider couplings
of the form
\begin{subequations}
\begin{equation}
  \label{sHH}
  \sigma\tr{\vH^\dagger \vH}
\end{equation}
or
\begin{equation}
  \label{sDHDH}
  \sigma\tr{(\vD_\mu \vH)^\dagger (\vD^\mu \vH)}.
\end{equation}
\end{subequations}
The former operator is of lower dimension and might therefore be
considered the dominant contribution.  It is part of the Higgs
potential and influences Higgs mixing and production processes.  In
the present work, we assume that the scalar state has been broken down
in terms of the SM Higgs doublet and further states, which themselves
arrange as multiplets.  Since the SM Higgs couplings in the
lowest-order EFT, the pure SM, saturate the Higgs couplings to SM
particles and are fixed by definition, residual mixing and potential
terms arrange into higher-dimensional operators.  In particular, a
resonance coupled to Nambu-Goldstone bosons is represented by the 
term~(\ref{sDHDH}), while the lower-dimensional term (\ref{sHH}) does
not enter.   We therefore do not consider~(\ref{sHH}) and concentrate
on the dimension-five coupling~(\ref{sDHDH}).

This leads to a current for the scalar isoscalar resonance of the form 
\begin{align}
  J_{\sigma} &= F_\sigma
  \tr{ \left ( \vD_\mu \vH \right )^\dagger \vD^\mu \vH} \; .
\end{align}


\subsection{The Isotensor Representation}

While the description of an isoscalar is simple, we have to look at the
interactions of the isotensor more carefully.  For simplicity, we will
first restrict ourselves to a scalar field multiplet.

A resonance with chiral $SU(2)_L\times SU(2)_R$ quantum numbers $(1,1)$ has 
nine scalar degrees of freedoms. In the chiral representation these nine
degrees of freedom can be represented as the tensor $\Phi^{ab}$ with
the indices $a,b \in \{1,2,3\}$. Therefore, the Lagrangian
describing an isotensor resonance in the Nambu-Goldstone/Higgs
boson sector can be written as
\begin{align}
  \label{eq:Lagrangian_chiraltensor}
  \LL_{\Phi}= \frac{1}{2} \partial_\mu \Phi^{ab} \partial^{\mu}\Phi^{ab}
  - \frac{m_\Phi^2}{2} \Phi^{ab}\Phi^{ab}
  + J^{ab}_{\Phi}\Phi^{ab} \,
\end{align}
where the current has a $SU(2)_L$ and a $SU(2)_R$ index
\begin{align}
  \label{eq:Isotensor-current}
  J^{ab}_\Phi =
  F_\phi \tr{\left(\vD_\mu \vH\right)^\dagger \tau^a \vD^\mu \vH \tau^b}\, .
\end{align}
Analogously to the isoscalar case, the coupling $F_\phi$ is suppressed
by a new physics scale $\Lambda$. To expose the coupling structure to the
Nambu-Goldstone/Higgs boson sector, the current can be expanded in
the gaugeless limit 
\begin{align}
  \label{eq:isotensor_current_expanded}
  \begin{aligned}
    \tr{\left(\vD_\mu \vH\right)^\dagger \tau^a \vD_\nu \vH \tau^b}
    =& \frac{1}{2} \left(\partial_\mu h \partial_\nu h - \partial_\mu
      w^i \partial_\nu w^i \right) \delta^{ab}  
    -  \frac{1}{2}
    \left (\partial_\mu w^i \partial_\nu h+\partial_\nu
      w^i \partial_\mu h  \right)\varepsilon^{abi} \\ 
    &+\frac{1}{2} \left(\partial_\mu w^a \partial_\nu w^b
      + \partial_\mu w^b \partial_\nu w^a \right)
  \end{aligned}
\end{align}
Here, the decomposition into isotensor, isovector and isoscalar
is already manifest. The resonance $\Phi^{ab}$ can be represented in
a basis constructed from tensor products of $SU(2)$ generators by the
Clebsch-Gordon decomposition 
\begin{align}
  \mathbf{1} \otimes \mathbf{1} = \mathbf{2} + \mathbf{1} + \mathbf{0} \, .
\end{align}
Using the basis in the appendix~\ref{appendix:isospin-basis}, 
the resonance $\Phi^{ab}$ is
rewritten into its $SU(2)_C$ components 
\begin{align}
  \Phi^{ab} \rightarrow \Phi_t + \Phi_v + \Phi_s
\end{align}
with
\begin{subequations}
  \begin{align}
    \Phi_t &= \phi_t^{++}\tau_t^{++}+\phi_t^{+}\tau_t^{+}+
    \phi_t^{0}\tau_t^{0}+\phi_t^{-}\tau_t^{-}+\phi_t^{--}\tau_t^{--} \, ,\\
    \Phi_v &= \phi_v^{+}\tau_v^{+} +
    \phi_v^{0}\tau_v^{0}+\phi_v^{-}\tau_v^{-} \, ,\\
    \Phi_s &=		\phi_s\tau_s \, .
  \end{align}
\end{subequations}
The Lagrangian~\eqref{eq:Lagrangian_chiraltensor}
can be written in terms of the $SU(2)_C$ basis
\begin{subequations}
\label{eq:Lagrangian-Isotensor-full}
\begin{align}
  \LL_\phi =& \frac{1}{2}\sum_{i=s,v,t} \tr{(\partial_\mu \Phi_i)^\dagger
                                          \partial^\mu \Phi_i
    -m_\Phi^2 \Phi_i^2} 
  +\tr{\left(\Phi_t + \frac{1}{2}\Phi_v - \frac{2}{5}\Phi_s \right)J_\phi }\\
  J_{\phi} =& F_\phi
  \left ( 
    \left ( \vD_\mu \vH \right )^\dagger \otimes \vD^\mu \vH 
    +\frac {1}{8} \tr{\left ( \vD_\mu \vH \right )^\dagger \vD^\mu \vH }
  \right ) (\tau^a\otimes \tau^a)
\end{align}
\end{subequations}
In absence of the Higgs boson, the coefficient of the second term
is chosen in such a way, that the trace of the current vanishes.
In this scenario, the isovector and isoscalar degree of freedoms
decouple from the model and only the isotensor is needed to describe 
this resonance. However, including a Higgs the Lagrangian
\eqref{eq:Lagrangian-Isotensor-full} guarantees the amplitude 
relation between the Higgs and Nambu-Goldstone bosons that will be
introduced in section~\ref{section:gaugeless}. 
The crossing relations are manifest in the 
scattering amplitudes for the Nambu-Goldstone/Higgs boson,
which can be determined most easily in the gaugeless limit.

One prominent example for such scalar isotensor resonances appears in
the context of composite Higgs models of the type Little Higgs,
particularly in the so called Littlest Higgs model~\cite{LHM}.
These resonances predominantly couple to the (electro)weak gauge sector of
the SM.


\subsection{The Tensor Current}

We now construct the effective current that is coupled to a tensor
resonance multiplet.  By assumption, the resonance should be produced
in VBS processes.  We have to consider independent couplings to the
gauge and Higgs/Nambu-Goldstone sectors.  The gauge-sector couplings
should vanish in the gaugeless limit, so we are led to consider
the Higgs-sector coupling.  

For a tensor isoscalar resonance, the
lowest-dimensional current consists of two terms,
\begin{align}
  J^{\mu \nu}_f&=
  F_f \left (
    \tr{ \left ( \vD^\mu \vH \right )^\dagger \vD^\nu \vH}
    - \frac{c_f}{4} g^{\mu \nu}
    \tr{ \left ( \vD_\rho \vH \right )^\dagger \vD^\rho \vH}
  \right) .
\end{align}
The second term actually couples to the trace of the tensor field,
which vanishes on-shell.  It is therefore part of the non-resonant
continuum and can alternatively be replaced by higher-dimensional
operators in the EFT.  Nevertheless, it is required if, for instance,
we want to construct a traceless current.  For now, we leave the
coefficient $c_f$ undetermined.

The tensor-field coupling then reads
\begin{equation}
  f_{\mu \nu}J_f^{\mu \nu}
\end{equation}
in the Fierz-Pauli formulation (section~\ref{sec:fierz-pauli}), and
\begin{equation}
    f_{\mu \nu} J_f^{\mu \nu}
	    -  \sigma  {J_f}^{\mu}_{\, \mu}
		 - \frac{1}{m}A_\mu \partial_\nu J_f^{\mu \nu} 
		 + \frac{2}{m^2} 	\sigma	\partial_\mu \partial_\nu J_f^{\mu \nu}
\end{equation}
in the St\"uckelberg formulation (section~\ref{sec:stueckelberg}).  In the
second version, the momentum factors in the propagator have been
turned into derivatives that act on the current.  There is also a
coupling to the trace of the current.

The formally dominant high-energy ($s\to\infty$) behavior of the
amplitude thus is given by the exchange of St\"uckelberg vector and
scalar.  The contribution would vanish if the current was conserved.
Evaluating the divergence of first and second order, using
\eqref{eq:Def-FieldStrengthTensor} and \eqref{eq:eqofmotionH} in the
Appendix,
\begin{align}
\label{eq:derivative_Tensor_current}
\begin{aligned}
  \partial_\mu J_f^{\mu \nu} 
  =&
  F_f \tr{ \left ( \vD^2 \vH \right )^\dagger \vD^\nu \vH}
  +\frac{F_f}{4} \left ( c_f +2  \right )
  \tr{ \left ( \vD_\mu \vH \right )^\dagger\left [ \vD^\mu,\vD^\nu
      \right ] \vH} \\ 
  &-\frac{F_f}{4} \left (c_f -2  \right )
  \tr{ \left ( \vD_\mu \vH \right )^\dagger\left \{ \vD^\mu,\vD^\nu
    \right \} \vH} \\ 
  =&
  - F_f \lambda \tr{\widehat{\vH^\dagger \vH}} \tr{ \vH ^\dagger \vD^\nu \vH}  \\
  &-\ii g F_f
  \tr{ \left ( \vD_\mu \vH \right )^\dagger\vW^{\mu\nu} \vH}
  -\ii g^\prime F_f 
  \tr{\vH \vB^{\mu\nu} \left(\vD_\mu \vH\right)^\dagger }\, ,
\end{aligned}
\end{align}

\begin{align}
\label{eq:2derivative_Tensor_current}
\begin{aligned}
  \partial_{\nu}\partial_\mu J_f^{\mu \nu} 
  =& F_f \tr{ \left(\vD_\mu \vH \right)^\dagger 
    \left( 
    \vD_\nu \vD^\mu \vD^\nu \vH + \vD^\mu \vD^2 \vH 
    - \frac{c_f}{2} \vD^2 \vD^\mu \vH
    \right)}\\
  &+F_f \tr{ \left(\vD^2 \vH \right)^\dagger \vD^2 \vH}
  + F_f \tr{ \left(\vD_\mu \vD_\nu \vH \right)^\dagger \vD^\nu \vD^\mu \vH}\\
  &- \frac{c_f}{2} F_f \tr{ \left(\vD_\mu \vD_\nu \vH \right)^\dagger \vD^\mu \vD^\nu \vH}\\
  =&
  - F_f \lambda \tr{\widehat{\vH^\dagger \vH}} \tr{\vD_\mu \vH
    ^\dagger \vD^\mu \vH} 
  - F_f \lambda \tr{\widehat{\vH^\dagger \vH}} \tr{ \vH ^\dagger \vD^2
    \vH}   \\ 
  &- 2F_f \lambda \tr{{\vH^\dagger \vD_\mu \vH}} \tr{ \vH ^\dagger \vD^\mu \vH}\\
  &+\frac{g^2  F_f}{2}
  \left(\tr{ \left ( \vD_\mu \vH \right )^\dagger \vH \left(\vD^{ \mu}
    \vH \right )^\dagger\vH} 
  -\tr{ \left ( \vD_\mu \vH \right )^\dagger \left(\vD^{ \mu} \vH \right )
    \vH^\dagger\vH} 
  \right)\\
  &+\frac{{g^\prime}^2  F_f}{2} 
  \left(\tr{ \left ( \vD_\mu \vH \right )^\dagger \vH \left(\vD^{ \mu}
    \vH \right )^\dagger\vH} 
  -\tr{ \left ( \vD_\mu \vH \right )^\dagger \left(\vD^{ \mu} \vH \right )
    \vH^\dagger\vH} 
  \right)\\
  &+\frac{ g^2 F_f}{2} 
  \tr{  \vH^\dagger\vW_{\mu\nu} \vW^{\mu\nu} \vH}
  +\frac{ {g^\prime}^2 F_f}{2} 
  \tr{ \vH \vB_{\mu\nu} \vB^{\mu\nu} \vH^\dagger} \\
  &+ g{g^\prime} F_f
  \tr{ \vH^\dagger \vW_{\mu\nu} \vH \vB^{\mu\nu}} \\
  &-\ii g F_f
  \tr{ \left ( \vD_\mu \vH \right )^\dagger\vW^{\mu\nu}   \vD_\nu \vH }\\
  &-\ii g F_f
  \tr{ \left(\vD_\mu \vH\right) \vB^{\mu\nu} \left(\vD_\mu
    \vH\right)^\dagger } \, , 
\end{aligned}
\end{align}
we observe that the current is not conserved.
However, none of the nonvanishing
terms contributes to the $VV\to VV$ process at high energy.  The
St\"uckelberg fields effectively decouple, and the high-energy
behavior can be calculated from the propagator~(\ref{eq:proptensor}).

If we take EWSB into account, we do get a nonvanishing divergence also
at the two-particle level.  New terms arise that are proportional to
powers of $v$, and thus to the $W$, $Z$, and Higgs masses.  The
St\"uckelberg vector transmits, via EWSB mixing, a coupling to
transversal vector bosons.  In amplitudes, these factors are
accompanied by factors of $1/m$.  In the limit of a heavy resonance,
the St\"uckelberg terms are thus parametrically suppressed and become
relevant only for energies significantly beyond the resonance mass.
Conversely, if the resonance mass is comparable to the electroweak
scale, the St\"uckelberg terms are significant.

The remainder of the amplitude that corresponds to the genuine tensor
propagator~\eqref{eq:proptensor} does not contain momentum factors.
Nevertheless, the interaction is of dimension five, so we expect
contributions that rise with energy.  This occurs for external
longitudinally polarized vector bosons which carry a momentum factor.
We obtain a factor $s^2$ in the numerator that asymptotically cancels
with the denominator, so the effective rise is proportional to
$s/m^2$.  Qualitatively, this is the same result as for the case of a
scalar resonance, or for a Higgs-less theory.

We conclude that we can unitarize the amplitude uniformly for all
spin-isospin channels, starting from the gaugeless Nambu-Goldstone boson
limit, without having to account for transversal gauge bosons or
higher powers of $s$ beyond the resonance.  The algorithm can be taken
unchanged from the pure-EFT case~\cite{Kilian:2014zja}.  However, we
have to restrict the allowed values of resonance masses and couplings
such that the St\"uckelberg terms discussed above remain numerically
small within some finite energy range.  Outside this range, we can no
longer separate the Higgs/Nambu-Goldstone sector of the theory but are
sensitive to unknown strong interactions that involve all channels of
longitudinal, transversal, and Higgs exchange simultaneously.  While
the unitarization scheme of~\cite{Kilian:2014zja} is also applicable in
that situation, it becomes technically more involved; we defer this
case to future work.

\subsection{Complete model definition}

We now list the effective Lagrangians that we consider in the
subsequent calculations.  In all cases, the basic theory is the SM
EFT, i.e., the SM with the observed light Higgs boson in linear
representation, extended by higher-dimensional operators.  We add
four different resonance multiplets, corresponding to all combinations
of spin and isospin 0 and 2, respectively.  The Lagrangians can be
combined.  

The spin-two Lagrangian is presented in the St\"uckelberg gauge.
Regarding the resonance fields, we should further select electroweak
quantum numbers, as discussed in section~\ref{sec:multiplets}, by
defining the precise form of the covariant derivative acting on the
resonance field in the kinetic operator.  However, as long as we are
not interested in EW radiative corrections, we may work with a simple
partial derivative and omit the gauge couplings to $W$, $Z$, and photon.

The Lagrangian for the isoscalar-scalar $\sigma$,
the isotensor-scalar $\phi$, the isoscalar-tensor $f$
and the isotensor-tensor $X$ are given by
\begin{subequations}
\label{eq:appendix_lagrangian}
\begin{align}
  \LL_\sigma =& 
  \frac{1}{2} \partial_\mu \sigma \partial^\mu \sigma
  -\frac{1}{2} m_\sigma^2 \sigma^2
  + \sigma J_\sigma \, \\
  \LL_\phi =& \frac{1}{2}\sum_{i=s,v,t} \tr{\partial_\mu \Phi_i \partial^\mu \Phi_i
    -m_\Phi^2 \Phi_i^2} 
  +\tr{\left(\Phi_t + \frac{1}{2}\Phi_v - \frac{2}{5}\Phi_s \right)J_\phi }\, ,\\
  \LL^{}_{f} =
  & \frac{1}{2}  f_{f\mu\nu}\left (-\partial^2 -  m_f^2 \right )f_f^{ \mu \nu} 
  + \frac{1}{2}  f^{ \mu}_{f\mu}\left (-\frac{1}{2} \left (-\partial^2
  - m_f^2 \right ) \right )f^{\nu}_{f\nu} \notag \\ 
  &+ \frac{1}{2} A_{f\mu} \left ( -\partial^2 -  m_f^2 \right )  A_f^\mu
  +\frac{1}{2} \sigma_f \left (  -\partial^2 - m_f^2 \right ) \sigma_f 
  \notag \\
  &  + \left ( f_{f\mu \nu} -  \frac{1}{\sqrt{6}} g_{\mu \nu} \sigma +
  \frac{1}{\sqrt{2}m_f}\left (\partial_\mu A_\nu + \partial_\nu A_\mu \right )
  + \frac{\sqrt{2}}{\sqrt{3}m_f^2} 		\partial_\mu \partial_\nu \sigma
  \right )J_f^{\mu \nu} \, , \\
  \LL^{}_{X} =
  & \frac{1}{2}\sum_{i=s,v,t} \operatorname{tr} \Big[     		
    X^{}_{i\mu\nu}\left (-\partial^2 -  m_X^2 \right )X_i^{{} \mu \nu} 
    +  X^{{} \mu}_{i\mu}\left (-\frac{1}{2} \left (-\partial^2 - m_X^2
    \right ) \right )X^{{} \nu}_{i\nu} \notag \\ 
    & \phantom{ \frac{1}{2}\sum_{i=s,v,t} \operatorname{tr} \Big[ }
      + A_{i\mu} \left ( -\partial^2 -  m_X^2 \right )  A_i^\mu
      + \sigma_i \left (  -\partial^2 - m_X^2 \right ) \sigma_i
      \Big]
    \notag \\
    &  + \operatorname{tr} \Bigg[  
      \left ( X^{}_{t\mu \nu} -  \frac{g_{\mu \nu}}{\sqrt{6}}  \sigma_t +
      \frac{\partial_\mu A_{t\nu} + \partial_\nu A_{t\mu} }{\sqrt{2}m_X}
      + \frac{\sqrt{2}}{\sqrt{3}m_X^2} 		\partial_\mu
      \partial_\nu \sigma_t 
      \right )J_X^{\mu \nu} \notag \\
      &\phantom{+ \operatorname{tr} \Big[  }
	+\frac{1}{2}{ \left ( X^{}_{v\mu \nu} -  \frac{g_{\mu
              \nu}}{\sqrt{6}}  \sigma_v + 
	  \frac{\partial_\mu A_{v\nu} + \partial_\nu A_{v\mu} }{\sqrt{2}m_X}
	  + \frac{\sqrt{2}}{\sqrt{3}m_X^2} 		\partial_\mu
          \partial_\nu \sigma_v 
	  \right )J_X^{\mu \nu} } \notag \\
	&\phantom{+ \operatorname{tr} \Big[  } -\frac{2}{5} { \left (
            X^{}_{s\mu \nu} -  \frac{g_{\mu \nu}}{\sqrt{6}}  \sigma_s
            + 
	    \frac{\partial_\mu A_{s\nu} + \partial_\nu A_{s\mu} }{\sqrt{2}m_X}
	    + \frac{\sqrt{2}}{\sqrt{3}m_X^2} 		\partial_\mu
            \partial_\nu \sigma_s 
	    \right )J_X^{\mu \nu} } \Bigg] \, ,
\end{align}
\end{subequations}
respectively, where the tensor resonances are formulated in the
St\"uckelberg formalism with associated fields $\sigma_f$, $A_f$ and $f_f$
denoting the scalar, vector and tensor degrees of freedom,
respectively.
The corresponding St\"uckelberg fields for the isotensor-tensor
receive extra indices $\{s,v,t\}$
which represent the isoscalar, isovector and isotensor 
fields of the $SU(2)_C$ multiplet, respectively.
The couplings to the Nambu-Goldstone boson current in each 
case is given by
\begin{subequations}
  \begin{alignat}{2}
    J_{\sigma} &= F_\sigma&&
    \tr{ \left ( \vD_\mu \vH \right )^\dagger \vD^\mu \vH} \, ,\\
    J_{\phi} &= F_\phi&&
    \left ( 
      \left ( \vD_\mu \vH \right )^\dagger \otimes \vD^\mu \vH 
      +\frac {1}{8} \tr{\left ( \vD_\mu \vH \right )^\dagger \vD^\mu \vH }
    \right )\tau^{aa} \, , \\
    J^{\mu \nu}_f&=
    F_f &&\left (
      \tr{ \left ( \vD^\mu \vH \right )^\dagger \vD^\nu \vH}
      - \frac{c_f}{4} g^{\mu \nu}
      \tr{ \left ( \vD_\rho \vH \right )^\dagger \vD^\rho \vH}
    \right)  \, , \\
    J^{\mu \nu}_{X }&=
    F_X& & \Bigg [
    \frac{1}{2} \left (
      \left ( \vD^\mu \vH \right )^\dagger \otimes \vD^\nu \vH
      + \left (  \vD^\nu \vH \right )^\dagger \otimes \vD^\mu \vH
    \right )
    - \frac{c_X}{4} g^{\mu \nu}
    \left ( \vD_\rho \vH \right )^\dagger \otimes \vD^\rho \vH \notag \\
    &&&
    +\frac{1}{8} \left (
      \tr{\left ( \vD^\mu \vH \right )^\dagger \vD^\nu \vH}
      -  \frac{c_X}{4} g^{\mu \nu}
      \tr{\left ( \vD_\rho \vH \right )^\dagger \vD^\rho \vH}
    \right )
    \Bigg ] \tau^{aa} \, .
  \end{alignat}
\end{subequations}


\section{Unitary amplitudes for VBS at the LHC}
\label{sec:amplitudes}

\subsection{Gaugeless limit}
\label{section:gaugeless}

For a first estimate of the impact of generic resonances 
to vector-boson scattering processes at the LHC, we
study the on-shell Nambu-Goldstone boson scattering amplitudes.
When treating vector-boson scattering as $2 \rightarrow 2$
process of massless scalars at high energies, it is convenient to
describe kinematic dependencies using Mandelstam variables $s,t,u$.
Using custodial symmetry and crossing symmetries, the different
$2\rightarrow 2$ Nambu-Goldstone boson scattering amplitudes are
determined by the master amplitudes $\mathcal{A} \left (w^+w^-
\rightarrow zz \right )$. In the gaugeless
limit, the amplitudes for the resonance multiplets $\sigma$, $\phi$,
$f$, and $X$ are calculated in the gaugeless limit via the Feynman
rules given in appendix~\ref{appendix:Feynman_Rules}.

\subsubsection{Isoscalar-Scalar}
\begin{subequations}
\label{eq:sigma_amplitudes}
\begin{align}
  \mathcal{A}_\sigma \left (w^\pm w^\pm \rightarrow w^\pm w^\pm \right )&=
  - \frac{1}{4} {F_\sigma}^2 \left(\frac{t^2}{t-m_\sigma^2}+
  \frac{u^2}{u-m_\sigma^2} \right) \, ,\\ 
  \left. \begin{aligned}
    \mathcal{A}_\sigma& \left (w^\pm z \rightarrow w^\pm z \right )\\
    \mathcal{A}_\sigma& \left (hw^\pm \rightarrow hw^\pm \right )	\\
    \mathcal{A}_\sigma& \left (hz \rightarrow hz \right )\\
  \end{aligned} \right \} &=
  - \frac{1}{4} {F_\sigma}^2 \frac{t^2}{t-m_\sigma^2} \, ,\\
  \mathcal{A}_\sigma \left (w^\pm w^\mp \rightarrow w^\pm w^\mp \right )&=
  - \frac{1}{4} {F_\sigma}^2 \left (\frac{s^2}{s-m_\sigma^2}+ \frac{t^2}{t-m_\sigma^2} \right) \, ,\\
  \left. \begin{aligned}
    \mathcal{A}_\sigma& \left (w^\pm w^\mp \rightarrow zz \right )\\
    \mathcal{A}_\sigma& \left (hh \rightarrow w^\pm w^\mp \right )	\\
    \mathcal{A}_\sigma& \left (hh \rightarrow zz \right )\\
  \end{aligned} \right \}&=
  - \frac{1}{4} {F_\sigma}^2 \frac{s^2}{s-m_\sigma^2} \, ,\\
  \left. \begin{aligned}
    \mathcal{A}_\sigma& \left (zz \rightarrow zz \right )\\
    \mathcal{A}_\sigma& \left (hh \rightarrow hh \right )	
  \end{aligned} \right \} &=
  - \frac{1}{4} {F_\sigma}^2 \left( \frac{s^2}{s-m_\sigma^2}+ \frac{t^2}{t-m_\sigma^2} +\frac{u^2}{u-m_\sigma^2} \right) \, . 
\end{align}
\end{subequations}
\subsubsection{Isotensor-Scalar}

\begin{subequations}
\label{eq:phi_amplitudes}
\begin{align}
  \mathcal{A}_\phi \left (w^\pm w^\pm \rightarrow w^\pm w^\pm \right )&=
  -\frac{{F_\phi}^2}{8} \left(2 \frac{s^2}{s-m_\phi^2}
  + \frac{1}{2} \frac{u^2}{u-m_\phi^2}
  + \frac{1}{2} \frac{t^2}{t-m_\phi^2} \right)\, ,\\
  \left. \begin{aligned}
    \mathcal{A}_\phi& \left (w^\pm z \rightarrow w^\pm z \right )\\
    \mathcal{A}_\phi& \left (hw^\pm \rightarrow hw^\pm \right )	\\
    \mathcal{A}_\phi& \left (hz \rightarrow hz \right )\\
  \end{aligned} \right \} &=
  \frac{{F_\phi}^2}{8} \left( \frac{1}{2} \frac{t^2}{t-m_\phi^2}
  -  \frac{u^2}{u-m_\phi^2}
  -  \frac{s^2}{s-m_\phi^2} \right) \, ,\\
  \mathcal{A}_\phi \left (w^\pm w^\mp \rightarrow w^\pm w^\mp \right )&=
  -\frac{{F_\phi}^2}{8} \left( \frac{1}{2} \frac{s^2}{s-m_\phi^2}
  + 2 \frac{u^2}{u-m_\phi^2}
  + \frac{1}{2} \frac{t^2}{t-m_\phi^2} \right) \, ,\\
  \left. \begin{aligned}
    \mathcal{A}_\phi& \left (w^\pm w^\mp \rightarrow zz \right )\\
    \mathcal{A}_\phi& \left (hh \rightarrow w^\pm w^\mp \right )	\\
    \mathcal{A}_\phi& \left (hh \rightarrow zz \right )\\
  \end{aligned} \right \}&=
  \frac{{F_\phi}^2}{8} \left( \frac{1}{2} \frac{s^2}{s-m_\phi^2}
  -  \frac{u^2}{u-m_\phi^2}
  -  \frac{t^2}{t-m_\phi^2} \right)\, ,\\
  \left. \begin{aligned}
    \mathcal{A}_\phi& \left (zz \rightarrow zz \right )\\
    \mathcal{A}_\phi& \left (hh \rightarrow hh \right )	
  \end{aligned} \right \} &=
  -\frac{3{F_\phi}^2}{16} \left(  \frac{s^2}{s-m_\phi^2}
  +  \frac{u^2}{u-m_\phi^2}
  +  \frac{t^2}{t-m_\phi^2} \right) \, . 
\end{align}
\end{subequations}

\subsubsection{Isoscalar-Tensor}
\begin{subequations}
\label{eq:tensor_amplitudes}
\begin{align}
  \mathcal{A}_f \left (w^\pm w^\pm \rightarrow w^\pm w^\pm \right )=&
  - \frac{1}{24} {F_f}^2 \left(\frac{t^2}{t-m_f^2}P_2(t,s,u)+ \frac{u^2}{u-m_f^2}P_2(u,s,t) \right) \, ,\\
  \left. \begin{aligned}
    \mathcal{A}_f& \left (w^\pm z \rightarrow w^\pm z \right )\\
    \mathcal{A}_f& \left (hw^\pm \rightarrow hw^\pm \right )	\\
    \mathcal{A}_f& \left (hz \rightarrow hz \right )\\
  \end{aligned} \right \} =&
  - \frac{1}{24} {F_f}^2 \frac{t^2}{t-m_f^2}P_2 \left(t,s,u \right) \, ,\\
  \mathcal{A}_f \left (w^\pm w^\mp \rightarrow w^\pm w^\mp \right )=&
  - \frac{1}{24} {F_f}^2 \left (\frac{s^2}{s-m_f^2}P_2(s,t,u)+
  \frac{t^2}{t-m_f}P_2(t,s,u) \right) \, ,\\ 
  \left. \begin{aligned}
    \mathcal{A}_f& \left (w^\pm w^\mp \rightarrow zz \right )\\
    \mathcal{A}_f& \left (hh \rightarrow w^\pm w^\mp \right )	\\
    \mathcal{A}_f& \left (hh \rightarrow zz \right )\\
  \end{aligned} \right \}=&
  - \frac{1}{24} {F_f}^2 \frac{s^2}{s-m_f^2}P_2 \left(s,t,u \right) \, ,\\
  \left. \begin{aligned}
    \mathcal{A}_f& \left (zz \rightarrow zz \right )\\
    \mathcal{A}_f& \left (hh \rightarrow hh \right )	
  \end{aligned} \right \} =& 
  - \frac{1}{24} {F_f}^2 \Bigg(
  \frac{s^2}{s-m_f^2}P_2(s,t,u)
  +\frac{t^2}{t-m_f^2}P_2(t,s,u)\notag\\
  &\phantom{- \frac{1}{24} {F_f}^2 \Bigg(} +\frac{u^2}{u-m_f^2}P_2(u,s,t) \Bigg) \, . 
\end{align}
\end{subequations}
Here and in the following, $P_2(s,t,u) = [3(t^2 + u^2) - 2
s^2]/s^2$ is the second order Legendre polynomial in terms of the
Mandelstam variables.
\subsubsection{Isotensor-Tensor}
\begin{subequations}
\label{eq:isotensor_tensor_amplitudes_gaugeless}
\allowdisplaybreaks
\begin{align}
  \mathcal{A}_X \left (w^\pm w^\pm \rightarrow w^\pm w^\pm \right )=&
  -\frac{{F_X}^2}{96}  \Bigg(
  \frac{4s^2}{s-m_X^2}P_2 \left(s,t,u \right) 
  +\frac{t^2}{t-m_X^2}P_2 \left(t,s,u \right) \notag\\
  &\phantom{-\frac{{F_X}^2}{96}  \Bigg(}
  +\frac{u^2}{u-m_X^2}P_2 \left(u,s,t \right)
  \Bigg)\,  ,\\
  \left. \begin{aligned}
    \mathcal{A}_X& \left (w^\pm z \rightarrow w^\pm z \right )\\
    \mathcal{A}_X& \left (hw^\pm \rightarrow hw^\pm \right )	\\
    \mathcal{A}_X& \left (hz \rightarrow hz \right )\\
  \end{aligned} \right \} =&
  \frac{{F_X}^2}{96}  \Bigg (
  -\frac{2s^2}{s-m_X^2}P_2 \left(s,t,u \right)
  +\frac{t^2}{t-m_X^2}P_2 \left(t,s,u \right)\notag\\
  &\phantom{\frac{{F_X}^2}{96}  \Bigg (}
  -\frac{2u^2}{u-m_X^2}P_2 \left(u,s,t \right)
  \Bigg)\,  ,
  \\
  \mathcal{A}_X \left (w^\pm w^\mp \rightarrow w^\pm w^\mp \right )=&
  -\frac{{F_X}^2}{96}  \Bigg(
  \frac{s^2}{s-m_X^2}P_2 \left(s,t,u \right) 
  +\frac{t^2}{t-m_X^2}P_2 \left(t,s,u \right) \notag\\
  &\phantom{-\frac{{F_X}^2}{96}  \Bigg(}
  +\frac{4u^2}{u-m_X^2}P_2 \left(u,s,t \right)
  \Bigg)\, 
  ,
  \\
  \left. \begin{aligned}
    \mathcal{A}_X& \left (w^\pm w^\mp \rightarrow zz \right )\\
    \mathcal{A}_X& \left (hh \rightarrow w^\pm w^\mp \right )	\\
    \mathcal{A}_X& \left (hh \rightarrow zz \right )\\
  \end{aligned} \right \}=&
  \frac{{F_X}^2}{96}  \Bigg(
  \frac{s^2}{s-m_X^2}P_2 \left(s,t,u \right) 
  -\frac{2t^2}{t-m_X^2}P_2 \left(t,s,u \right)\notag \\
  &\phantom{\frac{{F_X}^2}{96}  \Bigg(}
  -\frac{2u^2}{u-m_X^2}P_2 \left(u,s,t \right)
  \Bigg)\, ,\\
  \left. \begin{aligned}
    \mathcal{A}_X& \left (zz \rightarrow zz \right )\\
    \mathcal{A}_X& \left (hh \rightarrow hh \right )	
  \end{aligned} \right \} =& 
  -\frac{1}{32} {F_X}^2 \Bigg(
  \frac{s^2}{s-m_X^2}P_2 \left(s,t,u \right) 
  +\frac{t^2}{t-m_X^2}P_2 \left(t,s,u \right) \notag \\
  &\phantom{-\frac{1}{32} {F_X}^2 \Bigg(} +
  \frac{u^2}{u-m_X^2}P_2 \left(u,s,t \right)
  \Bigg) \, . 
\end{align}
\end{subequations}


\subsection{Decomposition of eigenamplitudes}
\label{sec:IlinWHIZARD}

Since the leading-order amplitudes as listed above are unbounded both at
the pole and at high energy, we use the T-matrix
scheme~\cite{Kilian:2014zja} to restore unitarity.  In order to
implement the scheme in~\cite{Kilian:2014zja}, we decompose the
amplitudes into isospin-spin eigenamplitudes (the $S$-wave, $P$-wave
and $D$-wave kinematic functions $\mathcal{S}_i$, $\mathcal{P}_i$ and
$\mathcal{D}_i$ can be found in appendix~\ref{appendix:partialwaves}):  

\subsubsection{Isoscalar-Scalar}
\begin{subequations}
\label{eq:isospin-spin-scalar}
\begin{align}
  \amp_{00}&=
  F_\sigma^2
  \left (
    -\frac{3}{4} \frac{s^2}{s-m_\sigma^2}
    -\frac{1}{2} \mathcal{S}_0
  \right)
  ,\\
  \amp_{02}&=
  -\frac{1}{2} F_\sigma^2 \mathcal{S}_2
  ,\\
  \amp_{11}&=
  -\frac{1}{2} F_\sigma^2 \mathcal{S}_1
  ,\\
  \amp_{13}&=
  -\frac{1}{2} F_\sigma^2 \mathcal{S}_3
  ,\\
  \amp_{20}&=
  -\frac{1}{2} F_\sigma^2 \mathcal{S}_0
  ,\\
  \amp_{22}&=
  -\frac{1}{2} F_\sigma^2 \mathcal{S}_2
\end{align}
\end{subequations}
\subsubsection{Isotensor-Scalar}
\begin{subequations}
\label{eq:isospin-spin-isotensor}
\begin{align}
  \amp_{00}&=
  F_\phi^2
  \left (
    -\frac{1}{16} \frac{s^2}{s-m_\phi^2}
    -\frac{7}{8} \mathcal{S}_0
  \right)
  ,\\
  \amp_{02}&=
  -\frac{7}{8} F_\phi^2 \mathcal{S}_2
  ,\\
  \amp_{11}&=
  \frac{3}{8} F_\phi^2 \mathcal{S}_1
  ,\\
  \amp_{13}&=
  \frac{3}{8} F_\phi^2 \mathcal{S}_3
  ,\\
  \amp_{20}&=
  F_\phi^2
  \left (
    -\frac{1}{4} \frac{s^2}{s-m_\phi^2}
    -\frac{1}{8} \mathcal{S}_0
  \right)
  ,\\
  \amp_{22}&=
  -\frac{1}{8} F_\phi^2 \mathcal{S}_2
\end{align}
\end{subequations}
\subsubsection{Isoscalar-Tensor}
\begin{subequations}
\label{eq:isospin-spin-tensor}
\begin{align}
  \amp_{00}&=
  -\frac{1}{12}F_f^2\mathcal{D}_0
  ,\\
  \amp_{02}&=
  -\frac{1}{40}F_f^2 \frac{s^2}{s-m_f^2}
  -\frac{1}{12}F_f^2 
  \left (1+ 6 \frac{s}{m_f^2}+ 6 \frac{s^2}{m_f^4} \right )
  \mathcal{S}_2
  ,\\
  \amp_{11}&=
  -\frac{1}{12}F_f^2 \mathcal{D}_1
  ,\\
  \amp_{13}&=
  -\frac{1}{12}F_f^2 
  \left (1+ 6 \frac{s}{m_f^2}+ 6 \frac{s^2}{m_f^4} \right )
  \mathcal{S}_3
  ,\\
  \amp_{20}&=
  -\frac{1}{12}F_f^2 \mathcal{D}_0
  ,\\
  \amp_{22}&=
  -\frac{1}{12}F_f^2 
  \left (1+ 6 \frac{s}{m_f^2}+ 6 \frac{s^2}{m_f^4} \right )
  \mathcal{S}_2
\end{align}
\end{subequations}
\subsubsection{Isotensor-Tensor}
\begin{subequations}
\label{eq:isospin-spin-isotensor-tensor}
\begin{align}
  \amp_{00}&=
  -\frac{7}{48} F_X^2 \mathcal{D}_0
  ,\\
  \amp_{02}&=
  -\frac{1}{480}F_X^2 \frac{s^2}{s-m_X^2}
  -\frac{7}{48} F_X^2\left(1 + 6 \frac{s}{m_X^2}+ 6 \frac{s^2}{m_X^4}
  \right) \mathcal{S}_2 
  ,\\
  \amp_{11}&=
  \frac{1}{16} F_X^2 \mathcal{D}_1
  ,\\
  \amp_{13}&=
  \frac{1}{16} F_X^2 \left(1 + 6 \frac{s}{m_X^2}+ 6 \frac{s^2}{m_X^4}
  \right) \mathcal{S}_3 
  ,\\
  \amp_{20}&=
  -\frac{1}{48} F_X^2 \mathcal{D}_0
  ,\\
  \amp_{22}&=
  -\frac{1}{120}F_X^2 \frac{s^2}{s-m_X^2}
  -\frac{1}{48} F_X^2\left(1 + 6 \frac{s}{m_X^2}+ 6 \frac{s^2}{m_X^4}
  \right) \mathcal{S}_2 \, . 
\end{align}
\end{subequations}


\subsection{Width}
\label{sec:width}

As argued below in section~\ref{sec:tensor-unitary-gauge}, for the numerical
off-shell calculation of scattering processes we will need approximate
values for the resonance decay widths.  If suffices to compute those
in the gaugeless limit. Contributions
proportional to the masses of the vector bosons and the Higgs boson are 
assumed to be small at high resonance masses and are therefore neglected. 
\begin{subequations}
\label{eq:gamma_scalar}
\begin{align}
  \Gamma_\sigma
  &= \frac{m_\sigma^3}{32\pi}F_\sigma^2 \, ,
  \\
  \Gamma_\phi
  &= \frac{m_\phi^3}{128\pi}F_\phi^2 \, ,
  \\
  \Gamma_f
  &= \frac{m_f^3}{960\pi}F_f^2 \, ,
  \\
  \Gamma_X 
  &= \frac{m_X^3}{3840\pi}F_X^2 \, .
\end{align}
\end{subequations}

\subsection{Matching to the low-energy EFT}
\label{sec:matching}
    
For later convenience, we compute the coefficients of the effective
dimension-eight operators  $\LL_{S,0}$ and $\LL_{S,1}$~\cite{Kilian:2014zja},
 \begin{subequations}
 	\label{eq:dim8_operator_S}
			\begin{alignat}{4}
\label{eq:LL-S0}
	\LL_{S,0}&=
	 & &F_{S,0}\ &&
	  \tr{ \left ( \vD_\mu \vH \right )^\dagger \vD_\nu \vH}
		\tr{ \left ( \vD^\mu \vH \right )^\dagger \vD^\nu \vH},
\\
 \label{eq:LL-S1}
	\LL_{S,1}&=
	 & &F_{S,1}\ &&
	  \tr{ \left ( \vD_\mu \vH \right )^\dagger \vD^\mu \vH}
		\tr{ \left ( \vD_\nu \vH \right )^\dagger \vD^\nu \vH}.
\end{alignat}
\end{subequations}

 which result
from integrating out the resonances $\sigma,\phi,f,X$, one at a time.
\begin{alignat}{2}
  && F_{S,1} &= \frac{F_\sigma^2}{2 m_\sigma^2}\, , \\
  F_{S,0} &= \frac{F_\phi^2}{2m_\phi^2} \, , &\qquad F_{S,1} &=-
  \frac{F_\phi^2}{8 m_\phi^2}\, ,\\ 
  F_{S,0} &= \frac{F_f^2}{2 m_f^2}\, , &\qquad
  F_{S,1} &= - \frac{F_f^2}{6 m_f^2}\, , \\
  F_{S,0} &= \frac{F_X^2}{24 m_X^2} \, ,&\qquad
  F_{S,1} &= - \frac{7 F_X^2}{24 m_X^2} \,.
\end{alignat}

\subsection{Tensor exchange in unitary gauge}
\label{sec:tensor-unitary-gauge}
Beyond the resonance, the Nambu-Goldstone bosons scattering amplitudes
rise proportional to powers of the  
invariant mass of the scattering system.  They eventually violate
unitarity at a certain energy, depending on the resonance coupling.

Computing the $w^+w^-\to zz$ amplitude in the presence of an
isoscalar tensor resonance, for instance,
\begin{align}
   \amp_f \left (w^+w^- \rightarrow zz \right ) =&	 
   -\frac{F_f^2}{96}\left (c_f-2 \right )^2  \frac{s^3}{ m_f^4}
   -\frac{F_f^2}{48} \left (c_f-2 \right )c_f \frac{s^2}{ m_f^2} \notag\\
   & -\frac{F_f^2}{24} \left (3 \left (t^2+u^2 \right )-2s^2 \right )
   \frac{1}{s-m_f^2} \, ,
\end{align}
we observe that choosing
$c_f\neq 2$ results in a high degree of divergence.
This is due to contributions of the vector and scalar degree of freedoms
in the St\"uckelberg parameterization for the tensor
coupled to the derivatives of the current \eqref{eq:derivative_Tensor_current}
and \eqref{eq:2derivative_Tensor_current}.  As discussed above, such
terms can be written in a non-resonant form and should be interpreted as
coefficients of undetermined higher-dimensional local operators.  Setting
thus $c_f =2$, we
obtain an amplitude $\amp_f(s)$ which rises proportional
to $s$ beyond the resonance. 

However, the scalar and 
vector degree of freedoms provide additional contributions which 
are not manifest in the gaugeless limit.  A calculation
of the tensor scattering amplitude in the unitary
gauge is necessary. The longitudinal on-shell
$WW \rightarrow ZZ$ amplitude for $c_f=2$ is given by 
\begin{align}
\label{eq:amp_wwzz_tensor-scalar}
\amp_f \left(W_LW_L \rightarrow Z_LZ_L \right) 
=&-\frac{1}{24}\frac{F_f^2}{s-m_f^2} \Bigg[
	 \left(P_2\left[\cos (\theta) \right]  -1 \right)s^2
	+12 m_W^2 m_Z^2 \notag \\
	&\phantom{-\frac{1}{24}\frac{F_f^2}{s-m_f^2}} 
        \quad -12\frac{ m_W^2 m_Z^2}{m_f^2} \notag\\
        &\phantom{-\frac{1}{24}\frac{F_f^2}{s-m_f^2}} \notag
	\quad +\left(s-2 m_W^2\right) \left(s-2 m_Z^2\right) \\
        &\phantom{-\frac{1}{24}\frac{F_f^2}{s-m_f^2}} \quad \notag
	+4\frac{ m_W^2 m_Z^2}{m_f^4} s^2
+2\frac{\left( m_W^2+ m_Z^2 \right)s^2 -4 m_Z^2m_W^2 s}{m_f^2}
\Bigg] \, .
\end{align}
The first line represents the tensor contribution in the St\"uckelberg
parameterization. Due to its suppression by a power of $s$, the vector
part in the second line can be neglected 
for the longitudinal scattering amplitude. Besides the scalar contribution
originating from the trace of the current, additional contributions
related to the double derivative of the current and its mixing
with the trace part written in the fourth line will rise with energy.
However, they are suppressed by ${m_W^2}/{m_f^2}$ or ${m_W^4}/{m_f^4}$
and can be neglected if the mass of the tensor resonance is large
in comparison to the vector boson masses. In this case,
the longitudinal amplitude of the vector bosons calculated in the
unitary gauge coincides with the amplitude in gaugeless limit. 

Furthermore, due to the coupling to the derivatives of the
scalar and vector degrees of freedom, also amplitudes in channels with
transverse polarization
rise with the energy of the vector-boson scattering system.
A full list of these channels in the high-energy limit is displayed in 
Table \ref{table:Tensor-Polarization}.  We observe that all channels
which include at least one transversally polarized vector boson
are suppressed by ${m_W^2}/{m_f^2}$. Therefore, a calculation
within the gaugeless limit is sufficient to estimate the 
high-energy behavior for high masses of the tensor resonance.

\begin{table}[tb]
\begin{subequations}
  \begin{alignat*}{2}
    \hline
  \begin{array}{l l l l}
    (+,& \hspace{0ex}+,&\hspace{0ex}+,+ &\hspace{0ex}) \\[-3mm]
    (+,&\hspace{0ex}+,&\hspace{0ex}-,-&\hspace{0ex})   \\[-3mm]
    (-,&\hspace{0ex}-,&\hspace{0ex}+,+&\hspace{0ex})   \\[-3mm]
    (-,&\hspace{0ex}-,&\hspace{0ex}-,- &\hspace{0ex}) 
  \end{array}
  & \hspace{1 cm} & & 
  - \frac{c_f^2m_W^2 m_Z^2}{24m_f^2} F_f^2 s \\
  \hline
  \begin{array}{l l l l}
    (+,& \hspace{0ex}0,&\hspace{0ex}0,+ &\hspace{0ex}) \\[-3mm]
    (0,&\hspace{0ex}+,&\hspace{0ex}+,0&\hspace{0ex})   \\[-3mm]
    (0,&\hspace{0ex}-,&\hspace{0ex}-,0&\hspace{0ex})   \\[-3mm]
    (-,&\hspace{0ex}0,&\hspace{0ex}0,- &\hspace{0ex}) 
  \end{array}
  & \hspace{1 cm} & & 
  \frac{m_W m_Z}{8m_f^2} F_f^2 t	  \\
  \hline
  \begin{array}{l l l l}
    (+,& \hspace{0ex}0,&\hspace{0ex}+,0 &\hspace{0ex}) \\[-3mm]
    (0,&\hspace{0ex}+,&\hspace{0ex}0,+&\hspace{0ex})   \\[-3mm]
    (0,&\hspace{0ex}-,&\hspace{0ex}0,-&\hspace{0ex})   \\[-3mm]
    (-,&\hspace{0ex}0,&\hspace{0ex}-,0 &\hspace{0ex}) 
  \end{array}
  & \hspace{1 cm} & & 
  \frac{m_W m_Z}{8m_f^2} F_f^2 u \\
  \hline
  \begin{array}{l l l l}
    (+,& \hspace{0ex}0,&\hspace{0ex}-,0 &\hspace{0ex}) \\[-3mm]
    (0,&\hspace{0ex}+,&\hspace{0ex}0,-&\hspace{0ex})   \\[-3mm]
    (0,&\hspace{0ex}-,&\hspace{0ex}0,+&\hspace{0ex})   \\[-3mm]
    (-,&\hspace{0ex}0,&\hspace{0ex}+,0 &\hspace{0ex}) 
  \end{array}
  & \hspace{1 cm} & & 
  -\frac{m_W m_Z}{8m_f^2} F_f^2 t \\
  \hline
  \begin{array}{l l l l}
    (+,& \hspace{0ex}0,&\hspace{0ex}0,- &\hspace{0ex}) \\[-3mm]
    (0,&\hspace{0ex}+,&\hspace{0ex}-,0&\hspace{0ex})   \\[-3mm]
    (0,&\hspace{0ex}-,&\hspace{0ex}+,0&\hspace{0ex})   \\[-3mm]
    (-,&\hspace{0ex}0,&\hspace{0ex}0,+ &\hspace{0ex}) 
  \end{array}
  & \hspace{1 cm} & & 
  -\frac{m_W m_Z}{8m_f^2} F_f^2 u	  \\
  \hline
  \begin{array}{l l l l}
    (+,& \hspace{0ex}+,&\hspace{0ex}0,0 &\hspace{0ex}) \\[-3mm]
    (-,&\hspace{0ex}-,&\hspace{0ex}0,0&\hspace{0ex})   \\
  \end{array}
  & \hspace{1 cm} & & 
  \begin{aligned}
    &\frac{ m_f^2 + 2 m_Z^2  }{12 m_f^4} m_W^2 F_f^2 s
  \end{aligned} \\
  \hline
  \begin{array}{l l l l}
    (0,& \hspace{0ex}0,&\hspace{0ex}+,+ &\hspace{0ex}) \\[-3mm]
    (0,&\hspace{0ex}0,&\hspace{0ex}-,-&\hspace{0ex})   \\
  \end{array}
  & \hspace{1 cm} & & 
  \begin{aligned}
    &\frac{ m_f^2 +2 m_W^2  }{12 m_f^4} m_Z^2 F_f^2 s
  \end{aligned} \\
  \hline
  \begin{array}{l l l l}
    (0,& \hspace{0ex}0,&\hspace{0ex}0,0 &\hspace{0ex}) \\
  \end{array}
  & \hspace{1 cm} & & 
  \begin{aligned}
    \frac{F_f^2}{24} \frac{2s^2- 3t^2 - 3u^2 }{s}
    +\frac{ m_f^2 \left(m_W^2+m_Z^2 \right)
      +2m_W^2m_Z^2 }{12 m_f^4} F_f^2 s
  \end{aligned} \\
  \hline
\end{alignat*}
\end{subequations}
\caption{High energy limit 
of the $W^+W^-~\rightarrow~ZZ$ amplitude
for each polarization channel that
 rises with energy due to a
isoscalar-tensor resonance ($c_f=2$).}
\label{table:Tensor-Polarization}
\end{table}

For the tensor-isotensor amplitude, the analogous result with $c_X=2$ is
\begin{align}
  \amp_X \left (W^\pm_L W^\mp_L \rightarrow Z_LZ_L \right )=&
  \frac{{F_X}^2}{96}  \Bigg(
  \frac{s^2}{s-m_X^2}P_2 \left(s,t,u \right) 
  -\frac{2t^2}{t-m_X^2}P_2 \left(t,s,u \right) \notag 
  -\frac{2u^2}{u-m_X^2}P_2 \left(u,s,t \right)
  \Big) \notag \\
  &\qquad\quad+\frac{{F_X}^2}{24}\frac{m_{whz}^2}{m_X^2}
  \left(\frac{s^2}{s-m_X^2}-\frac{2t^2}{t-m_X^2}-\frac{2u^2}{u-m_X^2}
  \right)\notag\\ 
  &\qquad\quad-\frac{{F_X}^2}{48}\frac{\left(m_W^2-m_Z^2\right)^2}{m_X^4} 
  \left(\frac{t^2}{t-m_X^2}+\frac{u^2}{u-m_X^2} \right) 
  + \mathcal{O}\left(s^0 \right) \, ,
\end{align}
containing t-channel and u-channel contributions, as expected.

\subsection{Unitarized amplitudes}

The tree-level exchange amplitudes that directly result from
evaluating Feynman rules, exhibit two distinct sources of unitarity violation.
Firstly, the amplitude develops a pole at the resonance mass, on the
real axis.  Secondly, terms that rise with energy asymptotically
violate unitarity bounds.

In principle, the T-matrix unitarization scheme would be sufficient to
regulate both issues simultaneously.  At the pole, this boils down to
standard Dyson resummation, introducing the particle width as an
imaginary part in the denominator.  It can easily be verified that
this actually happens for the on-shell scattering amplitudes of
external Nambu-Goldstone bosons.  We obtain the correct value for the
resonance width in the gaugeless limit.  

However, we want to evaluate the amplitudes off-shell for physical $W$
and $Z$ bosons.  The simplified unitarization scheme that we describe
above is not exactly accurate as soon as we include finite corrections
due to transversal gauge bosons and finite $W/Z$ mass.  As a result,
there are contributions which are not cancelled on the resonance pole,
and a narrow but unbounded peak remains.

To avoid this problem, we simply insert an {\em a priori} width in the
resonant propagator.  We thus start from a \emph{complex} model
amplitude.  Therefore, we take the T-matrix scheme of~\cite{Kilian:2014zja}
at face value, and drop the reference to the usual K-matrix scheme
which implies an intermediate projection onto the real axis.  By
construction, in the gaugeless limit, the correct result is invariant
with respect to the introduction of this width, if it has the correct
on-shell value.  For finite gauge couplings and masses, the result
acquires a subleading dependence on this initial value since the model
amplitude is neither on the real axis nor exactly on the Argand
circle.  However, the amplitude after unitarization is now bounded
near the resonance pole, as required.

In the asymptotic regime, the simplified T-matrix scheme renders the
amplitude unitary at all energies, if the exchanged resonance is
scalar.  This enables us to compute cross sections and generate event
samples in this model for complete processes at the LHC
(cf.~section~\ref{sec:results}).

For a tensor resonance, in the St\"uckelberg approach, the genuine
tensor exchange terms are also regulated completely by this (simplified)
scheme.  The extra St\"uckelberg vector and scalar terms, however,
generate higher powers of $s$
which enter when trading Nambu-Goldstone bosons for physical vector
bosons in unitary gauge, suppressed by powers of $m_h,m_W,m_Z$.
Applying the unitarization framework for those extra terms would require a 
complete diagonalization of all vector-boson helicity amplitudes in
unitary gauge.  In
any case, parameter ranges where these terms play a role correspond to
a regime where all degrees of freedom of the SM interact strongly via
these couplings.  We therefore stay away from this range and choose
parameters where those terms are subleading within the accessible
energy range.

Computing the scale where the St\"uckelberg vector-scalar terms violate the
relevant unitarity bounds, we obtain the energy limit
\begin{align}
\label{eq:completebound_tensor-scalar_width}
	\sqrt{s} &\lesssim
  \sqrt{\frac{1}{5} \frac{ m_f}{\Gamma_f}} 
  \frac{m_f^2}{ m_{whz} }\, ,
\end{align}	
for the model which contains an isoscalar tensor, and
\begin{align}
\label{eq:completebound_isotensor-tensor-scalar_width}
	\sqrt{s} \lesssim \sqrt{\frac{1}{30}\frac{m_x}{\Gamma_x}}
	\frac{m_X^2}{m_{whz}} \,.
\end{align}
for the isotensor tensor multiplet.  Here, $m_{whz}$ indicates the
common mass scale of electroweak bosons $W,H,Z$.  Inserting the
accessible energy 
for the LHC collider, we can invert those relations to extract
parameter regions where the simplified models with a tensor resonance
are valid.  The numerical results in the following sections have been
obtained for parameter values that satisfy the bounds.

\section{Scenarios for VBS at the LHC}
\label{sec:results}

\subsection{Implementation}

In the previous section, we have derived the analytic expressions that
determine the on-shell VBS amplitudes in the presence of a resonance.
The amplitudes include correction terms that enforce
quantum-mechanical unitarity without altering the physical content of
the model.

Ultimately, we are interested in measurable effects in LHC data.  For
a complete calculation, the unitarized amplitudes that are originally
defined for on-shell VBS processes, have to be extrapolated off-shell
in a practically meaningful way.  As long as the kinematical
conditions are approximately met, we can evaluate the interactions in
unitary gauge, eliminating all explicit references to Nambu-Goldstone
bosons in favor of physical vector fields, and derive the Feynman
rules in that gauge.  The effective Feynman rules for the unitarity
corrections become momentum dependent and
involve theta functions that restrict the insertions to the $s$-channel
of VBS where partial-wave projection and unitarization is defined.

In the physical processes at the LHC,
\begin{equation}
  pp \to q q \to q q V V
\end{equation}
where $q$ generically denotes a quark and $V$ is either $W$ or $Z$,
the final-state quarks are detected as jets in the forward direction.
With suitable cuts, we can arrange that there is significant
contribution from the subprocess $VV\to VV$ where the initial-state
vector bosons are spacelike but approximately on-shell, in the limit
of high invariant $VV$ mass.  This subprocess, i.e., the associated
off-shell amplitude, obtains contributions from resonance exchange and
is affected by unitarization.

We have implemented this prescription as a model in the Monte-Carlo
integration and event generation package
WHIZARD~\cite{Kilian:2007gr,Moretti:2001zz,Kilian:2011ka,Kilian:2012pz}. This
is a universal event generator for simulations at hadron and lepton
colliders at leading order and next-to-leading (QCD)~\cite{WHIZARD_NLO}
order. Though interfaces to automated tools for beyond the SM models
exist~\cite{Christensen:2010wz}, they cannot be used for the
implementation of unitarization projections for operators and
resonances. The reason is the global structure of the unitarization
projection.  Therefore the models described in the current paper have
been manually added to the framework.

For each resonance type ($\sigma,\phi,f, X$), we can compute the relation of the
resonance width (section~\ref{sec:width}) to the operator coefficients in the
low-energy EFT (section~\ref{sec:matching}) which result when the
resonance is integrated out. These relations are listed in
Table~\ref{tab:gamma}. 

\begin{table}[htb]
\centering{%
  \begin{tabular}{c| M{2 cm}  M{2 cm}  M{2 cm} M{2 cm} N}
    & $\sigma$ & $\phi$  & $f$  &  $X$ & \\
    \hline
    $F_{S,0}$ & $\frac{1}{2}$ & $2$ & $15$& 5 &\\[4 ex]
    $F_{S,1}$ & -- & -$\frac{1}{2}$ & -$5$&  -35 &
  \end{tabular}
}
\caption{Relation of resonance width $\Gamma$ and mass $M$ to the
  corresponding $D=8$ operator coefficients in the low-energy EFT, for all
  resonance types considered in this paper.  The
  factors listed in the table have to be multiplied by $32 \pi \Gamma/M^5$.}
\label{tab:gamma}
\end{table}

The analysis of LHC run-I data by the ATLAS
experiment~\cite{Aad:2014zda} has been cast into 
bounds on the EFT parameters $F_{S,0}$ and $F_{S,1}$, namely 
\begin{equation}
  |F_{S,0}| < 480\;\TeV^{-4}
\qquad
  |F_{S,1}| < 480\;\TeV^{-4} \qquad ,
\end{equation}
where only one parameter was varied at a time.  This analysis covered the
same-sign leptonic decay channel of $W^+W^+$ and $W^-W^-$.  It was based on
the T-matrix unitarized version of the extrapolated EFT as its reference
model, with the pure SM as the limit for vanishing parameters.  A CMS
analysis can be found in~\cite{CMS:2015jaa}

\subsection{On-shell Invariant Mass Distributions}

In the following, we will present results both for on-shell $W/Z$
final states and for complete partonic final states.  On-shell vector
bosons cannot be detected directly but their distributions directly
reflect the actual features of the physical model.  Observable
distributions of fermions in the final state, which may be quarks
(jets), charged leptons, or neutrinos, are less directly linked to the
physical process and require detailed analysis along the lines
of~\cite{Aad:2014zda}.  This concerns, in particular, the separation of
signal and background based on detector data, which is beyond the
scope of the present paper.

We show results for particular parameter sets where we add one resonance at a
time on top of the SM, namely a scalar-isoscalar, tensor-isoscalar, or
scalar-isotensor resonance, respectively.  All extra higher-dimensional
operator coefficients are set to zero.  By varying the resonance parameters
within reasonable limits, this gives an overview of the expected
phenomenology.

For definiteness, we choose to plot the invariant mass of the vector-boson
pair system in the final state, which is the energy scale of the actual VBS
process.  The initial state is convoluted with the parton structure functions,
so the results hold for the LHC ($\sqrt{s}=14\;\TeV$), and
we apply standard VBS cuts to enhance the signal.  The final-state vector
bosons are taken on-shell. We show the distribution for the $W^+W^+$ and $ZZ$
final states, where the latter case as the \emph{golden channel} of VBS is
distinguished by the fact that the $ZZ$ invariant mass can be reconstructed
from the leptonic $Z$ decays.  This is not possible for $W^+W^+$, but the
corresponding same-sign lepton channel is distinguished by a favorable
signal-to-background ratio.  Note that in the on-shell plots, the vector-boson
decay branching ratios have not been included.

In all invariant-mass plots, we display the distribution for the unitarized
resonance model (blue curves) together with the pure SM prediction (black).
We also plot the unitarity bound for the appropriate partial wave,
extrapolated off-shell by the same algorithm, as a dashed curve (black).  For
illustrative purposes, we also display, in each case, the unitarized
extrapolation of the low-energy EFT (red, solid), where we choose the operator
coefficients equal to the formal result of integrating out the resonance.
Finally, we also display numerical results for the EFT without unitarization
(red, dashed) and the resonance with correct width but no further
unitarization (blue, dashed).

\subsubsection{Isoscalar-Scalar}

The simplest case is a scalar-isoscalar resonance.  This is a single isolated
resonance, as it could arise, e.g., as the extra scalar particle in a
singlet-doublet Higgs model or as a low-energy signal of a strongly
interacting Higgs sector that is neutral under the SM gauge group.

\begin{figure}[t]
    \includegraphics[width=0.48\linewidth]{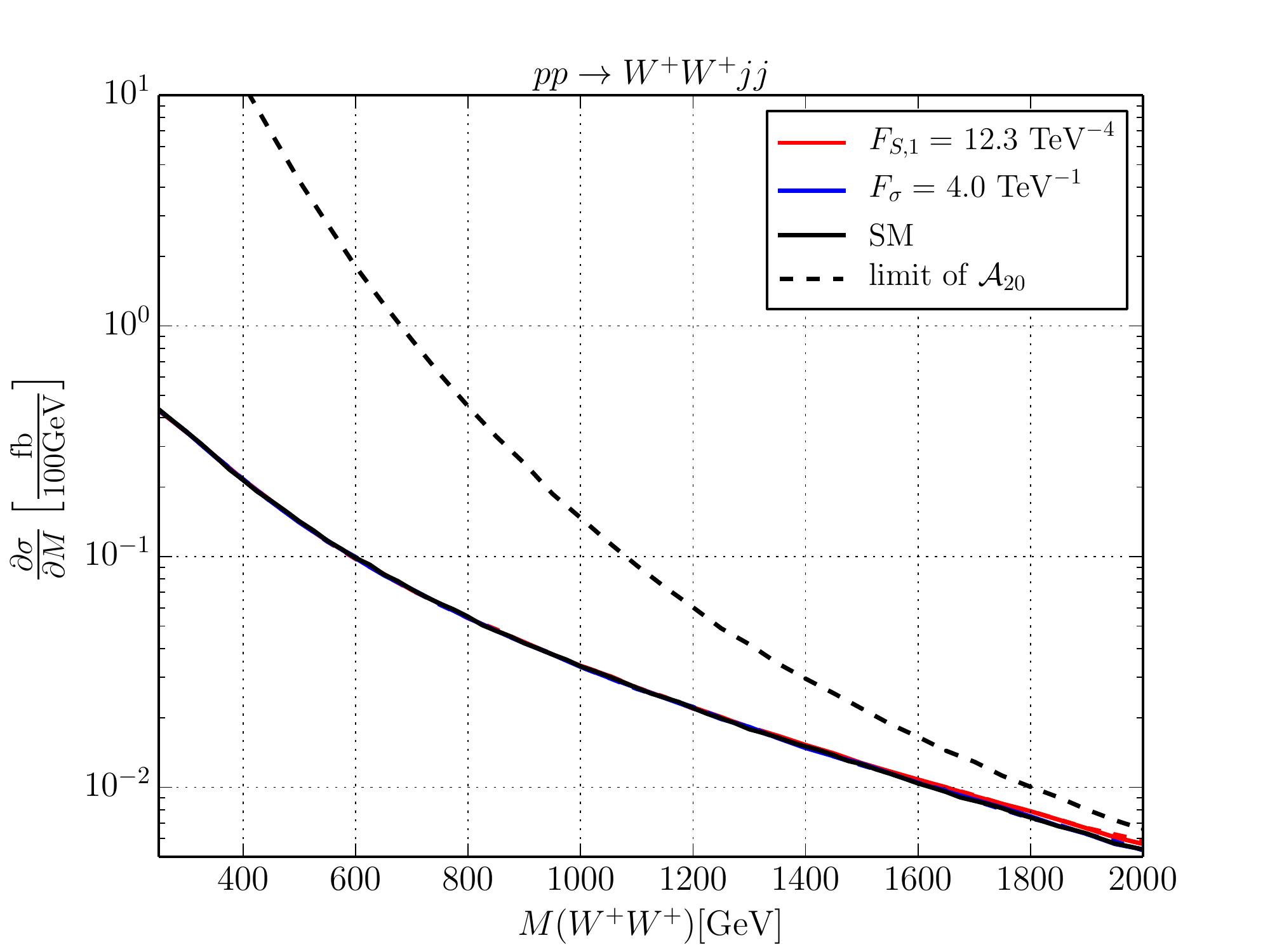}
    \includegraphics[width=0.48\linewidth]{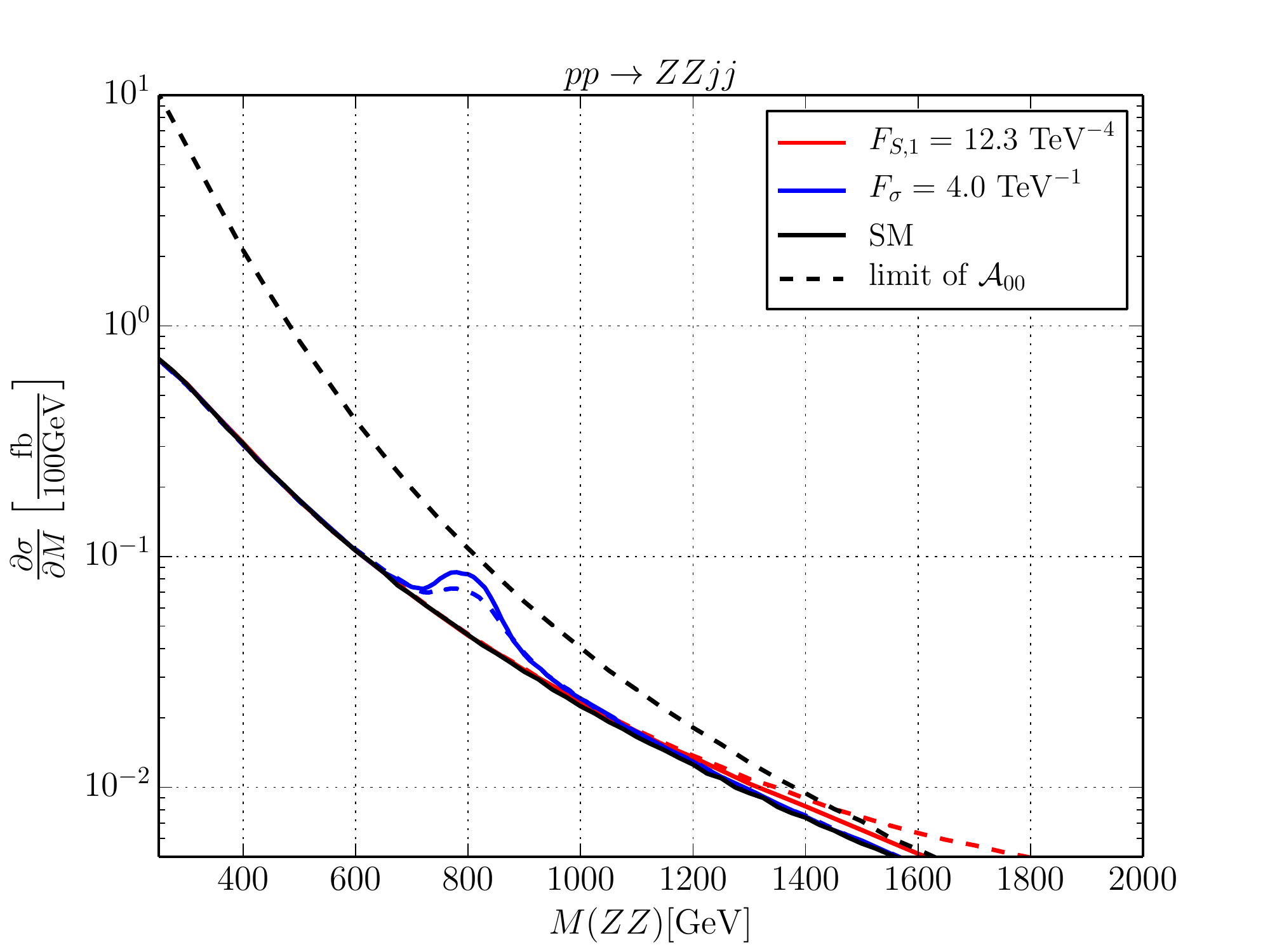}
    \\
    \includegraphics[width=0.48\linewidth]{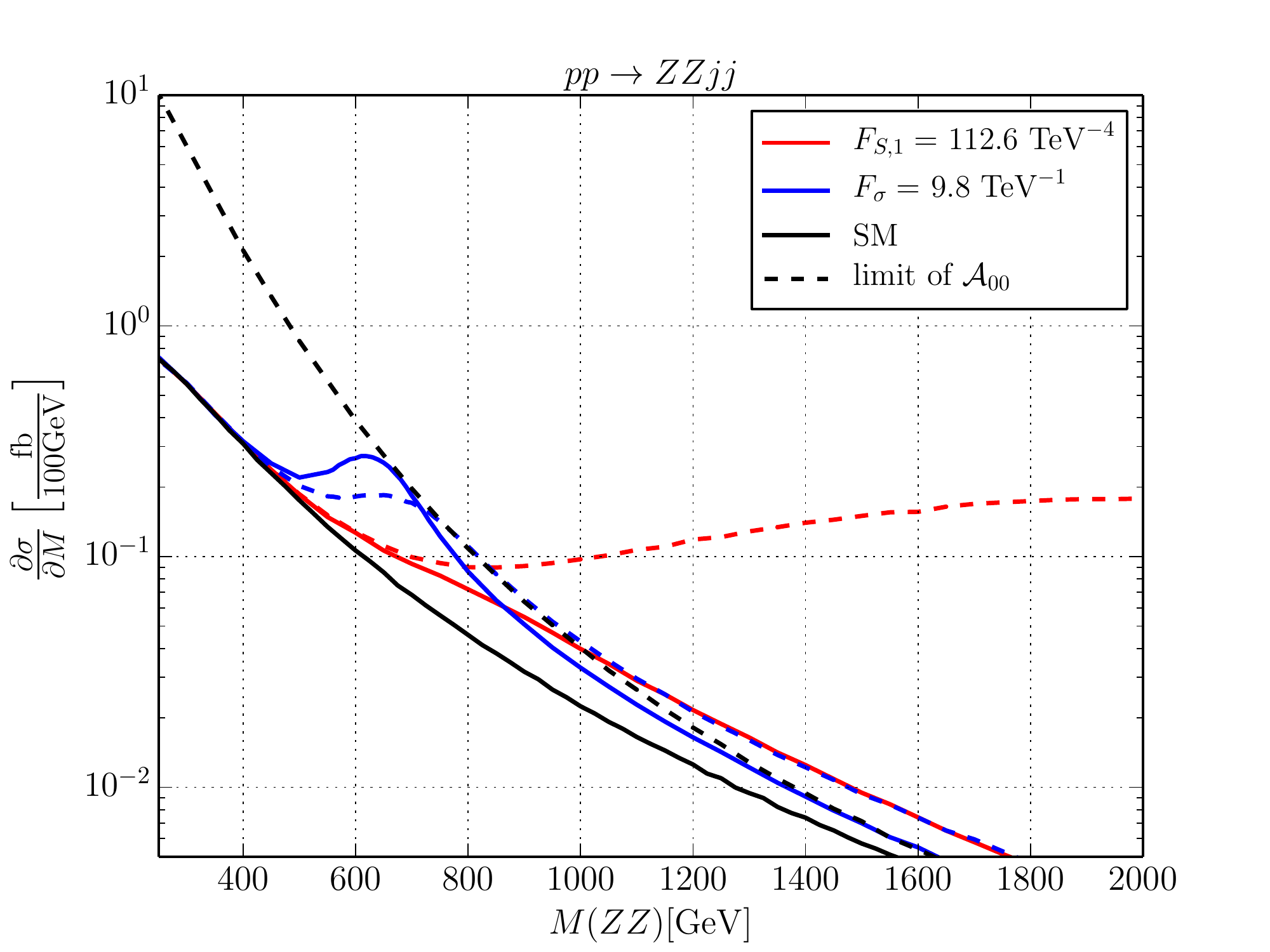}
  \caption{ 
    Differential cross sections for isoscalar scalar
    resonances. Upper plots show a weakly coupled isoscalar
    scalar with $m_\sigma=800 \, \mathrm{GeV}$ and $\Gamma_\sigma=80\,
    \mathrm{GeV}$, for the processes $pp\rightarrow W^+W^+jj$ (left)
    and $pp\rightarrow ZZjj$ (right), respectively. In the lower plot,
    there is a low lying isoscalar scalar with $m_\sigma=650 \,
    \mathrm{GeV}$ and $\Gamma_\sigma=260\, \mathrm{GeV}$ for the
    process $pp\rightarrow ZZjj$. Solid line: unitarized results,
    dashed lines: naive result, black dashed line: Limit of saturation
    of $\mathcal{A}_{20}$ $(W^+W^+)$ or $\mathcal{A}_{00}$ $(ZZ)$,
    respectiveluy. Cuts: $M_{jj} > 500$ GeV;  
    $\Delta\eta_{jj} > 2.4$; $p^j_T > 20$ GeV;
    $|\eta_j| > 4.5$. } 
  \label{fig:isoscalar-scalar}
\end{figure}

In Fig.~\ref{fig:isoscalar-scalar}, upper row, we have selected a
moderate mass of $800\;\GeV$ and a rather narrow width of $80\;\GeV$,
which corresponds to a weak coupling.  The isolated resonance is
clearly visible in the $ZZ$ channel, while the $W^+W^+$ channel is
barely affected.  For such weak coupling, the operator coefficient in
the EFT is small and more than one order of magnitude below the
current LHC run-I limit.  We can draw the conclusion that in this case
the resonance should be detectable for sufficient luminosity, but the
EFT approximation is not useful. 

Turning to a stronger coupling, we show the corresponding distribution in the
$ZZ$ channel for $m_\sigma=650\;\GeV$ and $\Gamma_\sigma =260\;\GeV$ in
Fig.~\ref{fig:isoscalar-scalar}, lower row.

Here, the EFT parameters are within the range that should become accessible at
LHC run II and beyond.  The EFT curve (red, solid) appears correctly
as the Taylor expansion of the resonance curve (blue) for low energy.
However, the energy region where the deviation from the SM becomes
sizable, already coincides with the resonance peak region, so the EFT
considerably underestimates the event yield.  Beyond the resonance,
the EFT misses the fact that the distribution falls down again,
approaching the SM prediction (black) from above. 

The result also demonstrates that the additional unitarization of the scalar
resonance beyond the Breit-Wigner approximation with constant width is
essential, as is seen by comparing the blue and blue-dashed curves.  The naive
EFT result without unitarization (red, dashed) grossly overshoots all
conceivable models, which should not cross the unitarity limit (black-dashed).


\subsubsection{Isoscalar-Tensor}

\begin{figure}[t]
    \includegraphics[width=0.48\linewidth]{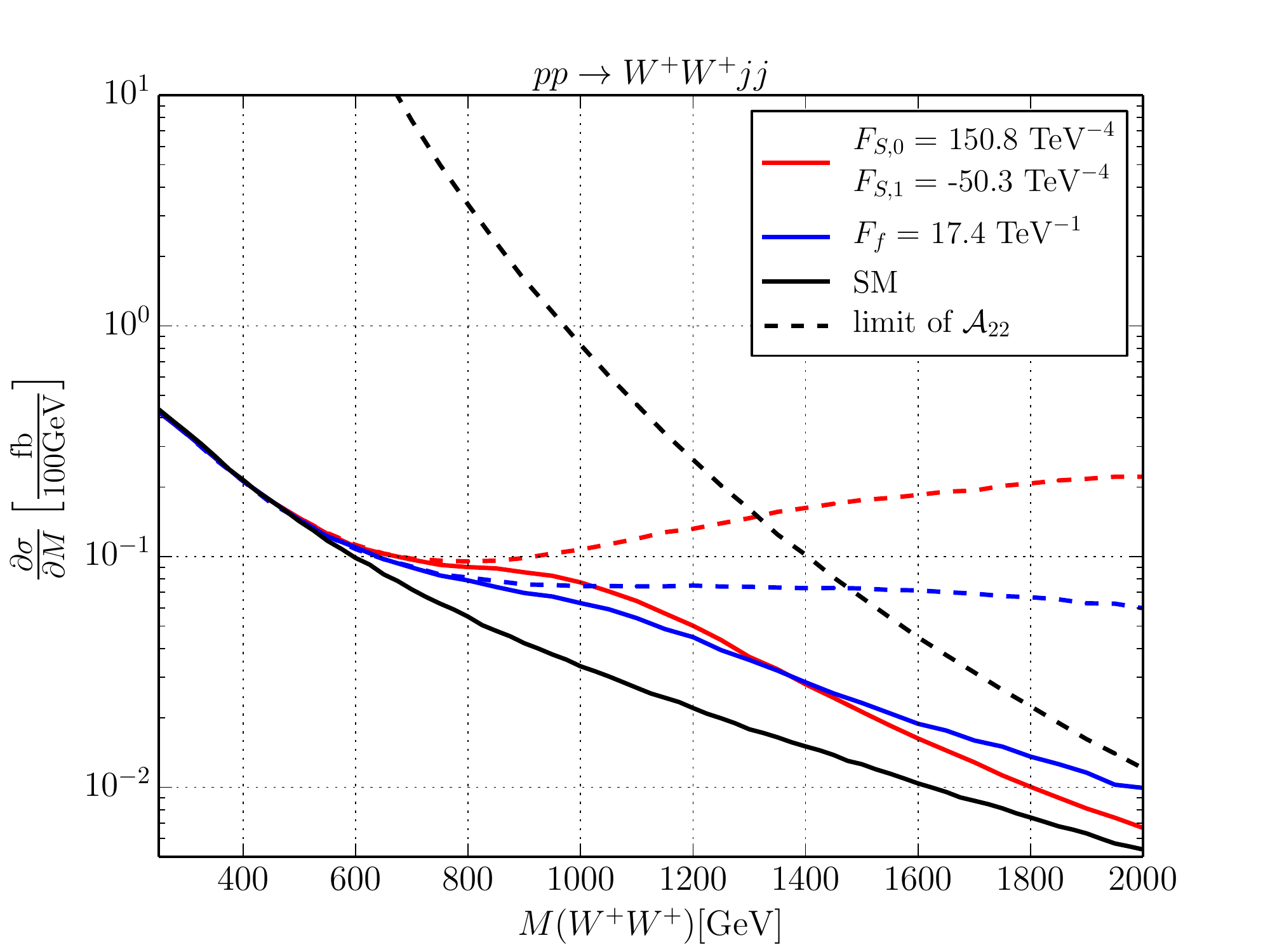}
    \includegraphics[width=0.48\linewidth]{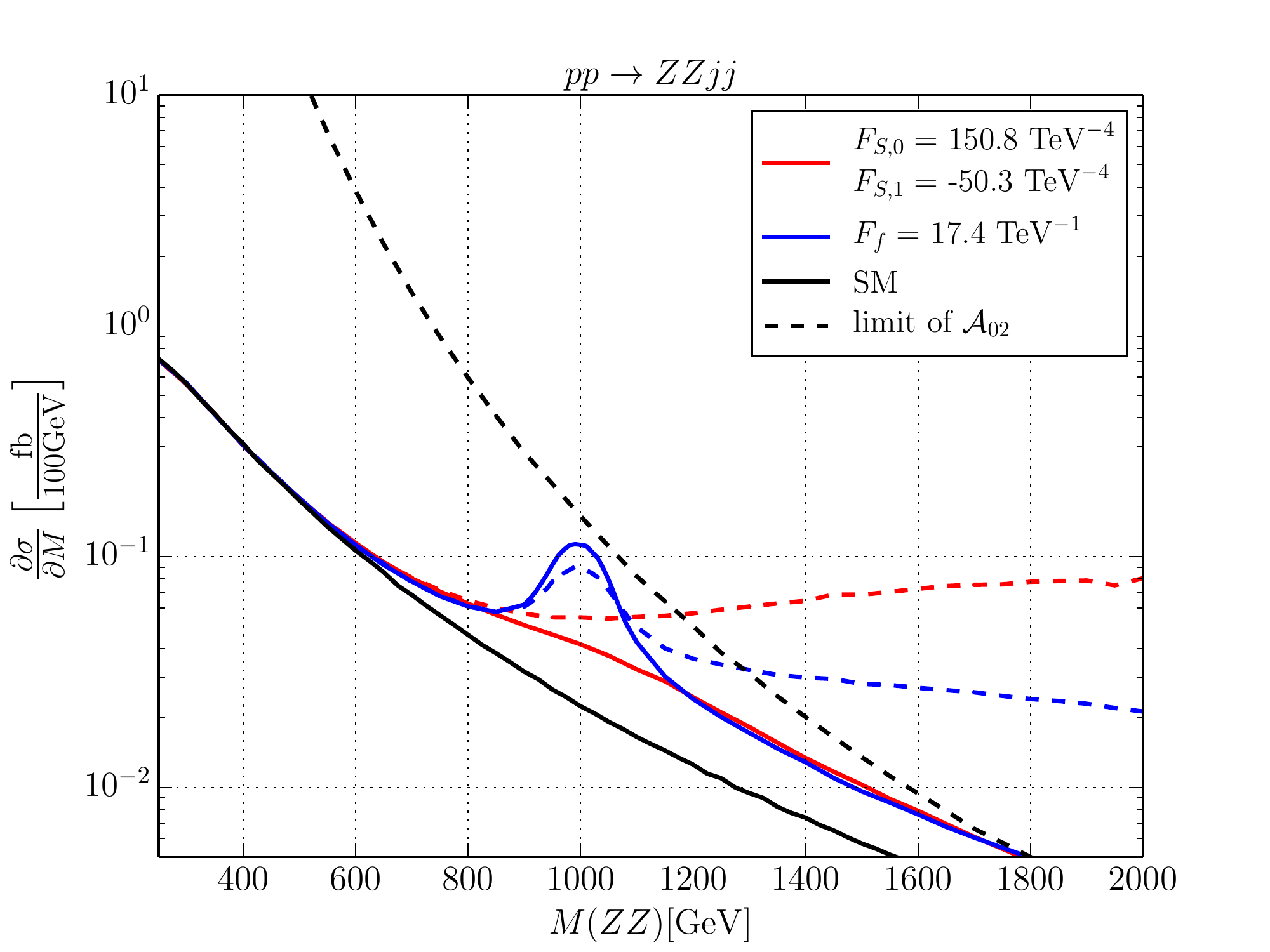} \\ 
    \includegraphics[width=0.48\linewidth]{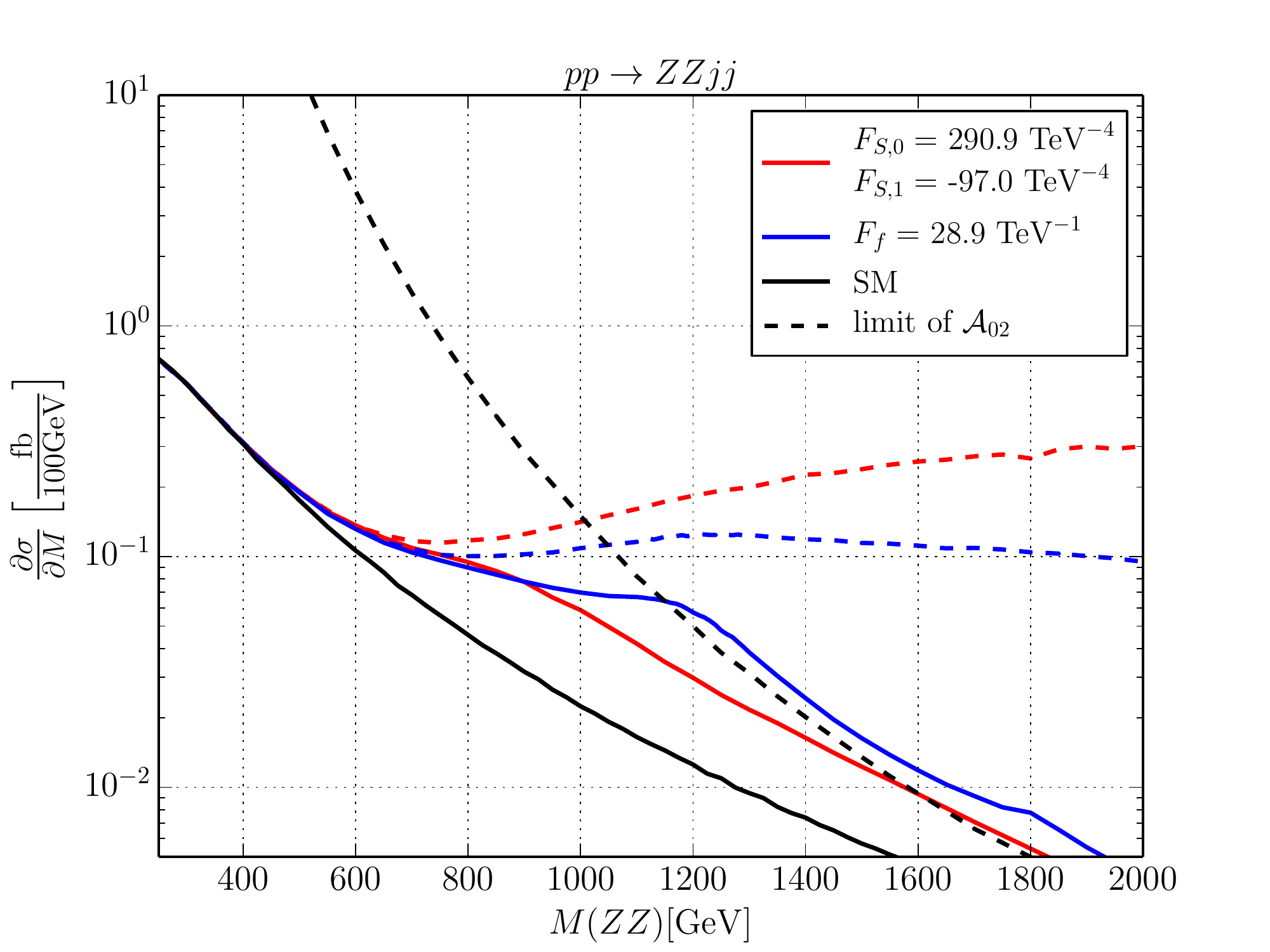}
  \caption{ 
    Differential cross sections of an isoscalar tensor
    resonance. Upper plots show a resonance with $m_f= 1000 \,
    \mathrm{GeV}$ and $\Gamma_f= 100 \, \mathrm{GeV}$ for the
    processes $pp\rightarrow W^+W^+jj$ (left), and $pp\rightarrow
    ZZjj$ (right), respectively. The lower plot is for a strongly
    interacting isoscalar tensor with $m_f= 1200 \, \mathrm{GeV}$ and
    $\Gamma_f= 480 \, \mathrm{GeV}$. 
    Solid line: unitarized results, dashed lines: naive result,
    black dashed line: Limit of saturation of $\mathcal{A}_{22}$ $(W^+W^+)$ or
    $\mathcal{A}_{02}$ $(ZZ)$, respectively. 
    Cuts are the same as in Fig.~\ref{fig:isoscalar-scalar}.} 
  \label{fig:isoscalar-tensor}
\end{figure}

As can be observed from Table~\ref{tab:gamma}, a tensor resonance has a
stronger impact on the low-energy EFT than a scalar resonance of equal width.
In Fig.~\ref{fig:isoscalar-tensor}, upper row, we display the
distributions for a tensor isoscalar resonance with mass
$m_f=1000\;\GeV$ and width $\Gamma_f =100\;\GeV$. 

The resonance visibly modifies the distribution already at low energy, such
that the EFT analysis, given sufficient sensitivity, should catch the
deviation from the SM.  Nevertheless, the excess at the peak in the $ZZ$
channel is sizable.  Beyond the resonance, unitarization is essential in the
tensor case.  In the $W^+W^+$ final state the tensor enters only as t-channel
exchange , so there is no resonance but a broad enhancement.  This enhancement
is rather well described by the corresponding unitarized
EFT~\footnote{Tensor resonances resulting in peaks in diboson spectra
  to explain a recent excess in ATLAS data around 2 TeV can be found
  e.g. in~\cite{Fichet:2015yia}.}.  

As in the scalar case, the curves without unitarization do not provide a
useful phenomenological description.

In Fig.~\ref{fig:isoscalar-tensor}, lower row, we consider a heavy
tensor-isoscalar with strong coupling, $m_\phi=1200\;\GeV$ and
$\Gamma_\phi=480\;\GeV$.  The resonance peak appears as a broad enhancement, which
extends to both low and high energies.  The EFT approximation, with sizable
coefficients, is rather accurate in this case.  The actual resonance curve
shows a nontrivial threshold structure which corresponds to the interplay of
all partial waves which are excited by s-channel and t-channel exchange
contributions.  However, we should keep in mind that the prediction for such a
strong coupling is uncertain in any case and should not be taken too
seriously.

\newpage
\clearpage

\subsubsection{Isotensor-Scalar}

\begin{figure}[t]
    \includegraphics[width=0.48\linewidth]{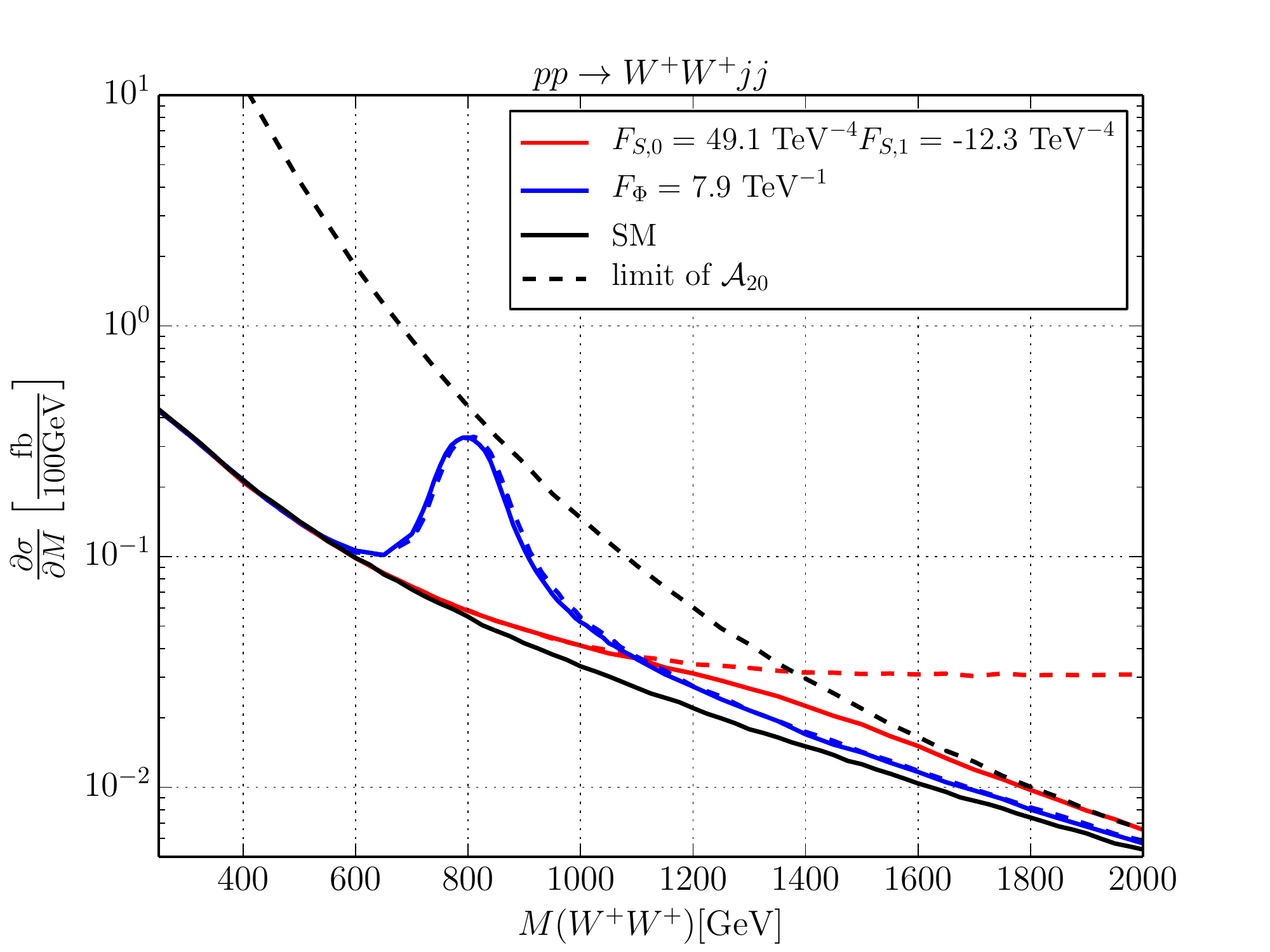}
    \includegraphics[width=0.48\linewidth]{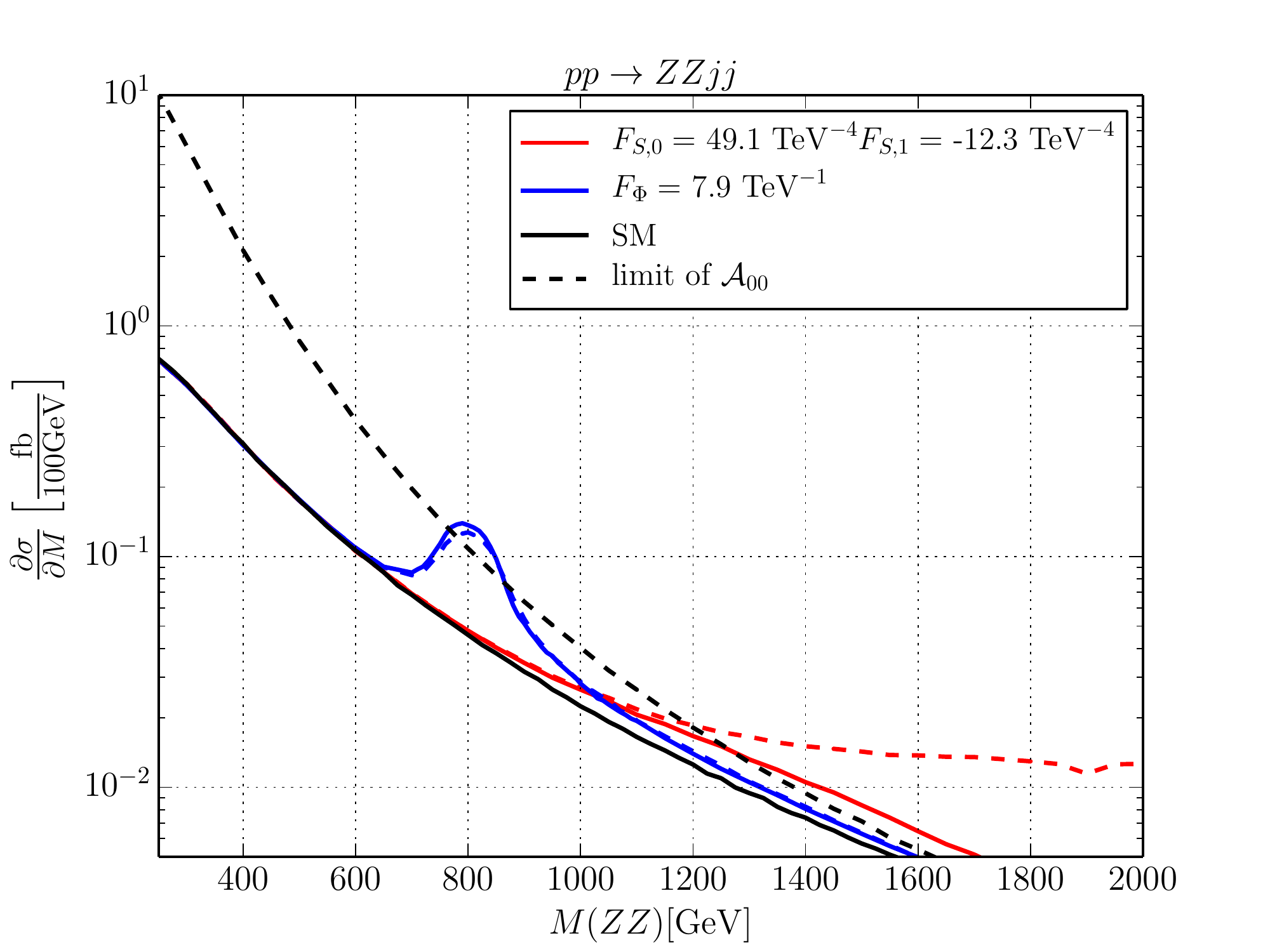} \\ 
    \includegraphics[width=0.48\linewidth]{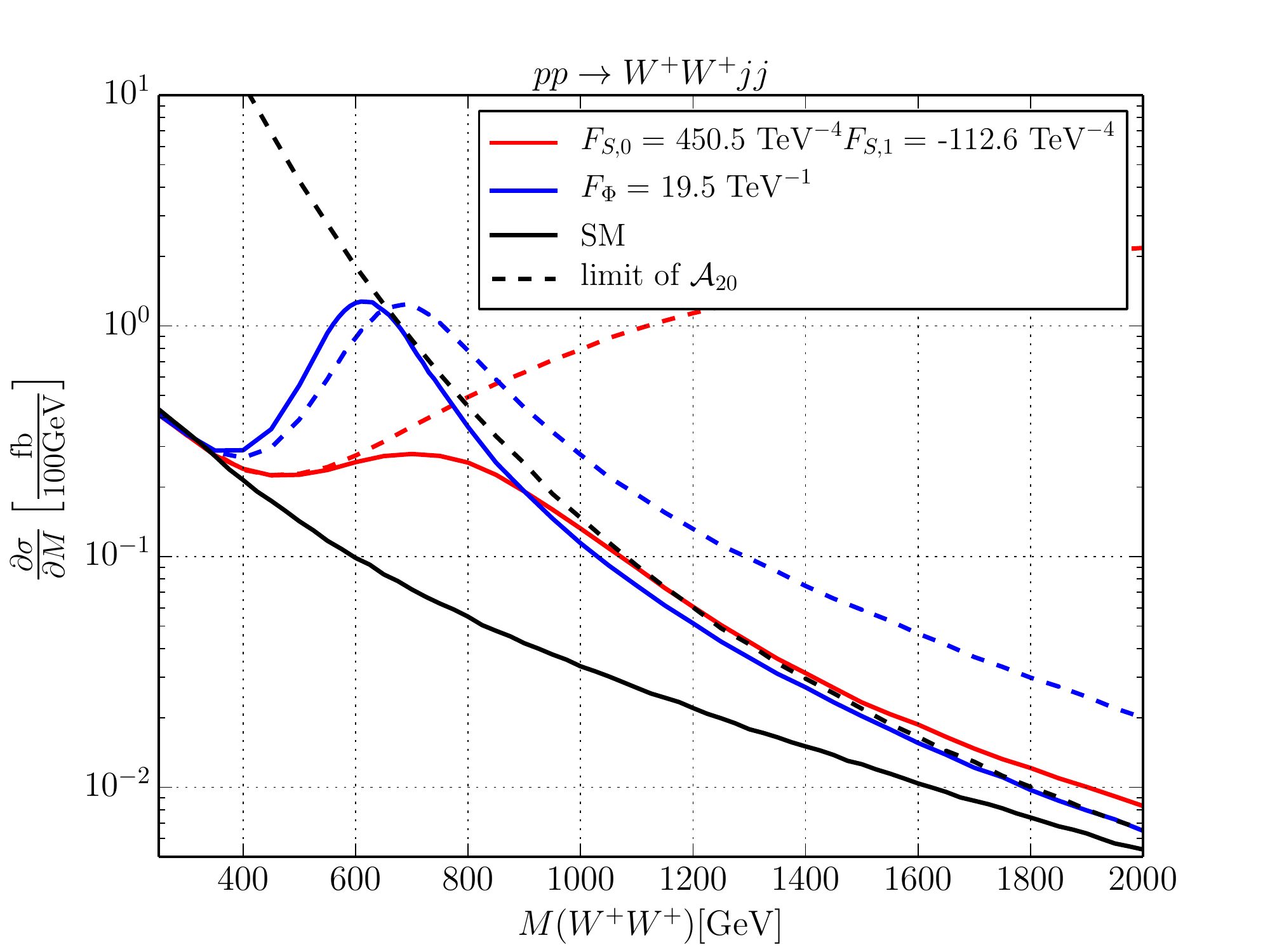}
  \caption{ 
    Differential cross sections of an isotensor scalar
    resonance. Upper plots show a resonance with $m_\phi=800 \,
    \mathrm{GeV}$ and $\Gamma_\phi=80\, \mathrm{GeV}$ for the
    processes $pp\rightarrow W^+W^+jj$ (left), and $pp\rightarrow
    ZZjj$ (right), respectively. The lower plot shows a low-lying
    isotensor scalar with $m_\phi=650 \, \mathrm{GeV}$ and
    $\Gamma_\phi=260\, \mathrm{GeV}$ for the process $pp\rightarrow
    W^+W^+jj$. Solid line: unitarized results, dashed lines: naive
    result, black dashed line: Limit of saturation of
    $\mathcal{A}_{20}$ $(W^+W^+)$ or $\mathcal{A}_{00}$ $(ZZ)$,
    respectively. Cuts are the same as in
    Fig.~\ref{fig:isoscalar-scalar}.}  
\label{fig:isotensor-scalar}
\end{figure}

Turning to the isotensor case, we now get a resonance in all final states
including $W^+W^+$.  This is illustrated by the plots in
Fig.~\ref{fig:isotensor-scalar} for $m_\phi=800\;\GeV$ and
$\Gamma_\phi=80\;\GeV$. 

Due to the large number of degrees of freedom (nine states which are
degenerate in mass), the peak is rather prominent while the low-energy
EFT parameters are again small.  We observe that the peak value is slightly
below ($W^+W^+$) and above ($ZZ$) the appropriate unitarity limit,
respectively.  This is the effect of $t$-channel exchange which also contributes
and can have either sign.

Contrary to the weakly interacting scenario, 
a non-unitarized low-lying and strongly interacting isotensor-scalar with
mass of $m_\phi= 650 \, \mathrm{GeV}$ and width 
$\Gamma_\phi = 260 \, \mathrm{GeV}$ 
violates the $\amp_{20}$ slightly above the resonance
as illustrated in Fig.~\ref{fig:isotensor-scalar}.
Therefore, a unitarization is needed 
for this strongly interacting resonance.
The low-energy effective field theory approach 
does only coincide in the unitarized case
at high energies, because
the eigenamplitudes of the isotensor-scalar
as well as the dimension-eight operators
are already saturated through the T-matrix formalism.

\newpage
\clearpage
\subsubsection{Isotensor-Tensor}

\begin{figure}[t]
    \includegraphics[width=0.48\linewidth]{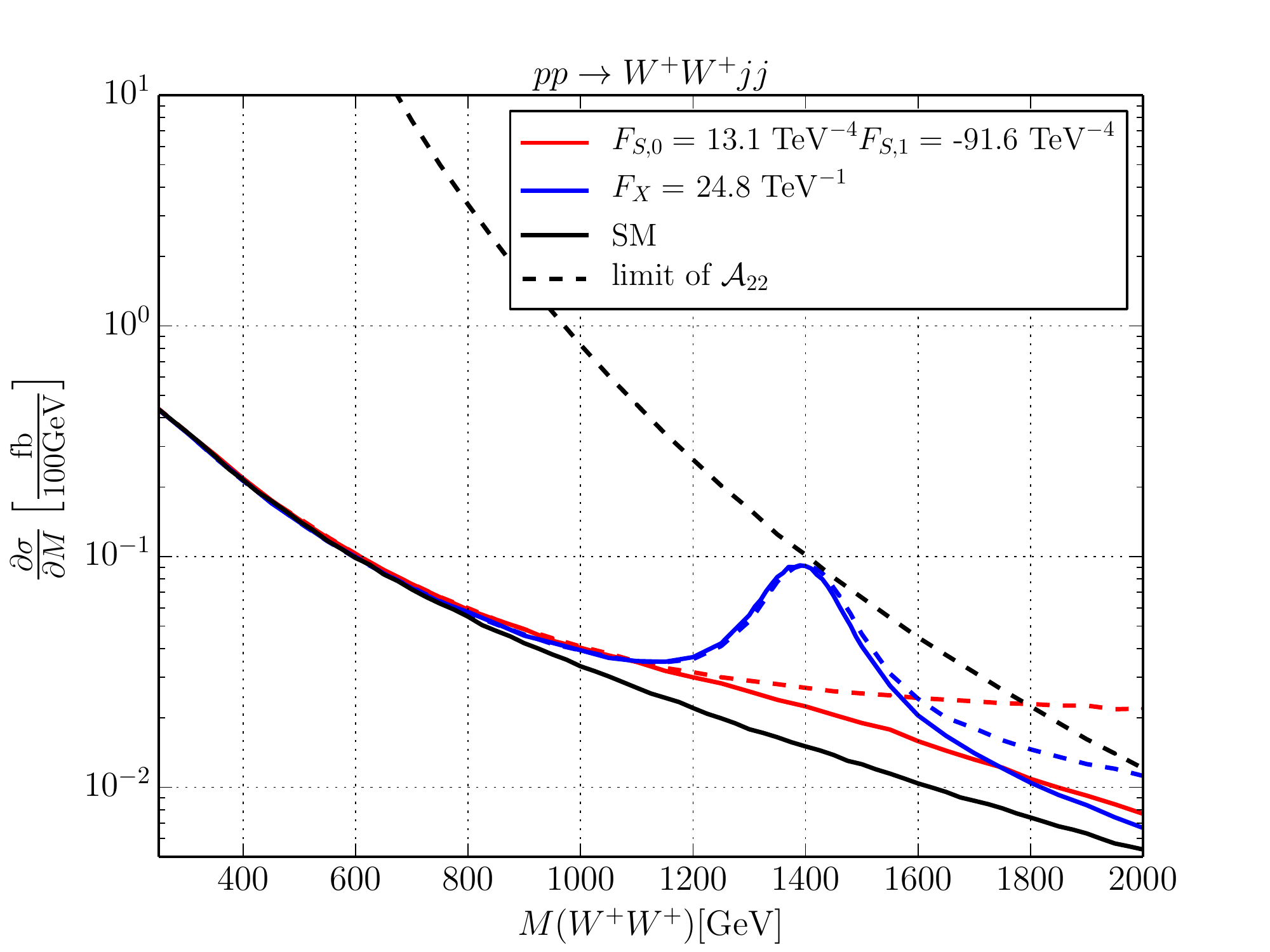}
    \includegraphics[width=0.48\linewidth]{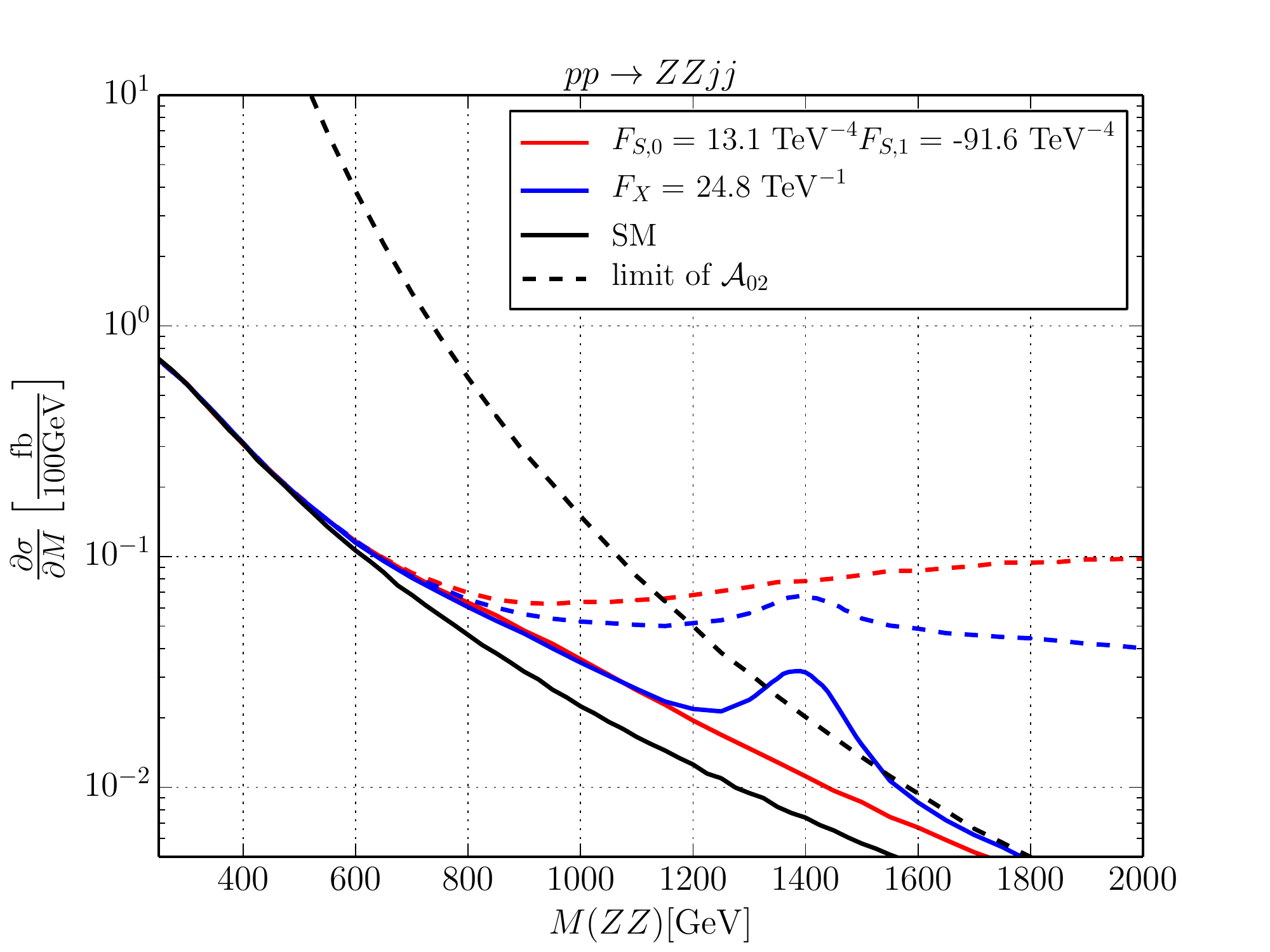} \\
    \includegraphics[width=0.48\linewidth]{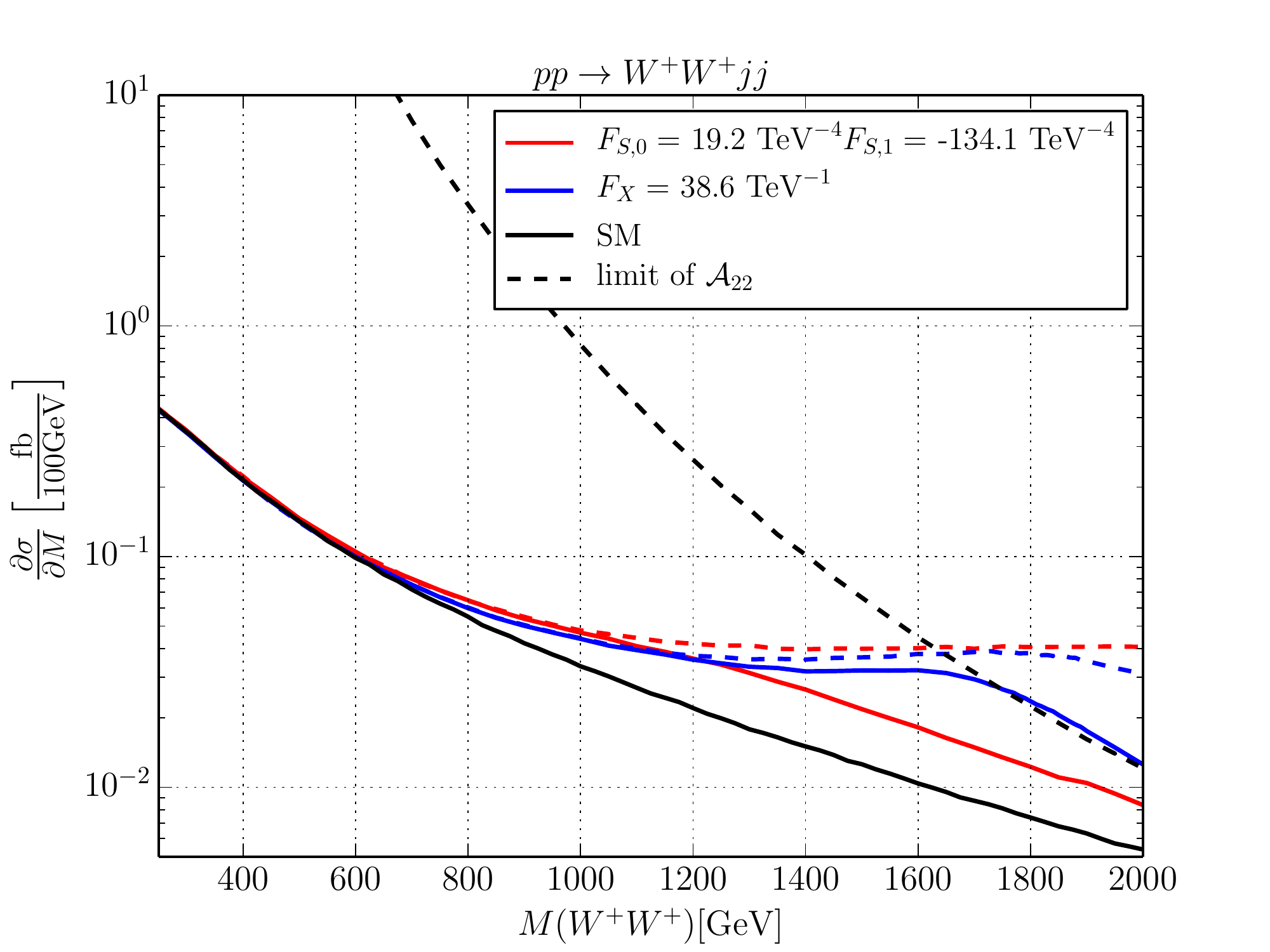}
  \caption{ 
    Differential cross sections of an isotensor tensor
    resonance. Upper plots show a resonance with $m_X= 1400 \,
    \mathrm{GeV}$ and $\Gamma_X= 140 \, \mathrm{GeV}$ for the
    processes $pp\rightarrow W^+W^+jj$ (left), and $pp\rightarrow
    ZZjj$ (right), respectively. The lower plot shows a strongly
    interacting isotensor tensor with $m_X= 1800 \, \mathrm{GeV}$ and
    $\Gamma_X= 720 \, \mathrm{GeV}$ for the process $pp\rightarrow
    W^+W^+jj$. Solid line: unitarized results,
    dashed lines: naive result,  black dashed line: Limit of saturation of 
    $\mathcal{A}_{22}$ $(W^+W^+)$ or $\mathcal{A}_{02}$ $(ZZ)$,
    respectively. Cuts are the same as in
    Fig.~\ref{fig:isoscalar-scalar}.}  
  \label{fig:isotensor-tensor}
\end{figure}

Similarly to the isotensor-scalar, every vector-boson scattering channel
receives a resonant contribution from the isotensor-tensor
multiplet. The $W^+W^+$ and $ZZ$ channel distributions of the
isotensor-tensor resonance with mass $m_X=1400 \, \mathrm{GeV}$ width
$\Gamma_X = 140 \, \mathrm{GeV}$ are plotted 
in Fig.~\ref{fig:isotensor-tensor}, upper row. Due to the bound of
equation \eqref{eq:completebound_isotensor-tensor-scalar_width}, the
mass  of the isotensor-tensor has to be chosen slightly higher than
the mass of the isoscalar-tensor in Fig.~\ref{fig:isoscalar-tensor}
when leaving the ratio of width and mass invariant. 

The effective field theory with the dimension-eight operators
coincides with the onset of the isotensor-tensor peak. 
Starting slightly below the resonance, the 
resonant cross section deviates from 
the effective field theory description. Analogously to the
isotensor-scalar, the very distinctive peak of the isotensor-tensor  
is not captured by the dimension-eight operators.
In the $W^+W^+$- channel, even the non-unitarized 
resonance contribution stays within the unitarity bound of
$\amp_{22}$. Contrary to the isotensor-scalar, the isotensor-tensor
needs unitarization  
for the $ZZ$ final state due to the large tensor contributions
in the $t$- and $u-$channel. The
non-unitarized amplitudes violate the $\amp_{02}$ unitarity
already below the mass of the resonance.
Even the resonance peak is hardly visible.
The unitarized 
resonance curve shows a peak, although it is slightly above the 
unitarity bound.

In a strongly interacting scenario ($\Gamma_X = 720 \, \mathrm{GeV}$ ),
the unitarized isotensor-tensor resonance peaks
below its actual mass at $m_X = 1800 \, \mathrm{GeV}$. 
This peak originates from the already saturated eigenamplitudes,
which then fall due to the parton distribution functions at high
energies. Besides the resonance peak, the low-energy effective field
theory coincides with the isotensor-tensor for both unitarized and
non-unitarized results. This is shown in the lower plot of 
Fig.~\ref{fig:isotensor-tensor}.

\newpage
\clearpage
\subsection{Results for Complete Processes}

The actual analysis of LHC data will have to exploit cross sections and
distributions for the complete final state which consists of the two tagging
jets and the decay products of the vector bosons.  In this paper, we only
investigate the $ZZ$ channel with its decay into four leptons, selecting the
$e^+e^-\mu^+\mu^-$ final state.  This process is straightforward to analyze,
but suffers from the low leptonic branching ratio, so for our simulation we
assume the high-luminosity mode of the LHC with integrated luminosity of
$3\;\ab^{-1}$.  We anticipate that by including also the leptonic $WW$ final
state and hadronic final states, the results can be considerably improved.

\begin{figure}[t]
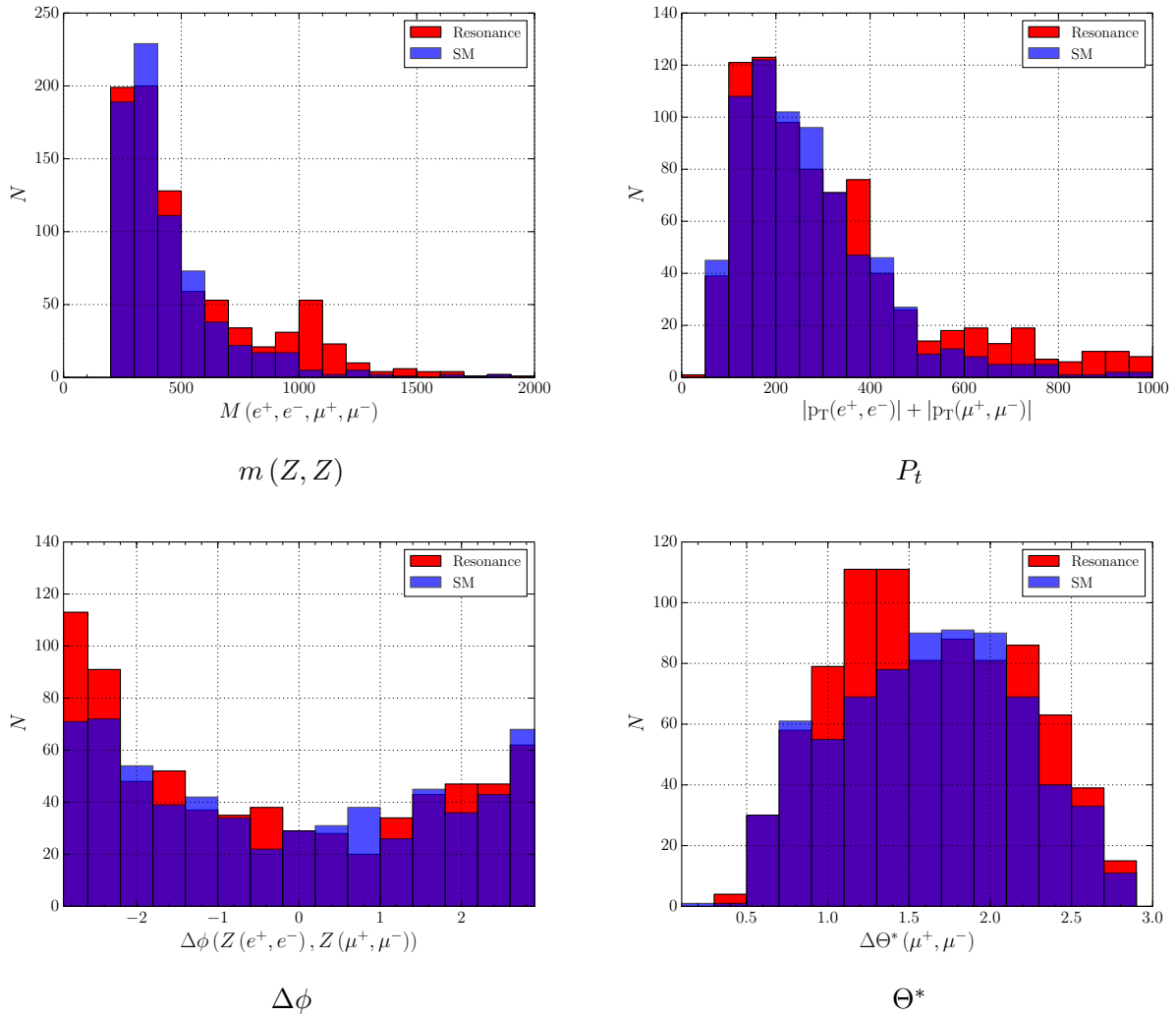

  \begin{tabular}{c c }
    \includegraphics[width=0.5\linewidth]{{{VBS_ppe2mu2_f_r-0.1_1000-mww}}}
    &
    \includegraphics[width=0.5\linewidth]{{{VBS_ppe2mu2_f_r-0.1_1000-pt}}}
    \\
    $m \left (Z,Z \right)$ 
    &
    $P_t$
    \\
    \includegraphics[width=0.5\linewidth]{{{VBS_ppe2mu2_f_r-0.1_1000-phi}}}
    &
    \includegraphics[width=0.5\linewidth]{{{VBS_ppe2mu2_f_r-0.1_1000-theta_star}}}
    \\
    $\Delta\phi$ & $\Theta^*$ \\
  \end{tabular}
  \caption{ $pp\rightarrow e^+ e^- \mu^+ \mu^- jj$ at $\sqrt{s} = 14
    \, \mathrm{TeV}$  with luminosity of $3000 \, \mathrm{fb}^{-1}$  
    with isoscalar tensor at $m_f=1000$ GeV and $\Gamma_f $=100 GeV.
    Cuts: $M_{jj} > 500$ GeV; 
    $\Delta\eta_{jj} > 2.4$;
    $p^j_T > 20$ GeV;
    $|\eta_j| > 4.5$;
    $100\; \mathrm{GeV} > M_{e^+e^-} > 80\; \mathrm{GeV}$;
    $100\; \mathrm{GeV} > M_{\mu^+\mu^-} > 80\; \mathrm{GeV}$.
  } 
  \label{fig:eemmjj}
\end{figure}

The simulation generates event samples for the complete process with all
Feynman graphs, so there is no restriction on resonant vector bosons
as the origin of the final-state leptons.  We apply standard VBS cuts and
compare, in Fig.~\ref{fig:eemmjj}, various distributions for the SM (blue),
resonance model with a single isoscalar-scalar (red), and the unitarized
low-energy EFT (purple).

The resonance with mass $m=1000\;\GeV$ and width $\Gamma=100\;\GeV$
appears, as expected, in the invariant mass distribution and, more
indirectly, in other plots. Clearly, this parameter set is at the
margin of observability in this single channel.  The situation
obviously improves if we consider resonances with lower mass, larger
coupling, in higher representions, and add other analysis channels.

\section{Conclusions}

The Higgs sector of the SM, after the discovery of a light Higgs, is a
new field of study for the experiments at the LHC, and beyond.  While
the SM yields precise predictions in accordance with the notion of a
weakly coupled theory, a thorough analysis of electroweak data
should be guided by reference simplified models which differ from the SM.
Extending the EFT by higher-dimensional operators is useful for
analyzing observables with bounded energy, but open scattering data
require enforcing unitarity and extrapolating into a region where
perturbation theory in the EFT is insufficient.

Without reference to any particular high-energy model, we have
augmented the EFT by resonances with even spin, namely scalar or
tensor.  Assuming exact $SU(2)_L\times U(1)_Y$ gauge invariance and, for
simplicity, approximate custodial symmetry both in the EFT and beyond,
we can distinguish four distinct resonance multiplets with a single
free mass and coupling parameter each.  This class of models includes
the decoupling limit of multi-Higgs models and certain aspects of
massive-graviton models.

The models are set up such that we need only take the interaction with
the Higgs sector into account, while couplings to the gauge and
fermion sectors occur only via mixing.  This is consistent with the
symmetry assumptions and with our knowledge about electroweak
precision data, although it is clearly not guaranteed.  The models
allow for arbitrary higher-dimensional operators in the EFT, unrelated
to resonance exchange, so we do not lose generality.

All amplitude calculations are meaningless unless we enforce
quantum-mechanical unitarity, since naive extrapolations yield event
rates in the high-energy region that can exceed the unitarity bounds
by orders of magnitude.  We have consistently implemented the T-matrix
unitarization scheme which works on the complex scattering matrix of
the model directly, simplified for the asymptotic range where
longitudinal and transveral degrees of freedom decouple.

We have studied the case of a tensor resonance in detail.  Since we do
not necessarily restrict ourselves to states that are related to
gravity, the model differs from 
the various massive-graviton models and studies that can be found in
the literature.  To our knowledge, the coupling of a generic tensor
resonance to the Higgs sector and the resulting predictions for the
LHC have not been considered in detail before.  We find that by
employing a St\"uckelberg procedure for the implementation in the
Lagrangian, instead of the classic Fierz-Pauli approach, we are able
to set up the extended EFT for an isolated tensor resonance manifestly
separated from non-resonant effects.  Scalar and tensor
resonances can be handled in close analogy. It turns out that it
is possible to extend an effective theory with an isolated tensor
resonance up to a cutoff of order $\Lambda\lesssim M^2/m_H$, where $M$
is the resonance mass, and $m_H$ is the physical Higgs mass.

We have implemented the models in the Monte-Carlo package WHIZARD and
computed exemplary distributions and simulated event samples for the
LHC.  The numerical results illustrate that resonances in VBS may be
detected at the LHC within a certain range of mass and coupling
values.  For a final verdict, it will be necessary to perform a
complete experimental study and analysis, based on exclusive event
samples in combination with background and detector description.  We
also find that the comparison with pure-EFT results can be misleading
if resonance and background cannot be clearly separated, as it is
typical for the situation at the LHC.  We conclude that data should be
analyzed on base of resonance models as well as pure-EFT simulations.
This holds, in particular, if limits or values are to be
combined between distinct final states or with data obtained at a
future lepton collider like the
ILC~\cite{Baer:2013cma,Behnke:2013lya}. There has 
been a first study similar to the one presented here, investigating
resonances of spins and isospins zero, one and two in 1 TeV lepton
collisions~\cite{Beyer:2006hx}, where issues of unitarization did not
play a role.

\subsection*{Acknowledgments}

WK and JRR want to thank for the hospitality at the Institute for High Energy
Physics (IHEP) and the Center for Future High Energy Physics (CFHEP)
of the Chinese Academy of Sciences at Beijing, China, where parts
of this work have been completed. MS acknowledges support 
by JSPS and DAAD and thanks for the
hospitality of the KEK theory group during his stay over summer 2015.

\newpage
\appendix
\section{Notation and conventions}

\subsection{Fields}
\label{appendix:fields}
\begin{align}
	  &&\vH &=
  \frac 1 2 
  \begin{pmatrix}
    v+ h -\ii w^3 & -\ii \sqrt{2} w^+ \\
		-\ii \sqrt{2} w^- & v + h + \ii w^3  \\
  \end{pmatrix}.
\end{align}
To avoid adding terms proportional to the vacuum
expectation value, when adding a Higgs pair,
we introduce
\begin{align}
	\tr{\vH^\dagger \vH} \rightarrow
	\tr{\widehat{\vH^\dagger \vH}}:= \tr{\vH^\dagger \vH - \frac{v^2}{4}}.
	\label{eq:HHhat}
\end{align}
\begin{equation}
  \begin{aligned}
    \vW^{\mu\nu} &\equiv W^{\mu\nu}_i \frac{\tau_i}{2}&= +
    \frac{\ii}{g} \com{D_W^\mu}{D_W^\nu} 
    &= \left (\partial^\mu W^\nu_k  - \partial^\nu W^\mu_k 
      + g \varepsilon_{ijk} W^\mu_iW^\nu_j	\right ) \frac{\tau_k}{2} \\
    &&&=\partial^\mu \vW^\nu - \partial^\nu \vW^\mu 
    - \ii g \left [ \vW^\mu , \vW^\nu \right ], \\
    \vB^{\mu\nu} &\equiv\frac{Y}{2} B^{\mu\nu}& = +
    \frac{\ii}{g^\prime} \com{D_B^\mu}{D_B^\nu}  
    &=\frac{Y}{2} \left (\partial^\mu B^\nu - \partial^\nu B^\mu \right )\\
    &&&=\partial^\mu \vB^\nu - \partial^\nu \vB^\mu 
  \end{aligned}
  \label{eq:Def-FieldStrengthTensor}
\end{equation}

The covariant derivative is defined via
\begin{equation}
  \label{eq:CD_ME_Higgs}
  \vD_\mu \vH = \partial_\mu \vH - i g \vW_\mu \vH
\end{equation}
and
\begin{equation}
  \label{eq:CD_ME_W}
  \vD_\mu \vW_\nu = \partial_\mu \vW_\nu - i g \vW_\mu \vW_\nu
\end{equation}

The equations of motion for the Standard Model yield
\begin{subequations}
	\begin{align}
  \left(\vD^2 \vH \right) &=  \mu^2\vH - \lambda \tr{\vH^\dagger \vH}\vH \, ,
  \label{eq:eqofmotionH}\\
  \left (\vD^2 \vH \right)^\dagger &=  \mu^2 \vH^\dagger - \lambda \tr{\vH^\dagger \vH}\vH^\dagger \, ,
  \label{eq:eqofmotionHdagger} \\
    \partial_\mu \vB^{\mu\nu} &= -\ii \frac{g^\prime}{2} \left( 
    \vH^\dagger \vD^\nu \vH - \left (\vD^\nu \vH \right )^\dagger \vH
  \right ) \, ,
  \\
    \vD_\mu \vW^{\mu\nu} &= -\ii \frac{g}{2} \left( 
    \vD^\nu \vH \vH^\dagger -\vH \left(\vD^\nu \vH \right )^\dagger
  \right )
\end{align}
\end{subequations}


\subsection{$SU(2)$ Tensor Products}
\label{appendix:isospin-basis}

The tensor products of Pauli matrices for the isospin quintet $\tau_t$,
the isospin vector $\tau_v$, and the isospin scalar $\tau_s$ are
defined, respectively, as 
\begin{subequations}
\label{eq:isospin-eigenvectors}
\begin{alignat}{3}
	\tau_t^{++} &= && && \tau^+ \otimes \tau^+, \\
	\tau_t^+ &=&&& \frac{1}{2}& \left (
		\tau^+ \otimes \tau^3 + \tau^3 \otimes \tau^+
		\right), \\
	\tau_t^0 &= &&&\frac{1}{\sqrt{6}}& \left (
		\tau^3 \otimes \tau^3 - \tau^+ \otimes \tau^- - \tau^- \otimes \tau^+
		\right ), \\
	\tau_t^- &=&&& \frac{1}{2} &\left (
		\tau^- \otimes \tau^3 + \tau^3 \otimes \tau^-
		\right), \\
	\tau_t^{--} &=&& && \tau^- \otimes \tau^- \, , \\
	\tau_v^+ &=&&& \frac{\ii}{2}& \left (
		\tau^+ \otimes \tau^3 - \tau^3 \otimes \tau^+
		\right), \\
	\tau_v^0 &=&&& \frac{\ii}{\sqrt{2}}& \left (
		\tau^+ \otimes \tau^- - \tau^- \otimes \tau^+
		\right ), \\
	\tau_v^- &= &&&-\frac{\ii}{2} &\left (
		\tau^- \otimes \tau^3 - \tau^3 \otimes \tau^-
		\right), \\
		\tau_s&= && &
		\frac{1}{2\sqrt{3}} &\left( \tau^3 \otimes \tau^3 + 2 \tau^+ \otimes \tau^- + 2 \tau^- \otimes \tau^+ \right) \, ,
\end{alignat}
\end{subequations}
where the Pauli matrix for the isospin singlet is related to
\begin{align}
	\tau^{aa} \equiv \tau^a \otimes \tau^a = {2\sqrt{3}} \tau_s \, .
\end{align}

All nonzero traces of a product of two tensor products are normalized
\begin{align}
	\tr{\tau_t^{++}\tau_t^{--}}= \tr{\tau_t^+\tau_t^-}=\tr{\tau_t^0\tau_t^0}=
	\tr{\tau_v^+\tau_v^-}=\tr{\tau_v^0\tau_v^0}=
	\tr{\tau_s\tau_s}=1 \, .
\end{align}
From the properties of the tensor product
\begin{equation}
  (A\otimes B) (C\otimes D) = AC\otimes BD
\end{equation}
and the trace
\begin{equation}
  \tr{A\otimes B} = \tr{A} \tr{B}
\end{equation}
we find
\begin{align}
  \tr{\left (A\otimes B\right ) \left ( C \otimes D \right) } &= \tr{A
    C} \tr {B D}\,. 
\end{align}
This reduces the trace of an isospin singlet
\begin{align}
  \tr{ \left ( A\otimes B \right ) \tau^{aa}}=2 \tr{A B} - \tr {A} \tr {B}.
\end{align}
Multiplying the two Pauli matrices related to the isospin singlet leads to
\begin{align}
  \tau^{aa}\tau^{bb} = 3 \,\cdot\, \mathbf{1}\otimes \mathbf{1} -2 \, \tau^{aa} .
\end{align}


\section{Feynman Rules}
\label{appendix:Feynman_Rules}
The Feynman rules which are used to calculate the vector-boson scattering
amplitudes are summarized in this appendix. Focusing only on weak
vector-boson scattering, the Feynman rules are determined from the
Lagrangian, where gluons, photons and fermions are omitted.

\subsection{Lagrangian}
All Lagrangians are defined within the Higgs matrix realization whose
definition can be found in appendix~\ref{appendix:fields}.
The Standard Model Lagrangian is given by
\begin{align}
  \LL_{\text{SM}}=&-\frac{1}{2}\tr{\vW_{\mu\nu}\vW^{\mu\nu}}
  -\frac{1}{2}\tr{\vB_{\mu\nu}\vB^{\mu\nu}} \notag \\
  &+\tr{ \left ( \vD_\mu \vH \right )^\dagger
    \vD^\mu \vH }
  + \mu^2\tr{\vH^\dagger \vH}
  -\frac{\lambda}{2}\left( \tr{\vH^\dagger \vH} \right)^2.
\end{align}
Dimension-six and -eight operators affecting only the
Higgs/Nambu-Goldstone boson sector are discussed in sections
\ref{sec:matching} and are given by  
\begin{subequations}
\begin{alignat}{3}
  \LL_{HD}&= && F_{HD} && \tr{\vH^\dagger \vH - \frac{v^2}{4}} \cdot
  \tr{ \left ( \vD_\mu \vH \right )^\dagger \vD^\mu \vH} \, ,
  \\
  \LL_{S,0}&=
  & &F_{S,0}\ &&
  \tr{ \left ( \vD_\mu \vH \right )^\dagger \vD_\nu \vH}
  \cdot \tr{ \left ( \vD^\mu \vH \right )^\dagger \vD^\nu \vH} \, ,
  \label{eq:LL_S0}
  \\
    \label{eq:LL_S1}
  \LL_{S,1}&=
  & &F_{S,1}\ &&
  \tr{ \left ( \vD_\mu \vH \right )^\dagger \vD^\mu \vH}
  \cdot \tr{ \left ( \vD_\nu \vH \right )^\dagger \vD^\nu \vH} \, .
\end{alignat}
\end{subequations}
As an extension to model generic new physics, additional resonances
are introduced. The scalar resonance $\sigma$ and the tensor resonance
$f^{\mu\nu}$ represent singlets of the chiral symmetry group, whereas
$\Phi$ has the quantum numbers $1 \otimes 1$ under $SU(2)_L \times
SU(2)_R$. $\Phi$ is referred to as isotensor for historical reasons, 
but it actually includes an isovector $\Phi_v$ and isoscalar $\Phi_s$ besides
the isotensor $\Phi_t$. Also the Fierz-Pauli tensor $f$ can be reformulated
into a tensor $f_f$, a vector $A_f$ and a scalar $\sigma_f$ such that
canonical propagators can be used for each degree of freedom
separately instead of the complicated tensor propagator
\begin{subequations}
\begin{align}
  \Delta_{\mu\nu,\rho\sigma} (f) 
    &=  \frac{\ii}{k^2-m^2+\ii \epsilon}P_{\mu\nu,\rho\sigma}(k,m) \, ,\\
  \Delta_{\mu\nu,\rho\sigma} (f^\prime) 
  &= \frac{\ii}{k^2 - m^2+\ii \epsilon}
     \left(\frac{1}{2} g_{\mu\rho}g_{\nu\sigma} 
       +\frac{1}{2}g_{\mu\sigma}g_{\nu\rho}
       - \frac{1}{2} g_{\mu\nu}g_{\rho\sigma} 
     \right ) \, ,
  \\
  \Delta_{\mu\nu}  (A) 
  &= \frac{-\ii}{k^2 - m^2+\ii \epsilon} g_{\mu\nu} \, ,
  \\
  \Delta  (\sigma) 
  &= \frac{\ii}{k^2 - m^2+\ii \epsilon} \, ,
\end{align}
\end{subequations}
where the projection operator of spin-two states can be written in
terms of the spin-one projection operator,
\begin{align}
  \begin{aligned}
    P^{\mu_1\mu_2,\nu_1\nu_2}(k,m)
    &=
    \frac{1}{2}\Bigl[ P^{\mu_1\nu_1}(k,m)P^{\mu_2\nu_2}(k,m) +
    P^{\mu_1\nu_2}(k,m)P^{\mu_1\nu_2}(k,m) \Bigr]
    \\
    &\qquad\quad
    - \frac{1}{3} P^{\mu_1\mu_2}(k,m)P^{\nu_1\nu_2}(k,m),
  \end{aligned}
\end{align}

with
\begin{align}
  P^{\mu\nu}(k,m)
  &=
  \sum_\lambda \bar{\varepsilon}_{(\lambda)}^{\mu}(k,m) 
               \varepsilon_{(\lambda)}^{\nu}(k,m) 
  = g^{\mu\nu}-\frac{k^\mu k^\nu}{m^2}.
\end{align}


\subsection{Unitary Gauge}
The Feynman rules in unitary gauge of the Lagrangians defined in this
paper are listed in this section. Only the relevant vertices for the
vector-boson scattering process are shown. In other words, vertices
above four fields for effective operators and above three fields for
resonances are neglected. 
\subsubsection{Standard Model}
\label{appendix_rules_unitary_sm}
\begin{subequations}
  \label{eq:FR_unitary_sm}
  \begin{alignat}{2}
    A_{\mu_1}W^+_{\mu_2}W^-_{\mu_3} :& \hspace{0.5 cm} &
    -\ii e&
    \left [
      \left (p_{1 \, \mu_3}-p_{2 \, \mu_3} \right ) g_{\mu_1\mu_2} \right.
      + \left (p_{3 \, \mu_2}-p_{1 \, \mu_2} \right ) g_{\mu_1\mu_3} \notag \\
      &&& \hspace{2ex} \left. +\left (p_{2 \, \mu_1}-p_{3 \, \mu_1} \right ) g_{\mu_2\mu_3}
      \right ] \, ,\\
    Z_{\mu_1}W^+_{\mu_2}W^-_{\mu_3} :& \hspace{0.5 cm} & 
    -\ii c_w g & 
    \left [ 
      \left (p_{1 \, \mu_3}-p_{2 \, \mu_3} \right ) g_{\mu_1\mu_2}
      + \left (p_{3 \, \mu_2}-p_{1 \, \mu_2} \right ) g_{\mu_1\mu_3} \right.\notag\\
      &&& \hspace{2ex}\left. +\left (p_{2 \, \mu_1}-p_{3 \, \mu_1} \right ) g_{\mu_2\mu_3}\right ] \, ,\\
    h W^+_{\mu_2}W^-_{\mu_3} :& \hspace{0.5 cm} & 
    \ii  m_W g&g_{\mu_2 \mu_3} \, ,\\
    h Z_{\mu_2}Z_{\mu_3} :& \hspace{0.5 cm} & 
    \ii  m_Z g&g_{\mu_2 \mu_3} \, ,\\
    W^+_{\mu_1}W^+_{\mu_2}W^-_{\mu_3}W^-_{\mu_4} :& \hspace{0.5 cm} &
    - \ii g^2 &\left (g_{\mu_1\mu_4}g_{\mu_2\mu_3} +g_{\mu_1\mu_3}g_{\mu_2\mu_4}-2 g_{\mu_1\mu_2}g_{\mu_3\mu_4}  \right )\, ,\\
    Z_{\mu_1}Z_{\mu_2}W^+_{\mu_3}W^-_{\mu_4} :& \hspace{0.5 cm} &
    \ii c_w^2 g^2 &\left (g_{\mu_1\mu_4}g_{\mu_2\mu_3} +g_{\mu_1\mu_3}g_{\mu_2\mu_4}-2 g_{\mu_1\mu_2}g_{\mu_3\mu_4}  \right )
    \, ,\\
    A_{\mu_1}A_{\mu_2}W^+_{\mu_3}W^-_{\mu_4} :& \hspace{0.5 cm} &
    \ii e^2 &\left (g_{\mu_1\mu_4}g_{\mu_2\mu_3} +g_{\mu_1\mu_3}g_{\mu_2\mu_4}-2 g_{\mu_1\mu_2}g_{\mu_3\mu_4}  \right )
    \, ,\\
    A_{\mu_1}Z_{\mu_2}W^+_{\mu_3}W^-_{\mu_4} :& \hspace{0.5 cm} &
    \ii e c_wg &\left (g_{\mu_1\mu_4}g_{\mu_2\mu_3} +g_{\mu_1\mu_3}g_{\mu_2\mu_4}-2 g_{\mu_1\mu_2}g_{\mu_3\mu_4}  \right ) 
    \, ,\\
    h h W^+_{\mu_3}W^-_{\mu_4} :& \hspace{0.5 cm} &
    \frac{\ii}{2} g^2& g_{\mu_3\mu_4}
    \, ,\\
    h h Z_{\mu_3}Z_{\mu_4} :& \hspace{0.5 cm} &
    \frac{\ii}{2} \frac{g^2}{c_w^2} &g_{\mu_3\mu_4}\, .
  \end{alignat}
\end{subequations}
\subsubsection{\texorpdfstring{$\LL_{HD}$}{Lg}}
\label{appendix_rules_unitary_hd}
\begin{subequations}
  \label{eq:FR_unitary_hd}
  \begin{alignat}{2}
    hW^+_\mu W^-_\nu:& \hspace{1 cm} &
    \frac{\ii g^2 v^3}{4} F_{HD} & g_{\mu\nu} 
    \, ,\\
    hZ_\mu Z_\nu:& \hspace{1 cm} &
    \frac{\ii g^2 v^3}{4 s_w^2} F_{HD} & g_{\mu\nu} 
    \, ,\\
    h(p_1)h(p_2)h(p_3):& \hspace{1 cm} &
    -\ii v  F_{HD} &
    \left ( p_1\cdot p_2 + p_1\cdot p_3 + p_2\cdot p_3
    \right) 
    \, ,\\
    hhW^+_\mu W^-_\nu:& \hspace{1 cm} &
    \frac{5\ii g^2 v^2}{4} F_{HD}&  g_{\mu\nu} 
    \, ,\\
    hhZ_\mu Z_\nu:& \hspace{1 cm} &
    \frac{5\ii g^2 v^2}{4 s_w^2} F_{HD}&  g_{\mu\nu} 
    \, ,\\
    h(p_1)h(p_2)h(p_3)h(p_4):& \hspace{1 cm} &
    -\ii  F_{HD}& 
    \left (  p_1\cdot p_2 + p_1\cdot p_3 +p_1\cdot p_4 \right.
    \notag\\
    &&&\hspace{2 ex}\left. + p_2\cdot p_3 + p_2\cdot p_4 + p_3 \cdot p_4
    \right) \, .
  \end{alignat}
\end{subequations}

\subsubsection{\texorpdfstring{$\LL_S$}{Lg}}
\label{appendix_rules_unitary_s}
\begin{subequations}
  \label{eq:FR_unitary_s}
  \begin{alignat}{2}
    W^+_{\mu_1}W^+_{\mu_2}W^-_{\mu_3}W^-_{\mu_4} :& \hspace{0.5 cm} &
    \frac{ \ii g^4 v^4}{16}& \left [ 
      \left (F_{S,0} + 2F_{S,1} \right )  \left (
      g_{\mu_1\mu_3}g_{\mu_2\mu_4}+g_{\mu_1\mu_4}g_{\mu_2\mu_3}
      \right ) \right. \notag\\
      &&& \hspace{2ex} \left.  + 2 F_{S,0}  g_{\mu_1\mu_2}g_{\mu_3\mu_4}
      \right ]
    \, ,\\
    Z_{\mu_1}Z_{\mu_2}W^+_{\mu_3}W^-_{\mu_4} :& \hspace{0.5 cm} &
    \frac{ \ii g^4 v^4}{16 c_w^2} & \left [ 
      F_{S,0} \left (
      g_{\mu_1\mu_3}g_{\mu_2\mu_4}+g_{\mu_1\mu_4}g_{\mu_2\mu_3}
      \right ) \right. \notag   \\
      &&& \hspace{2ex}\left. +2 F_{S,1}  g_{\mu_1\mu_2}g_{\mu_3\mu_4} \right ]
    \, ,\\
    Z_{\mu_1}Z_{\mu_2}Z_{\mu_3}Z_{\mu_4} :& \hspace{0.5 cm} &
    \frac{ \ii g^4 v^4}{8 c_w^4}  &\left ( F_{S,0} +F_{S,1} \right ) \left [
      g_{\mu_1\mu_2}g_{\mu_3\mu_4} +  g_{\mu_1\mu_3}g_{\mu_2\mu_4}
      \right . \notag \\
      &&& \hspace{2ex}\phantom{\left ( F_{S,0} +F_{S,1} \right )} \left. +g_{\mu_1\mu_4}g_{\mu_2\mu_3}
      \right ] \, ,\\
    h({p_1})h({p_2})W^+_{\mu_3}W^-_{\mu_4} :& \hspace{0.5 cm} &
    -\frac{ \ii g^2 v^2}{4} & \left [ 
      F_{S,0} \left (
      p_{1 \,\mu_3}p_{2 \,\mu_4}+ p_{1 \,\mu_4}p_{2 \,\mu_3} \right ) 
      \right. \notag \\
      &&& \hspace{2ex} \left. +2 F_{S,1} g_{\mu_3\mu_4} p_1 \cdot p_2    \right ]
    \, ,\\
    h({p_1})h({p_2})Z_{\mu_3}Z_{\mu_4} :& \hspace{0.5 cm} &
    -\frac{ \ii g^2 v^2}{4 c_w^2} & \left [ 
      F_{S,0} \left (
      p_{1 \,\mu_3}p_{2 \,\mu_4}+ p_{1 \,\mu_4}p_{2 \,\mu_3} \right )  
      \right. \notag \\
      &&& \hspace{2ex} \left. +2 F_{S,1} g_{\mu_3\mu_4} p_1 \cdot p_2    \right ]
    \, ,\\
    h({p_1})h({p_2})h({p_3})h({p_4}) :& \hspace{0.5 cm} &
    2 \ii &\left ( F_{S,0} +F_{S,1} \right )  \left [ 
      \left ( p_1 \cdot p_2 \right ) \left ( p_3 \cdot p_4 \right )  \right.
      \notag \\
      &&&  \phantom{\left ( F_{S,0} +F_{S,1} \right ) }
      \hspace{2ex} + \left ( p_1 \cdot p_3 \right ) \left ( p_2 \cdot p_4 \right ) \notag \\ 
      &&& \phantom{\left ( F_{S,0} +F_{S,1} \right ) }
      \hspace{2ex} \left. + \left ( p_1 \cdot p_4 \right ) \left ( p_2 \cdot p_3 \right )
      \right ]\, .
  \end{alignat}
\end{subequations}
\subsubsection{\texorpdfstring{$\LL_\sigma$}{Lg}}
\label{appendix_rules_unitary_sigma}
\begin{subequations}
\label{eq:FR_unitary_sigma}
	\begin{alignat}{2}
  \sigma W^+_\mu W^-_\nu:& \hspace{1 cm} &
  \frac{\ii g^2 v^2}{4} F_\sigma & g_{\mu\nu} \, ,\\
  \sigma Z_\mu Z_\nu:& \hspace{1 cm} &
  \frac{\ii g^2 v^2}{4 c_w^2} F_\sigma & g_{\mu\nu} \, , \\
      \sigma h \left (p_1 \right ) h\left (p_2 \right )
  :& \hspace{1 cm} &
  -\ii  F_{\sigma} & p_1 \cdot p_2 \, .
	\end{alignat}
\end{subequations}

\subsubsection{\texorpdfstring{$\LL_\phi$}{Lg}}
\label{appendix_rules_unitary_phi}
\begin{subequations}
\label{eq:FR_unitary_phi}
	\begin{alignat}{2}
  \phi_t^{\pm \pm} W^{\mp}_\mu W^{\mp}_\nu:& \hspace{1 cm} &
  \frac{\ii g^2 v^2}{4 } F_\phi & g_{\mu\nu} \, ,\\
  \phi_t^{\pm } W^{\mp}_\mu Z_\nu:& \hspace{1 cm} &
  \frac{\ii g^2 v^2}{4 \sqrt{2}c_w} F_\phi & g_{\mu\nu} \, ,\\
  \phi_t^{0} W^{\mp}_\mu W^{\pm}_\nu:& \hspace{1 cm} &
  -\frac{\ii g^2 v^2}{4 \sqrt{6}} F_\phi & g_{\mu\nu} \, ,\\
    \phi_t^{0} Z_\mu Z_\nu:& \hspace{1 cm} &
  \frac{\ii g^2 v^2}{2 \sqrt{6}c_w^2} F_\phi & g_{\mu\nu} \, ,\\
  \phi_s W^{\mp}_\mu W^{\pm}_\nu:& \hspace{1 cm} &
  \frac{\ii g^2 v^2}{8 \sqrt{3}} F_\phi & g_{\mu\nu} \, ,\\
    \phi_s Z_\mu Z_\nu:& \hspace{1 cm} &
  \frac{\ii g^2 v^2}{8 \sqrt{3}c_w^2} F_\phi & g_{\mu\nu} \, ,\\
    \phi_v^{\pm } h \left(p \right ) W^\mp_{\mu}:& \hspace{1 cm} &
  -\frac{ g v}{2 \sqrt{2}} F_\phi & {p}_{\mu} \, ,\\
  \phi_v^{\pm } h \left(p \right ) Z_{\mu}:& \hspace{1 cm} &
  \frac{ g v}{2 \sqrt{2}c_w} F_\phi & {p}_{\mu} \, ,\\
   \phi_s h \left (p_1 \right ) h \left (p_2 \right ):& \hspace{1 cm} &
   \frac{\sqrt{3}   }{2 }\ii F_\phi & p_1 \cdot p_2 \, .
	\end{alignat}
\end{subequations}

\subsubsection{\texorpdfstring{$\LL_f$}{Lg}}
\label{appendix_rules_unitary_f}
	\begin{subequations}
	\label{eq:FR_unitary_f}
	\begin{alignat}{2}
  f_{\mu \nu} W^+_\rho W^-_\sigma:& \hspace{1 cm} &
  \frac{\ii g^2 v^2}{8} F_f & 
  \left[
  g_{\mu \sigma } g_{\nu \rho }
			+ g_{\mu \rho } g_{\nu \sigma }
			-\frac{c_f}{2} g_{\mu \nu} g_{\rho \sigma }
   \right]  \, ,\\
  f_{\mu \nu} Z_\rho Z_\sigma:& \hspace{1 cm} &
  \frac{\ii g^2 v^2}{8 c_w^2} F_f & 
  \left[
  g_{\mu \sigma } g_{\nu \rho }
			+ g_{\mu \rho } g_{\nu \sigma }
			-\frac{c_f}{2} g_{\mu \nu} g_{\rho \sigma }
   \right]
   \, ,\\
      f_{\mu \nu} h \left (p_1 \right ) h\left (p_2 \right )
  :& \hspace{1 cm} &
      -\frac{\ii}{2}  F_f 
  & \left [p_{1\, \mu} p_{2\, \nu} + p_{1\, \nu} p_{2\, \mu} 
   -\frac{c_f}{2} g_{\mu\nu}p_1\cdot p_2 \right] \, .
\end{alignat}
	\end{subequations}

\subsubsection{\texorpdfstring{$\LL_f$ in St\"uckelberg formalism}{Lg}}
\label{appendix_rules_unitary_f_stueck}
	
	\begin{subequations}
	\begin{alignat}{2}
  f_{f\mu \nu} W^+_\rho W^-_\sigma:& \hspace{1 cm} &
  \frac{\ii g^2 v^2}{8} F_f & 
  \left[
  g_{\mu \sigma } g_{\nu \rho }
			+ g_{\mu \rho } g_{\nu \sigma }
			-\frac{c_f}{2} g_{\mu \nu} g_{\rho \sigma }
   \right]  \, ,\\
  f_{f\mu \nu} Z_\rho Z_\sigma:& \hspace{1 cm} &
  \frac{\ii g^2 v^2}{8 c_w^2} F_f & 
  \left[
  g_{\mu \sigma } g_{\nu \rho }
			+ g_{\mu \rho } g_{\nu \sigma }
			-\frac{c_f}{2} g_{\mu \nu} g_{\rho \sigma }
   \right]
   \, ,\\
      f_{f\mu \nu} h \left (p_1 \right ) h\left (p_2 \right )
  :& \hspace{1 cm} &
      -\frac{\ii}{2}  F_f 
  & \left [p_{1\, \mu} p_{2\, \nu} + p_{1\, \nu} p_{2\, \mu} 
   -\frac{c_f}{2} g_{\mu\nu}p_1\cdot p_2 \right]\, .
\end{alignat}
	\end{subequations}

Because of $\partial_\nu J_f^{\mu \nu} \neq 0$:

	\begin{alignat}{2}
  A_{f \,\mu} \left ( p \right ) W^+_\rho W^-_\sigma :& \hspace{1 cm} &
	    \frac{ g^2 v^2}{4 \sqrt{2}m_f} F_f &
	    \left (p_\rho g_{\mu \sigma} + p_\sigma g_{\mu \rho} 
	    - \frac{c_f}{2} p_\mu g_{\sigma \rho}  \right ) \, ,\\
  A_{f \, \mu}\left ( p \right )  Z_\rho  Z_\sigma :& \hspace{1 cm} &
	    \frac{ g^2 v^2}{4 c_w^2 \sqrt{2}m_f} F_f &
	    \left ( p_\rho g_{\mu \sigma} + p_\sigma g_{\mu \rho} 
	    - \frac{c_f}{2} p_\mu g_{\sigma \rho}  \right ) \, ,\\
  A_{f \, \mu} h \left (p_1 \right ) h\left (p_2 \right )
  :& \hspace{1 cm} &
   \frac{1}{\sqrt{2}m_f}  F_f &
	  \left [ 
	  p_1^2 p_{2 \,\mu} +p_2^2 p_{1 \,\mu} \right . \\
	  &&& \hspace{2ex} \left .+\frac{1}{2} \left ( 2 - c_f \right )
	  p_1\cdot p_2 
	  \left (p_1 + p_2 \right )_\mu	  
	  \right ] \, .
\end{alignat}

Because of $\partial_\mu \partial_\nu J_f^{\mu \nu} \neq 0$ and $ {J_f}^{\mu}_\mu \neq 0$:
\begin{subequations}
	\begin{alignat}{2}
  \sigma_f \left ( p \right ) W^+_\rho W^-_\sigma :& \hspace{0.5 cm} &
	   \frac{\ii g^2 v^2}{4\sqrt{6}} F_f & 
	  \left [\left ( c_f - 1 \right ) g_{\rho \sigma} 
	  -\frac{1}{m_f^2}\left ( 2 k_\rho k_\sigma - \frac{c_f}{2} k^2 g_{\rho\sigma} \right )
	  \right ] \, ,\\
  \sigma_f\left ( p \right )  Z_\rho  Z_\sigma :& \hspace{0.5 cm} &
	  \frac{\ii g^2 v^2}{4\sqrt{6} c_w^2} F_f & 
	  \left [\left ( c_f - 1 \right ) g_{\rho \sigma} 
	  -\frac{1}{m_f^2}\left ( 2 k_\rho k_\sigma - \frac{c_f}{2} k^2 g_{\rho\sigma} \right )
	  \right ] \, ,\\
  \sigma_f h \left (p_1 \right ) h\left (p_2 \right )
  :& \hspace{0.5 cm} &
    -\frac{\ii}{\sqrt{6}}  F_f 
  & \, \Big[ 
    	    \left ( c_f - 1 \right)  \left ( p_1 \cdot p_2 \right ) \notag \\
    &&& \hspace{2ex} - \frac{1}{m_f^2}  \Big ( 2 p_1\cdot\left (p_1 +p_2 \right )
    p_2 \cdot \left (p_1 +p_2 \right ) \notag\\
      &&& \left. \hspace{2ex} \phantom{- \frac{1}{m_f^2}  \Big (}\hspace{2ex}
      - \frac{c_f}{2}  p_1\cdot p_2 \left (p_1 +p_2 \right )^2 \Big ) 
     \right] \, .
\end{alignat}
\end{subequations}

\subsubsection{\texorpdfstring{$\LL_X$}{Lg}}
\label{appendix_rules_unitary_X}
\begin{subequations}
\label{eq:FR_unitary_X}
	\begin{alignat}{2}
  X_{t\mu\nu}^{\pm \pm} W^{\mp}_\rho W^{\mp}_\sigma:& \hspace{1 cm} &
  \frac{\ii g^2 v^2}{8 } F_X & \left[
  g_{\mu \sigma } g_{\nu \rho }
			+ g_{\mu \rho } g_{\nu \sigma }
			-\frac{c_X}{2} g_{\mu \nu} g_{\rho \sigma }
   \right] \, , \\ 
  X_{t\mu\nu}^{\pm } W^{\mp}_\rho Z_\sigma:& \hspace{1 cm} &
  \frac{\ii g^2 v^2}{8 \sqrt{2}c_w} F_X & \left[
  g_{\mu \sigma } g_{\nu \rho }
			+ g_{\mu \rho } g_{\nu \sigma }
			-\frac{c_X}{2} g_{\mu \nu} g_{\rho \sigma }
   \right] \, , \\ 
  X_{t\mu\nu}^{0} W^{\mp}_\rho W^{\pm}_\sigma:& \hspace{1 cm} &
  -\frac{\ii g^2 v^2}{8 \sqrt{6}} F_X & \left[
  g_{\mu \sigma } g_{\nu \rho }
			+ g_{\mu \rho } g_{\nu \sigma }
			-\frac{c_X}{2} g_{\mu \nu} g_{\rho \sigma }
   \right] \, , \\ 
    X_{t\mu\nu}^{0} Z_\rho Z_\sigma:& \hspace{1 cm} &
  \frac{\ii g^2 v^2}{4 \sqrt{6}c_w^2} F_X & \left[
  g_{\mu \sigma } g_{\nu \rho }
			+ g_{\mu \rho } g_{\nu \sigma }
			-\frac{c_X}{2} g_{\mu \nu} g_{\rho \sigma }
   \right] \, , \\ 
  X_{s\mu\nu} W^{\mp}_\rho W^{\pm}_\sigma:& \hspace{1 cm} &
  \frac{\ii g^2 v^2}{16 \sqrt{3}} F_X & \left[
  g_{\mu \sigma } g_{\nu \rho }
			+ g_{\mu \rho } g_{\nu \sigma }
			-\frac{c_X}{2} g_{\mu \nu} g_{\rho \sigma }
   \right] \, , \\ 
    X_{s\mu\nu} Z_\rho Z_\sigma:& \hspace{1 cm} &
  \frac{\ii g^2 v^2}{16 \sqrt{3}c_w^2} F_X & \left[
  g_{\mu \sigma } g_{\nu \rho }
			+ g_{\mu \rho } g_{\nu \sigma }
			-\frac{c_X}{2} g_{\mu \nu} g_{\rho \sigma }
   \right] \, , \\ 
    X_{v\mu\nu}^{\pm } h \left(p \right ) W^\mp_{\rho}:& \hspace{1 cm} &
  -\frac{ g v}{4 \sqrt{2}} F_X & 
  \left[ {p}_{\mu} g_{\nu\rho} + {p}_{\nu} g_{\mu\rho} - \frac{c_X}{2}{p}_{\rho} g_{\mu\nu}  \right]\, , \\ 
  X_{v\mu\nu} h \left(p \right ) Z_{\rho}:& \hspace{1 cm} &
  \frac{ g v}{4 \sqrt{2}c_w} F_X & 
    \left[ {p}_{\mu} g_{\nu\rho} + {p}_{\nu} g_{\mu\rho} - \frac{c_X}{2}{p}_{\rho} g_{\mu\nu}  \right]\, , \\ 
   X_{s\mu\nu} h \left (p_1 \right ) h \left (p_2 \right ):& \hspace{1 cm} &
   \frac{\sqrt{3}   }{4 }\ii F_X & \left [p_{1\, \mu} p_{2\, \nu} +
     p_{1\, \nu} p_{2\, \mu}  
   -\frac{c_X}{2} g_{\mu\nu}p_1\cdot p_2 \right] \, .
	\end{alignat}
\end{subequations}
\newpage

\subsection{Partial wave functions}
\label{appendix:partialwaves}

In this appendix we collect expressions appearing in the partial-wave
expansion of amplitudes.

\begin{subequations}
\label{eq:Partialwaves_Abbr}
\begin{align}
  \mathcal{S}_{0}\left(s, m \right) =& m^2+\frac{m^4}{s}\log \left (\frac{m^2}{s + m^2} \right )
  -\frac{s}{2}\, , \\ 
  \mathcal{S}_{1}\left(s, m \right) =& 2 \frac{m^4}{s}
  +\frac{m^4}{s^2} \left( 2 m^2 +s \right )
  \log \left (\frac{m^2}{s + m^2} \right )
  +\frac{s}{6}\, , \\ 
  \mathcal{S}_{2}\left(s, m \right) =&  \frac{m^4}{s^2} 
  \left (6 m^2 + 3s \right )
  +\frac{m^4}{s^3} \left( 6 m^4 +6m^2 s +s^2 \right )
  \log \left (\frac{m^2}{s + m^2} \right )\, , \\
  \mathcal{P}_{0}\left(s, m \right) =& 1+\frac{m^2+2s}{s}\log \left (\frac{m^2}{s + m^2} \right )\, , \\
  \mathcal{P}_{1}\left(s, m \right) =& \frac{m^2+2s}{s^2}\left(2s + \left({2m^2+s}\right) \log \left (\frac{m^2}{s + m^2} \right ) \right)\, \, , \\
  \mathcal{D}_{0}\left(s, m \right) =& m^2 + \frac{11}{2}s + \frac{1}{s}\left(m^4 + 6m^2s +6s^2\right)\log \left (\frac{m^2}{s + m^2} \right )\, , \\
  \mathcal{D}_{1}\left(s, m \right) =& 2\frac{m^4}{s}+12m^2+\frac{73}{6}s \notag \\
  &+ \frac{1}{s^2}\left(2m^2 +s\right)\left(m^4 + 6m^2s +6s^2\right)\log \left (\frac{m^2}{s + m^2} \right ) \, .
\end{align}
\end{subequations}

\section{T-matrix Counterterms}

In the T-matrix unitarization scheme, the
unitarization corrections are expressed as momentum-dependent
counterterms for the use as effective Feynman rules in the complete
amplitude evaluation.  Starting from the spin-isospin eigenamplitudes
in the gaugeless limit, section~\ref{sec:IlinWHIZARD}, the
straightforward application of the algorithm in~\cite{Kilian:2014zja}
yields $s$-dependent amplitude corrections $\Delta\amp_{IJ}(s)$.  The
insertion as effective Feynman rules proceeds in form of the following
expressions: 
\begin{subequations}
\label{eq:CT_FeynmanRules}
\begin{alignat}{3}
  W^{\pm}_{\mu_1}W^{\pm}_{\mu_2}\rightarrow W^{\pm}_{\mu_3}W^{\pm}_{\mu_4}
  &: \quad & & \frac{g^4 v^4}{4} & &\left [ 
  \left ( \Delta \amp_{20}(s)-10\Delta \amp_{22}(s) \right )
  \frac{g_{\mu_1\mu_2}g_{\mu_3\mu_4}}{s^2}
   \right .
  \notag\\
  &&&& & \left .+ 15 \Delta \amp_{22}(s)
  \frac{g_{\mu_1\mu_3}g_{\mu_2\mu_4}+ g_{\mu_1\mu_4}g_{\mu_2\mu_3}}{s^2} \right ],
  \\
  W^\pm_{\mu_1}W^\mp_{\mu_2}\rightarrow Z_{\mu_3}Z_{\mu_4}
  &:& & \frac{g^4 v^4}{4 c_w^2} & &\left [ 
  \left ( \frac{1}{3} \left ( \Delta \amp_{00}(s)-\Delta \amp_{20}(s)
  \right ) \right . \right . \notag \\
  &&&& &\left. \left . -\frac{10}{3}\left ( \Delta \amp_{02}(s)-\Delta \amp_{22}(s)
  \right )   \right )
  \frac{g_{\mu_1\mu_2}g_{\mu_3\mu_4}}{s^2}
   \right .
  \notag\\
  &&&& &\left . + 5 \left (\Delta \amp_{02}(s)-\Delta \amp_{22}(s) \right)
  \frac{g_{\mu_1\mu_3}g_{\mu_2\mu_4}+ g_{\mu_1\mu_4}g_{\mu_2\mu_3}}{s^2} \right ],
  \\
  W^\pm_{\mu_1}Z_{\mu_2}\rightarrow W^\pm_{\mu_3}Z_{\mu_4}
  &:& & \frac{g^4 v^4}{4c_w^2} & &\left [
  \left (\frac{1}{2}\Delta \amp_{20}(s)-5\Delta \amp_{22}(s) \right )
  \frac{g_{\mu_1\mu_2}g_{\mu_3\mu_4}}{s^2}
   \right .
  \notag\\
  &&&& &+ \left ( -\frac{3}{2}\Delta \amp_{11}(s)+\frac{15}{2}\Delta
    \amp_{22}(s) \right )\frac{g_{\mu_1\mu_3}g_{\mu_2\mu_4}}{s^2}
  \notag\\
  &&&& & \left .+  \left (\frac{3}{2}\Delta \amp_{11}(s)+\frac{15}{2}\Delta
    \amp_{22}(s) \right )\frac{g_{\mu_1\mu_4}g_{\mu_2\mu_3}}{s^2}
    \right ] ,
    \\
  W^\pm_{\mu_1}W^\mp_{\mu_2}\rightarrow W^\pm_{\mu_3}W^\mp_{\mu_4}
  &:& & \frac{g^4 v^4}{4} & &\left [
  \left ( \frac{1}{6} \left ( 2\Delta \amp_{00}(s)+\Delta \amp_{20}(s)
  \right ) 
  \right . \right . \notag\\ &&&& &  \left . \left .
  -\frac{5}{3}\left (2 \Delta \amp_{02}(s)+\Delta \amp_{22}(s)
  \right )
  \right ) \frac{g_{\mu_1\mu_2}g_{\mu_3\mu_4}}{s^2} \right .
  \notag\\
  &&&& &+ \left ( 5\Delta \amp_{02}(s)-\frac{3}{2}\Delta
    \amp_{11}(s)+\frac{5}{2} \Delta \amp_{22}(s) \right )\frac{g_{\mu_1\mu_3}g_{\mu_2\mu_4}}{s^2}
  \notag\\
  &&&& & \left . + \left ( 5\Delta \amp_{02}(s)+\frac{3}{2}\Delta
    \amp_{11}(s)+\frac{5}{2} \Delta \amp_{22}(s) \right )
  \frac{g_{\mu_1\mu_4}g_{\mu_2\mu_3}}{s^2}
    \right ],
\\
   Z_{\mu_1}Z_{\mu_2}\rightarrow Z_{\mu_3}Z_{\mu_4}
   &:& & \frac{g^4 v^4}{4c_w^4} & &\left [
  \left ( \frac{1}{3} \left ( \Delta \amp_{00}(s)+2\Delta \amp_{20}(s)
  \right )
	  \right . \right . \notag\\ &&&& &  \left . \left .
	  -\frac{10}{3}\left ( \Delta \amp_{02}(s)+2\Delta
    \amp_{22}(s) \right )
    \right ) \frac{g_{\mu_1\mu_2}g_{\mu_3\mu_4}}{s^2} \right .
  \notag\\
  &&&& & \left .+ 5 \left (\Delta \amp_{02}(s)+2\Delta \amp_{22}(s) \right )
  \frac{g_{\mu_1\mu_3}g_{\mu_2\mu_4}+ g_{\mu_1\mu_4}g_{\mu_2\mu_3}}{s^2} \right ].
\end{alignat}
These relations are the generalizations of the corresponding formulae
in reference~\cite{Kilian:2014zja} for the case of resonances.
Scattering processes involving a Higgs boson have a different
off-shell extrapolation. Therefore, the Higgs momentum is included in
the Feynman rules for the analogous effective vertices given by
\begin{alignat}{3}
  {W^\pm}^{\mu_1}{W^\mp}^{\mu_2}\rightarrow 
  h h 
  &:& & \;-g^2 v^2 & &\left [ 
  \left ( \frac{1}{3} \left ( \Delta \amp_{00}(s)-\Delta \amp_{20}(s)
  \right ) \right . \right . \notag \\
  &&&& &\left. \left . -\frac{10}{3}\left ( \Delta \amp_{02}(s)-\Delta \amp_{22}(s)
  \right )   \right )
  \frac{g^{\mu_1\mu_2} \left(k_3\cdot k_4\right)}{s^2}
   \right .
  \notag\\
  &&&& &\left . + 5 \left (\Delta \amp_{02}(s)-\Delta \amp_{22}(s) \right)
  \frac{k_3^{\mu_1}k_4^{\mu_2}+ k_4^{\mu_1}k_3^{\mu_2}}{s^2} \right ],
  \\
  Z^{\mu_1}Z^{\mu_2}\rightarrow 
  h h 
  &:& & \;-\frac{g^2 v^2}{c_w^2} & &\left [ 
  \left ( \frac{1}{3} \left ( \Delta \amp_{00}(s)-\Delta \amp_{20}(s)
  \right ) \right . \right . \notag \\
  &&&& &\left. \left . -\frac{10}{3}\left ( \Delta \amp_{02}(s)-\Delta \amp_{22}(s)
  \right )   \right )
  \frac{g^{\mu_1\mu_2} \left(k_3\cdot k_4\right)}{s^2}
   \right .
  \notag\\
  &&&& &\left . + 5 \left (\Delta \amp_{02}(s)-\Delta \amp_{22}(s) \right)
  \frac{k_3^{\mu_1}k_4^{\mu_2}+ k_4^{\mu_1}k_3^{\mu_2}}{s^2} \right ],
  \\
  {W^\pm}^{\mu_1}h\rightarrow {W^\pm}^{\mu_3}h
  &:& & \;-g^2 v^2 & &\left [
  \left (\frac{1}{2}\Delta \amp_{20}(s)-5\Delta \amp_{22}(s) \right )
  \frac{k_2^{\mu_1}k_4^{\mu_3}}{s^2}
   \right .
  \notag\\
  &&&& &+ \left ( -\frac{3}{2}\Delta \amp_{11}(s)+\frac{15}{2}\Delta
    \amp_{22}(s) \right )\frac{g^{\mu_1\mu_3}\left (k_2\cdot k_4 \right)}{s^2}
  \notag\\
  &&&& & \left .+  \left (\frac{3}{2}\Delta \amp_{11}(s)+\frac{15}{2}\Delta
    \amp_{22}(s) \right )\frac{k_4^{\mu_1}k_2^{\mu_3}}{s^2}
    \right ] ,
    \\
  {Z}^{\mu_1}h\rightarrow {Z}^{\mu_3}h
  &:& & \;-\frac{g^2 v^2}{c_w^2} & &\left [
  \left (\frac{1}{2}\Delta \amp_{20}(s)-5\Delta \amp_{22}(s) \right )
  \frac{k_2^{\mu_1}k_4^{\mu_3}}{s^2}
   \right .
  \notag\\
  &&&& &+ \left ( -\frac{3}{2}\Delta \amp_{11}(s)+\frac{15}{2}\Delta
    \amp_{22}(s) \right )\frac{g^{\mu_1\mu_3}\left (k_2\cdot k_4 \right)}{s^2}
  \notag\\
  &&&& & \left .+  \left (\frac{3}{2}\Delta \amp_{11}(s)+\frac{15}{2}\Delta
    \amp_{22}(s) \right )\frac{k_4^{\mu_1}k_2^{\mu_3}}{s^2}
    \right ] ,
    \\
   hh\rightarrow hh
   &:& &  & 4&\left [
  \left ( \frac{1}{3} \left ( \Delta \amp_{00}(s)+2\Delta \amp_{20}(s)
  \right )
	  \right . \right . \notag\\ &&&& &  \left . \left .
	  -\frac{10}{3}\left ( \Delta \amp_{02}(s)+2\Delta
    \amp_{22}(s) \right )
    \right ) \frac{\left( k_1 \cdot k_2\right )\left( k_3 \cdot k_4 \right ) }{s^2} \right .
  \\
  &&&& & \left .+ 5 \left (\Delta \amp_{02}(s)+2\Delta \amp_{22}(s) \right )
  \frac{\left( k_1 \cdot k_4\right )\left( k_2 \cdot k_3 \right )+ \left( k_1 \cdot k_4\right )\left( k_2 \cdot k_3 \right )}{s^2} \right ] \notag.
\end{alignat}
\end{subequations}
\bibliographystyle{unsrt}

\end{document}